\documentclass[]{jfm}

\usepackage{amsmath, amssymb} 
\usepackage{array}            
\usepackage{booktabs}  

\usepackage{graphicx}
\usepackage{newtxtext}
\usepackage{newtxmath}

\usepackage{makecell}
\usepackage{natbib}
\usepackage{hyperref}
\usepackage{wasysym}
\usepackage{subfigure}	
\usepackage{verbatim}
\usepackage{xcolor}
\hypersetup{
	colorlinks = true,
	urlcolor   = blue,
	citecolor  = black,
}

\newcommand{\RomanNumeralCaps}[1]

\title{Spatial instability analysis and mode transition of a viscoelastic jet in a co-flowing gas stream}

\author{Jiawei Li\aff{1}\thanks{The authors contribute equally in this work}, 
	Ming Wang\aff{2}\footnotemark[1],
	Kai Mu\aff{2}\corresp{\email{mukai@ustc.edu.cn}}, 
	Zhaodong Ding\aff{1,3}\corresp{\email{dingzhd@imu.edu.cn}}
    and Ting Si\aff{2}
}

\affiliation{\aff{1}School of Mathematical Science, Inner Mongolia University, Hohhot, Inner Mongolia 010021, PR China
	\aff{2}State Key Laboratory of High Temperature Gas Dynamics / Department of Modern Mechanics, University of Science and Technology of China, Hefei 230026, PR China
	\aff{3}Inner Mongolia Key Laboratory of Mathematical Modeling and Scientific Computing, Hohhot, Inner Mongolia 010021, PR China}

\begin{document}
\maketitle

\begin{abstract}
Spatial linear instability analysis is employed to investigate the instability of a viscoelastic liquid jet in a co-flowing gas stream. 
The theoretical model incorporates a non-uniform axial base profile represented by a hyperbolic tangent, capturing the shear layer. The Oldroyd-B model discretized with Chebyshev polynomials is employed, and energy budget analysis is used to interpret underlying mechanisms.
At low Weber numbers, the jet evolves axisymmetrically and the instability is governed by interfacial gas-pressure fluctuations; as the Weber number increases, the growing inertia drives a transition of the predominant mode from axisymmetric to helical. 
At weak elasticity, the instability is also primarily governed by gas-pressure fluctuations.
As elasticity increases, the predominant mode transitions from axisymmetric to helical.
This transition is accompanied by a migration of disturbance structures from the interface toward the jet interior and an enhanced coupling between velocity perturbation and the basic flow.
These trends reveal a new predominant instability mechanism---the elasticity-enhanced shear-driven instability (ESI)---which is distinct from capillary or Kelvin–Helmholtz instabilities in Newtonian jets.
A $\We$--$\El$ phase diagram delineates the boundary between predominant modes, and experimental results obtained in a flow-focusing configuration validate the theoretical predictions. 
Compared with temporal stability results, the spatial framework—by directly resolving the convective downstream amplification of disturbances—achieves quantitative agreement with experiments and highlight the superiority of spatial instability analysis in capturing the dynamics of strongly convective, non-parallel jet flows. 
These findings provide mechanistic insight into viscoelastic jet instabilities and offer guidance for applications involving droplet and fiber formation in co-flow systems.
\end{abstract}

\begin{keywords}
	viscoelastic jets, multiphase flow, spatial instability analysis
\end{keywords}

%

\section{Introduction}
\label{sec:introduction}

Among various strategies for generating microjets, flow focusing, a representative capillary-driven technique, allows for the stable formation of microjets that break into highly monodisperse droplets or microfibers \citep{GananCalvo2013}. In flow focusing, a high-speed laminar outer stream elongates the liquid meniscus into a cusp, from whose apex a slender microthread is steadily drawn once the aerodynamic suction overcomes interfacial stresses. The ensuing jet diameter and breakup route are set by a balance among capillary pressure, viscous dissipation, inertia, and the imposed pressure/flow-rate conditions. Within appropriate parameter regimes, this configuration supports dripping, jetting, or tip-streaming: the latter yields droplets far smaller than the nozzle scale and is stabilized by gentle interfacial tractions of the focusing flow \citep{Montanero2020}. In gas-focused flow focusing specifically, a robust steady microthread emerges when the pressure drop and focusing geometry suppress large gas-pressure fluctuations, leading to axisymmetric, nearly monodisperse sprays; increasing outer forcing promotes sinuous (non-axisymmetric) responses and polydispersity \citep{GananCalvo1998, Ganan-Calvo1999}. Owing to its simplicity, robustness, and resistance to clogging, flow focusing has become a versatile platform for fundamental studies of jet instability and droplet formation \citep{GananCalvo2013}.

The study of liquid jet instability has a long history, beginning with \citet{Plateau1873} and \citet{Rayleigh1878}. While Plateau’s experiments revealed that a cylindrical liquid column breaks up under surface-tension-driven disturbances, Rayleigh established the theoretical foundation through a linear instability analysis, showing that axisymmetric capillary disturbances grow most rapidly at a wavelength about $9$ times the jet radius ($\approx 4.5$ diameters). Viscous effects were introduced by \citet{Weber1931}, who demonstrated that viscosity stabilizes jets by damping surface perturbations, a result later incorporated into Chandrasekhar’s monograph \citep{Chandrasekhar1961}. These classical theories were formulated within a temporal framework, assuming disturbances grow in time at fixed positions, which is insufficient for free jets where instabilities convect downstream. \citet{Keller1973} were among the first to apply spatial instability theory, showing that although the most unstable wavelengths are close to those predicted by temporal instability analysis, only the spatial framework correctly describes downstream amplification. \citet{Garg1981} extended this approach to weakly non-parallel shear flows using local approximations, obtaining a critical Reynolds number consistent with experiments and thereby validating spatial theory in non-parallel configurations. \citet{Lin1987} applied spatial instability analysis to atomization, showing that breakup results from a competition between capillary instability and Kelvin--Helmholtz instability, with the dominant mechanism determined by the Weber and Reynolds numbers. \citet{Lin1998} further clarified the role of interfacial shear through energy budget analysis, showing that it suppresses Rayleigh modes but drives Taylor modes in confined coflowing jets.

The instability and mode transitions of Newtonian liquid jets under flow focusing have been widely explored, yielding a comprehensive understanding of jet modes, breakup dynamics, and parametric regimes. Early experiments demonstrated that a high-speed gas stream can stabilize slender axisymmetric microthreads, which eventually undergo regular Rayleigh-type breakup into nearly monodisperse droplets \citep{GananCalvo1998}. Subsequent studies mapped multiple flow modes in the parameter space defined by the liquid flow rate and the applied gas pressure drop, identifying transitions between dripping, axisymmetric jetting, and more complex non-axisymmetric states, with the breakup wavelength and length scales rationalized using temporal and spatio-temporal instability analyses \citep{Si2009}. Spatial theory has proven particularly valuable in capturing predominant mode boundaries: it reveals that non-axisymmetric disturbances may dominate over axisymmetric ones at sufficiently high Weber numbers, in better agreement with experimental observations than temporal theory \citep{Si2010}. The dripping--jetting transition has also been rationalized in terms of convective-to-absolute instability, with critical Weber numbers delineating stable jetting from dripping regimes in compound and coaxial liquid--gas configurations \citep{Herrada2010}.

In practice, many technologically relevant processes operate with non-Newtonian viscoelastic jets—for example, inkjet printing, solution spinning, electrospinning, and the fabrication of micro- and nanofibers. Unlike Newtonian liquids, which are characterized by a simple linear constitutive relation, viscoelastic fluids display richer rheological behavior owing to fluid elasticity. This elasticity originates from the conformational dynamics of polymer chains, most notably their stretching and relaxation under deforming flows \citep{Larson2005,Schroeder2018,Ewoldt2022}. The degree of elasticity is commonly quantified by the Weissenberg number ($\Wi$) or the Deborah number ($\De$), which compare the polymer relaxation time to a characteristic flow time scale, with $\Wi$ typically used for steady shear and extensional flows and $\De$ for unsteady or oscillatory flows. To capture these effects at the continuum level, a variety of constitutive models have been developed. The Oldroyd-B model is widely used for dilute polymer solutions, though it becomes limited at high deformation rates because it predicts unbounded extensional viscosity \citep{Larson1992}. The FENE-P model addresses this limitation by accounting for the finite extensibility of polymer chains \citep{Bird1987}, while the Phan-Thien-Tanner (PTT) model effectively describes shear thinning and finite extensional viscosity in concentrated polymer solutions and melts \citep{Thien1977}. A direct macroscopic manifestation of these constitutive relations is the emergence of finite normal stress differences (see \S\ref{sec:base flow}), in sharp contrast to the zero normal stresses characteristic of simple Newtonian fluids.

Linear instability theory was subsequently extended to viscoelastic jets. \citet{Goldin1969} conducted a linear instability analysis on viscoelastic jets and suggested that elasticity could accelerate breakup in weakly elastic solutions, while subsequent experiments by \citet{Gordon1973} indicated the opposite trend, showing delayed breakup. To reconcile this discrepancy, \citet{Goren1982} introduced the concept of unrelaxed axial tension, showing that stress history and finite relaxation times can stabilize jets and reconcile theory with experiments. \citet{Ruo2011} extended the linear instability analysis to three-dimensional disturbances and more complex rheological models, showing that viscoelasticity can induce oscillatory non-axisymmetric modes. More recent advances have emphasized the importance of non-uniform base flows and gas–liquid interactions. For instance, \citet{Ding2022} showed that unrelaxed stresses interact with perturbation velocities through coupling terms, leading to a non-monotonic dependence of growth rates on elasticity, while energy budget analyses have revealed the competing roles of surface tension, elastic tension, and aerodynamic forcing. 
Recent studies have also highlighted a distinct class of linear instabilities associated with polymer stress diffusion. \citet{Beneitez2023} showed that the inclusion of a small but finite stress-diffusion term in inertialess shear flows can destabilize the laminar state, giving rise to the so-called polymer diffusive instability (PDI). This instability typically manifests as short-wavelength, wall-localized modes and persists even in the singular limit of vanishing diffusivity. Follow-up work demonstrated that inertia can significantly amplify PDI and broaden its existence region, and may trigger elastic or elasto-inertial turbulence in simulations \citep{Couchman2024,Beneitez2024}.

\begin{figure}
	\centerline{\includegraphics[width=0.75\linewidth]{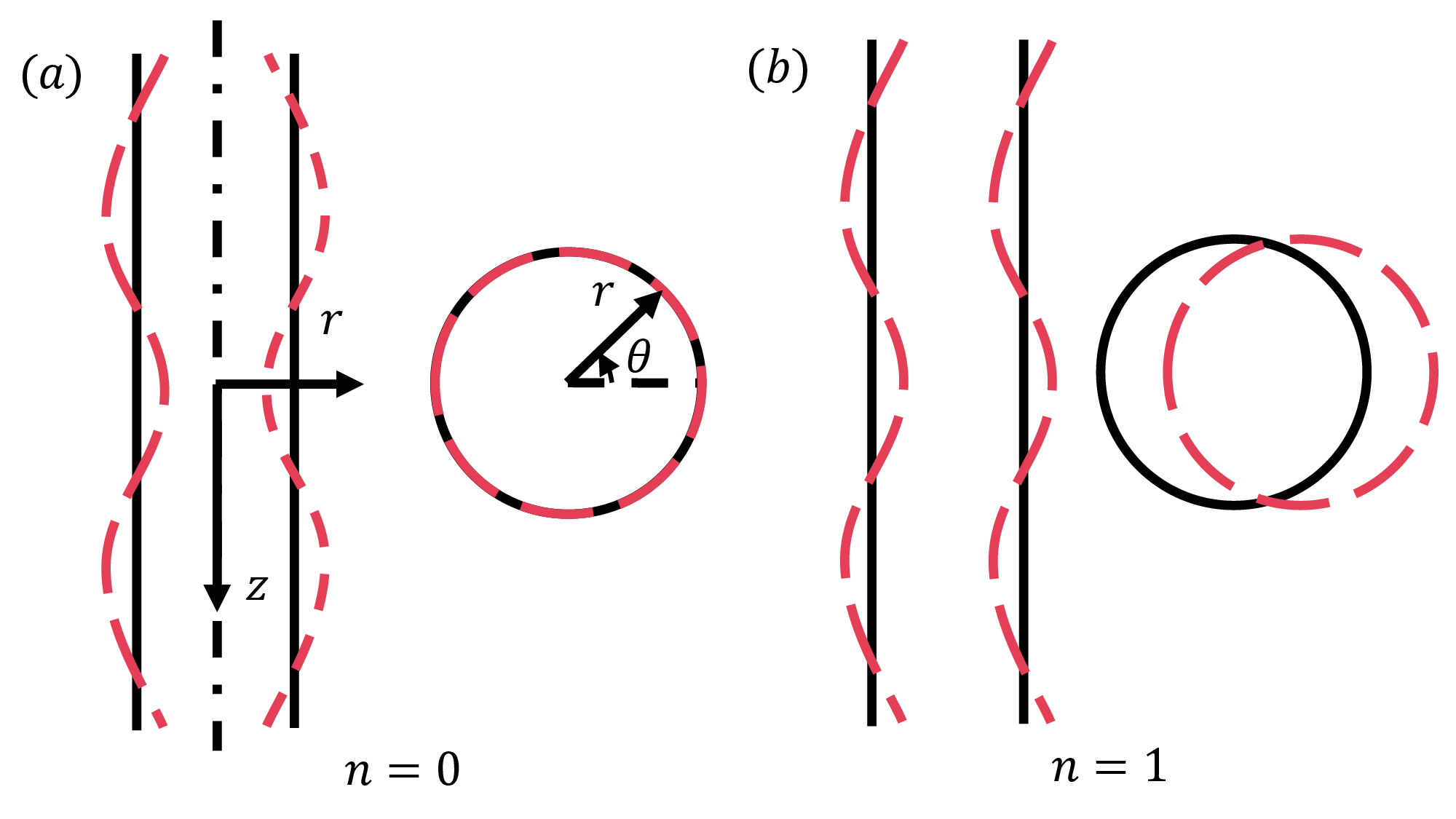}}
	\caption{Sketches of the azimuthal modes $n=0$ and $n=1$. 
		The black solid lines and red dotted lines represent the interface profiles before and after perturbation, respectively. (a) The axisymmetric ($n=0$) mode in the $r$--$z$ and $r$--$\theta$ planes. (b) The non-axisymmetric ($n=1$) mode in the $r$--$z$ and $r$--$\theta$ planes, which induces a sinuous or corkscrew-like displacement of the jet centerline and is hereafter referred to as the helical mode.}
	\label{fig:1}
\end{figure}

Despite extensive studies, most analyses of viscoelastic jet instability have been carried out within temporal frameworks, which are insufficient to describe the downstream development observed in experiments. In this work, we adopt a spatial instability analysis and focus on the influence of elasticity on the transition between different instability modes and the underlying mechanisms.
Representative jet profiles for the $n=0$ and $n=1$ modes are shown in figure~\ref{fig:1}: in the axisymmetric mode, disturbances act directly on the interface, producing varicose-like thickness fluctuations while preserving a circular cross-section along the axis; in contrast, the non-axisymmetric mode retains a nearly circular cross-section but induces a sinuous, corkscrew-like displacement of the jet centerline, manifesting as a helical trajectory. Hereafter, we refer to the non-axisymmetric mode ($n=1$) as the helical mode for brevity.

The paper is organized as follows.
Section~\ref{sec:theoretical model} formulates the viscoelastic liquid--gas coaxial jet, presenting the governing equations, base flow, numerical methodology, and the associated energy-budget analysis.
Section~\ref{Sec:results and discussion} discusses the effects of elasticity and Weber number, as well as the associated mode transitions, including comparisons with flow-focusing experiments.
Finally, Section~\ref{Sec:conclusion} summarizes the main findings and outlines future work.

\section{Theoretical model of viscoelastic liquid-gas coaxial jet}\label{sec:theoretical model}

\begin{figure}
	\centerline{\includegraphics[width=0.75\linewidth]{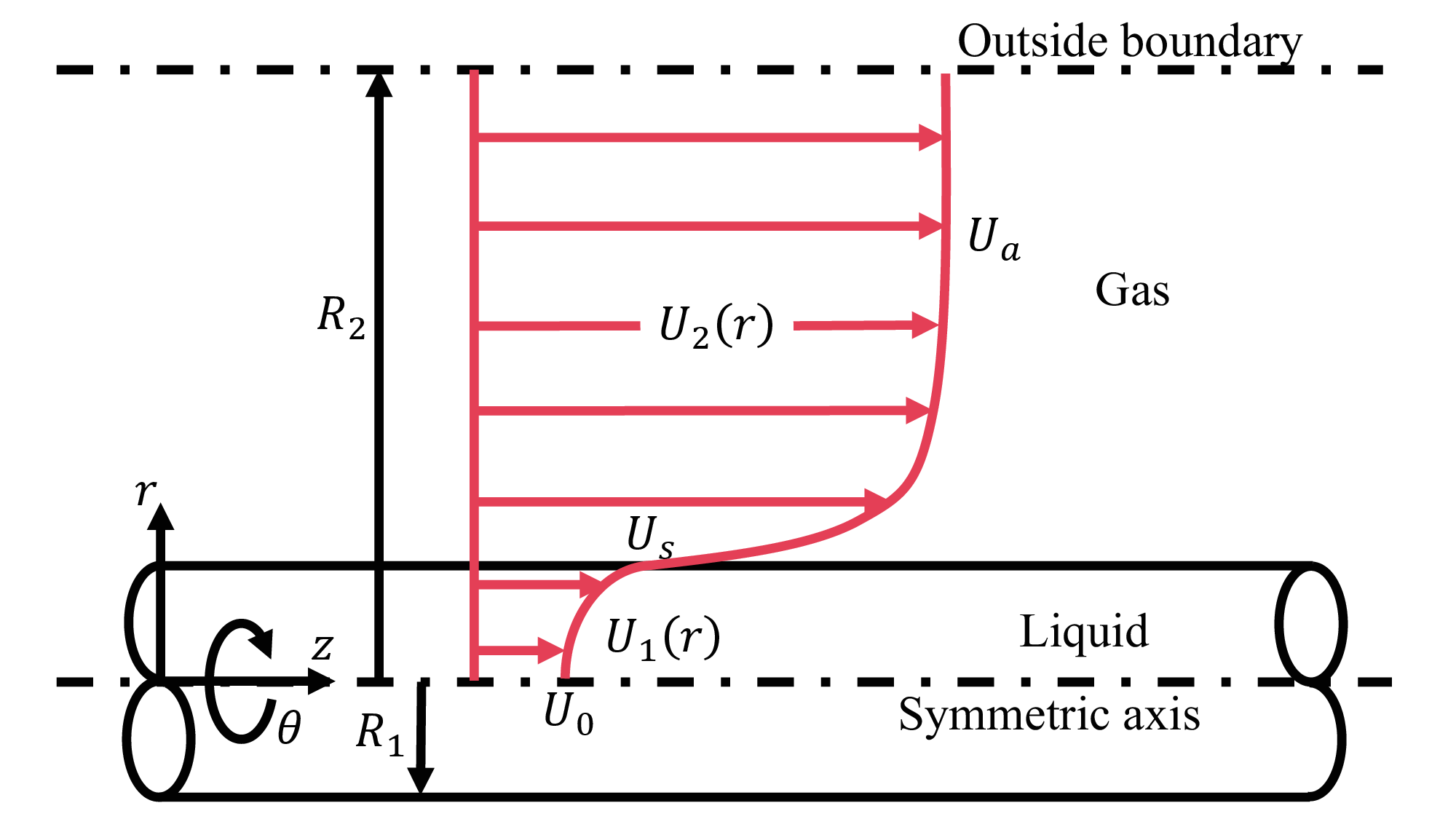}}
	\caption{Schematic of the theoretical model for viscoelastic liquid–gas coaxial jets, with the base flow represented by a hyperbolic-tangent velocity profile.}
	\label{fig:2}
\end{figure}

A theoretical model is developed for a viscoelastic liquid jet with radius $R_1$ and density $\rho_1$, surrounded by a coaxial gaseous phase of radius $R_2$ and density $\rho_2$. The system is formulated in cylindrical coordinates $(z, r, \theta)$, where $z$, $r$, and $\theta$ denote the axial, radial, and azimuthal directions, respectively. The theoretical model is sketched in figure~\ref{fig:2}. The viscoelastic behavior of the liquid jet is characterized by two viscosity components: the solvent viscosity $\mu_{\!s}$ and the polymer viscosity $\mu_{\!p}$, yielding a total effective viscosity $\mu_1 = \mu_{\!s} + \mu_{\!p}$. The viscosity of the surrounding gas is denoted by $\mu_2$.

To nondimensionalize the governing equations, the characteristic scales are selected as follows: length scale $R_1$, velocity scale $U_0$ (the base flow velocity at the jet centerline), time scale $R_1/U_0$, and pressure scale $\rho_1 U_0^2$. Based on these scales, the following dimensionless parameters are introduced: the Reynolds number $\Rey = \rho_1 U_0 R_1 / \mu_1$, representing the ratio of inertial to viscous forces; the Weber number $\We = \rho_1 U_0^2 R_1 / \sigma$, where $\sigma$ is the liquid-gas interfacial tension; the Weissenberg number $\Wi = \lambda U_0 / R_1$, reflecting the ratio of the polymer relaxation time $\lambda$ to the flow time scale; the density ratio $Q = \rho_2 / \rho_1$; the radius ratio $a = R_2 / R_1$; the viscosity ratio $N = \mu_2 / \mu_1$; and the solvent viscosity fraction $X = \mu_s / \mu_1$. Notably, when $Wi=0$ or $X=1$, the polymeric contribution vanishes and the model reduces to the Newtonian liquid jet case.

\subsection{Governing equations}

The motion of the viscoelastic liquid jet is governed by the dimensionless incompressible Navier-Stokes equations. The continuity and momentum equations for the liquid as following:
\begin{equation}\label{eq:2.1}
	\boldsymbol{\nabla} \cdot \boldsymbol{V}_1 = 0,
\end{equation}
\begin{equation}\label{eq:2.2}
	\frac{\partial \boldsymbol{V}_1}{\partial t} 
	+ \boldsymbol{V}_1 \cdot \boldsymbol{\nabla} \boldsymbol{V}_1
	= -\boldsymbol{\nabla} p_1 + \boldsymbol{\nabla}\cdot \boldsymbol{T}.
\end{equation}
Here, $\boldsymbol{V}_1 = u_1 \boldsymbol{e}_{\!z} + v_1 \boldsymbol{e}_{\!r} + w_1 \boldsymbol{e}_{\!\theta}$ and $p_1$ denote the velocity and pressure of the liquid jet.  
The extra-stress tensor $\boldsymbol{T}$ is decomposed as
\begin{equation}\label{eq:2.3}
	\boldsymbol{T} = \boldsymbol{T}_{\!S} + \boldsymbol{T}_{\!P},
\end{equation}
with the solvent stress
\begin{equation}\label{eq:2.4}
	\boldsymbol{T}_{\!S} = \frac{X}{\Rey}\left(\boldsymbol{\nabla}\boldsymbol{V}_1 
	+ \boldsymbol{\nabla}\boldsymbol{V}_1^{\mathrm{T}}\right).
\end{equation}
The polymeric stress $\boldsymbol{T}_{\!P}$ follows the Oldroyd-B constitutive equation \citep{Bird1987}:
\begin{equation}\label{eq:2.5}
	\boldsymbol{T}_{\!P} 
	+ \Wi\left(
	\frac{\partial \boldsymbol{T}_{\!P}}{\partial t} 
	+ \boldsymbol{V}_1 \cdot \boldsymbol{\nabla}\boldsymbol{T}_{\!P}
	- \boldsymbol{\nabla}\boldsymbol{V}_1^{\mathrm{T}} \cdot \boldsymbol{T}_{\!P}
	- \boldsymbol{T}_{\!P} \cdot \boldsymbol{\nabla}\boldsymbol{V}_1
	\right) 
	= \frac{1-X}{\Rey}\left(\boldsymbol{\nabla}\boldsymbol{V}_1 
	+ \boldsymbol{\nabla}\boldsymbol{V}_1^{\mathrm{T}}\right).
\end{equation}

The surrounding gas, treated as an incompressible Newtonian fluid, satisfies
\begin{equation}\label{eq:2.6}
	\boldsymbol{\nabla}\cdot \boldsymbol{V}_2 = 0,
\end{equation}
\begin{equation}\label{eq:2.7}
	\frac{\partial \boldsymbol{V}_2}{\partial t} 
	+ \boldsymbol{V}_2 \cdot \boldsymbol{\nabla}\boldsymbol{V}_2
	= -\frac{1}{Q}\,\boldsymbol{\nabla}p_2 
	+ \frac{N}{Q \Rey}\,\boldsymbol{\nabla}^2 \boldsymbol{V}_2,
\end{equation}
where $\boldsymbol{V}_2 = u_2 \boldsymbol{e}_{\!z} + v_2 \boldsymbol{e}_{\!r} + w_2 \boldsymbol{e}_{\!\theta}$,  
$p_2$ is the gas pressure, $Q=\rho_{2}/\rho_1$ is the density ratio, and $N=\mu_2/\mu_1$ the viscosity ratio.

Boundary conditions are required at the axis $r=0$, the outer boundary $r=a$ ($a=R_2/R_1$), and the liquid--gas interface $r=1+\eta$, where $\eta$ is the interfacial displacement. 

At $r=0$, the bounded condition and the single-valued condition should be satisfied as following:
\begin{equation}\label{eq:2.8}
	\boldsymbol{V}_1 < \infty, \quad
	\frac{\partial \boldsymbol{V}_1}{\partial \theta} = 0, \quad
	\frac{\partial p_1}{\partial \theta} = 0.
\end{equation}
It should be noted that different azimuthal modes require different treatments at $r=0$; the corresponding mode-dependent formulations are given in equation~(\ref{eq:A30}).

At $r=a$, the outer boundary conditions impose bounded velocity and pressure fields:
\begin{equation}\label{eq:2.9}
	\boldsymbol{V}_2 < \infty, \quad p_2 < \infty.
\end{equation}

At the liquid-gas interface, velocity continuity, kinematic, and dynamic conditions hold:
\begin{equation}\label{eq:2.10}
	\boldsymbol{V}_1 = \boldsymbol{V}_2,
\end{equation}
\begin{equation}\label{eq:2.11}
	v_1 = \left(\frac{\partial}{\partial t} + \boldsymbol{V}_1\cdot\boldsymbol{\nabla}\right)\eta,
\end{equation}
\begin{equation}\label{eq:2.12}
	\left(\boldsymbol{T}_2 - \boldsymbol{T}_1\right)\cdot \boldsymbol{n}
	= \frac{1}{\We}\,(\boldsymbol{\nabla}\cdot\boldsymbol{n})\,\boldsymbol{n},
\end{equation}
where $\boldsymbol{T}_1 = -p_1\boldsymbol{I} + \boldsymbol{T}_{\!S} + \boldsymbol{T}_{\!P}$ is the liquid stress, and 
$\boldsymbol{T}_2 = -p_2\boldsymbol{I} + N/\Rey\left(\boldsymbol{\nabla}\boldsymbol{V}_2 + \boldsymbol{\nabla}\boldsymbol{V}_2^{\mathrm{T}}\right)$ 
is the gas stress. Here, $\boldsymbol{I}$ denotes the identity tensor and $\boldsymbol{n}$ the outward-pointing unit normal to the interface.

\subsection{Basic velocity profiles and stress tensor}\label{sec:base flow}

To perform a linear instability analysis, it is first necessary to prescribe a physically reasonable base flow that captures the key shear interactions between the liquid and gas. We consider an incompressible, axisymmetric, unidirectional state, written as $\boldsymbol{U}_{\!i} = (0,0,U_{\!i}(r))$,
where $U_{\!i}(r)$ is the axial velocity profile in the radial direction $r$, with $i=1,2$ denoting the liquid and gas phases. Previous theoretical works have employed approximate profiles to describe the interfacial shear layer, including error-function forms \citep{Yecko2002,Matas2018,Xu2024} and hyperbolic-tangent profiles \citep{Ganan-Calvo2006,Si2009,Ding2022}.  

Here we adopt a hyperbolic-tangent base profile,
\begin{subequations}\label{eq:2.13}
	\begin{align}
		U_1(r) &= (U_{\!s}-1)\tanh\!\left[\frac{K}{U_{\!s}-1}(r-1)\right] + U_{\!s}, 
		\quad r \leq 1, \label{eq:2.13a}\\
		U_2(r) &= (U_{\!a}-U_{\!s})\tanh\!\left[\frac{K}{N(U_{\!a}-U_{\!s})}(r-1)\right] + U_{\!s}, 
		\quad r > 1. \label{eq:2.13b}
	\end{align}
\end{subequations}
where $U_{\!s}$ is the interfacial velocity, $U_{\!a}$ the axial velocity at the outer boundary $r=a$, and $K$ the slope of the liquid profile. In flow-focusing configurations, the gas-to-liquid momentum ratio is typically close to unity, giving the approximation $U_{\!a} \approx Q^{-1/2}$ \citep{Si2009}.  

By substituting (\ref{eq:2.13}) into (\ref{eq:2.5}), the base-state polymeric stress tensor $\boldsymbol{\Gamma}$ is obtained as
\begin{subequations}\label{eq:2.14}
	\begin{align}
		\Gamma_{\!rr} &= \Gamma_{\!r\theta} = \Gamma_{\!\theta\theta} = \Gamma_{\!\theta z} = 0, 
		\label{eq:2.14a}\\
		\Gamma_{\!rz} &= \frac{1-X}{\Rey}\,\frac{\mathrm{d}U_1}{\mathrm{d}r}, \label{eq:2.14b}\\
		\Gamma_{\!zz} &= \frac{2(1-X)}{\Rey}\,\Wi\left(\frac{\mathrm{d}U_1}{\mathrm{d}r}\right)^{2}. \label{eq:2.14c}
	\end{align}
\end{subequations}

As a consequence of the steady and axisymmetric base flow, most components of the polymeric stress tensor $\boldsymbol{\Gamma}$ vanish, leaving only two non-zero terms. Specifically, the unrelaxed elastic stress components in the axial direction, namely $\Gamma_{\!zz}$ and $\Gamma_{\!rz}$, remain finite, whereas the radial and azimuthal components ($\Gamma_{\!rr}$, $\Gamma_{\!\theta\theta}$, $\Gamma_{\!r\theta}$, and $\Gamma_{\!\theta z}$) vanish identically. Here, $\Gamma_{\!zz}$ represents the unrelaxed axial elastic tension associated with the stretching of polymer chains along the flow direction, whereas $\Gamma_{\!rz}$ corresponds to the unrelaxed shear elastic tension induced by the radial gradient of the axial velocity $U_1(r)$. These two non-zero components capture the essential elastic effects of the viscoelastic fluid and play a central role in determining the instability and dynamics of the jet.

\subsection{Linear instability analysis}\label{sec:linear_instability}

In the present analysis, each flow variable is decomposed into a steady base state and a small perturbation:
\begin{subequations}\label{eq:2.15}
	\begin{align}
		\boldsymbol{V}_{i} &= \boldsymbol{U}_{i} + \tilde{\boldsymbol{V}}_{i}, \label{eq:2.15a}\\
		p_{i} &= P_{i} + \tilde{p}_{i},
		\label{eq:2.15b}\\
		\boldsymbol{T}_{\!P} &= \boldsymbol{\Gamma} + \tilde{\boldsymbol{T}}_{\!P}, \label{eq:2.15c}
	\end{align}
\end{subequations}
where $\boldsymbol{U}_{i}$, $P_{i}$, and $\boldsymbol{\Gamma}$ represent the base flow velocity, pressure, and polymeric stress tensor, respectively.

The disturbance velocity vector in cylindrical coordinates is written as
\begin{equation}\label{eq:2.16}
	\tilde{\boldsymbol{V}}_{\!i} 
	= \tilde{u}_{i} \boldsymbol{e}_{\!z} 
	+ \tilde{v}_{i} \boldsymbol{e}_{\!r} 
	+ \tilde{w}_{i} \boldsymbol{e}_{\!\theta},
\end{equation}
and the perturbation polymeric stress tensor is expanded as
\begin{equation}\label{eq:2.17}
	\begin{aligned}
		\tilde{\boldsymbol{T}}_{\!P} 
		&= \tilde{\tau}_{rr}\,\boldsymbol{e}_{\!r}\boldsymbol{e}_{\!r} 
		+ \tilde{\tau}_{r\theta}\,(\boldsymbol{e}_{\!r}\boldsymbol{e}_{\!\theta} + \boldsymbol{e}_{\!\theta}\boldsymbol{e}_{\!r}) 
		+ \tilde{\tau}_{rz}\,(\boldsymbol{e}_{\!r}\boldsymbol{e}_{\!z} + \boldsymbol{e}_{\!z}\boldsymbol{e}_{\!r}) \\
		&\quad + \tilde{\tau}_{\theta\theta}\,\boldsymbol{e}_{\!\theta}\boldsymbol{e}_{\!\theta}
		+ \tilde{\tau}_{\theta z}\,(\boldsymbol{e}_{\!\theta}\boldsymbol{e}_{\!z} + \boldsymbol{e}_{\!z}\boldsymbol{e}_{\!\theta}) 
		+ \tilde{\tau}_{zz}\,\boldsymbol{e}_{\!z}\boldsymbol{e}_{\!z},
	\end{aligned}
\end{equation}
where $\tilde{u}_{i}$, $\tilde{v}_{i}$, and $\tilde{w}_{i}$ are the axial, radial, and azimuthal velocity perturbations, and each $\tilde{\tau}_{ij}$ denotes the corresponding component of the perturbation polymeric stress tensor. Substituting (\ref{eq:2.15}) into (\ref{eq:2.1})--(\ref{eq:2.12}) and neglecting nonlinear terms, we obtain a set of linearized equations describing the disturbance evolution, the detailed forms of which are given in Appendix~\ref{app:linearized_equations}.

To exploit the spatial homogeneity in the axial ($z$) and azimuthal ($\theta$) directions, each perturbation quantity is expressed in normal-mode form:
\begin{equation}\label{eq:2.18}
	\{\tilde{u}_i, \tilde{v}_i, \tilde{w}_i, \tilde{p}_i, \tilde{\boldsymbol{T}}_{\!P}, \eta\}
	= \{\hat{u}_i(r), \hat{v}_i(r), \hat{w}_i(r), \hat{p}_i(r), \hat{\boldsymbol{T}}_{\!P}(r), \hat{\eta}\}\,
	e^{\,\mathrm{i}(kz+n\theta-\omega t)},
\end{equation}
where $n$ is the prescribed azimuthal wavenumber characterizing the circumferential structure of the disturbance, $\omega$ is the imposed real frequency, and $k=k_{r}+\mathrm{i} k_i$ is the complex wavenumber to be solved as an eigenvalue. The real part $k_r$ determines the axial wavelength, while the imaginary part $-k_i$ measures the spatial amplification or decay. A negative imaginary part ($-k_i > 0$) indicates exponential downstream growth, whereas a positive one ($-k_i < 0$) corresponds to decay. In this study, attention is restricted to the dominant azimuthal modes $n=0$ (axisymmetric) and $n=1$ (helical), which play central roles in jet instability according to both theory and experiment.

It is worth recalling that hydrodynamic instability problems can be formulated either temporally or spatially. In a temporal framework, $k$ is prescribed real and $\omega$ is solved as complex, whereas in a spatial framework, $\omega$ is prescribed real and $k$ is solved as complex. The latter, adopted here, is more appropriate for open flows such as jets, since it directly captures the downstream evolution of disturbances and has been shown to yield predictions in good agreement with experiments \citep{Garg1981,Si2010}.

To systematically examine the effect of elasticity, we introduce the elasticity number \(\El = \Wi/\Rey = \lambda \mu_1 /\rho_1 R_1^2\), defined as the ratio of the elastic to viscous time scales \citep{Brenn2000,Khalid2021}. In the following, we formulate the instability problem by substituting Eq.~(\ref{eq:2.18}) into Eq.~(\ref{eq:A1})--Eq.~(\ref{eq:A15}), and their complete forms are provided in Appendix~\ref{app:linearized_equations}.

Equations (\ref{eq:A16})–(\ref{eq:A37}) constitute an eigenvalue problem for the disturbance fields at a
prescribed real frequency $\omega$ and azimuthal mode number $n$:
\begin{equation}\label{eq:2.19}
	[\boldsymbol{A}]\,\boldsymbol{x} = k\,[\boldsymbol{B}_1]\,\boldsymbol{x} + k^2\,[\boldsymbol{B}_2]\,\boldsymbol{x},
\end{equation}
where $\boldsymbol{x}=\left(\hat{u}_1,\hat{v}_1,\hat{w}_1,\hat{p}_1,\hat{\tau}_{rr},
\hat{\tau}_{r\theta},\hat{\tau}_{rz},\hat{\tau}_{\theta \theta},
\hat{\tau}_{\theta z},\hat{\tau}_{zz},\hat{u}_2,\hat{v}_2,\hat{w}_2,\hat{p}_2,\eta\right)^{\mathrm T}$,
and $\boldsymbol{A}$, $\boldsymbol{B}_1$, and $\boldsymbol{B}_2$ are the corresponding coefficient matrices.

Since equations (\ref{eq:2.19})
are linear in the frequency $\omega$ but quadratic in the wavenumber $k$, 
we introduce an auxiliary set of variables
\begin{equation}\label{eq:2.20}
	\boldsymbol{x}'=k\,\boldsymbol{x},
\end{equation}
which transform the quadratic eigenvalue problem (QEP) into an equivalent 
linear eigenvalue problem for wavenumber $k$:
\begin{equation}\label{eq:2.21}
	\begin{bmatrix}
		\boldsymbol{A} & \boldsymbol{0}\\
		\boldsymbol{0} & \boldsymbol{I}
	\end{bmatrix}
	\begin{bmatrix}
		\boldsymbol{x}\\
		\boldsymbol{x}'
	\end{bmatrix}
	=
	k
	\begin{bmatrix}
		\boldsymbol{B}_1 & \boldsymbol{B}_2\\
		\boldsymbol{I} & \boldsymbol{0}
	\end{bmatrix}
	\begin{bmatrix}
		\boldsymbol{x}\\
		\boldsymbol{x}'
	\end{bmatrix},
\end{equation}
where $\boldsymbol{I}$ denotes the identity matrix.

To numerically solve the eigenvalue problem 
(\ref{eq:2.21}), 
we employ a spectral collocation method based on Chebyshev polynomials. 
The Chebyshev spectral method is well known for its accuracy and efficiency 
in computing eigenvalues in hydrodynamic instability problems  \citep{Boyd1999,Schmid2001,Ye2017,Chaudhary2021}. A detailed description of the numerical formulation and validation 
is provided in Appendix~\ref{app:method_validation}.

\subsection{Energy budget}\label{sec:energy budget}

While eigenvalue analysis provides growth rates and frequencies of disturbances, 
it does not directly reveal the physical mechanisms responsible for instability. 
To gain deeper insight into the origin and evolution of disturbances, 
it is necessary to examine how perturbation kinetic energy is produced, transferred, 
and dissipated through various physical processes. 
For this purpose, the energy budget method has been widely applied 
in temporal instability analyses to clarify the evolution of disturbance energy 
\citep{Lin1998,Li2011,Mu2021}. In the present work, this approach is applied within the framework of spatial instability analysis.

The derivation proceeds as follows. 
The momentum equations are projected onto the dimensionless disturbance velocity 
$\tilde{\boldsymbol{V}}_1$, and the resulting expression is integrated over the control 
volume $V$ and one oscillation period $T = 2\pi/\omega$. 
An additional averaging over a single axial wavelength $\lambda = 2\pi/k_r$ and one period $T$ then yields the governing energy budget relation,

\begin{equation}\label{eq:2.22}
	\begin{aligned}
		\int\limits_{T} \int\limits_{V} 
		\left(\partial_t + \boldsymbol{U}_1 \cdot \boldsymbol{\nabla} \right)e 
		\,\mathrm{d}V \,\mathrm{d}t 
		= -\int\limits_{T} \int\limits_{V} 
		\boldsymbol{\tilde{V}}_1 \cdot 
		\left(\boldsymbol{\tilde{V}}_1 \cdot \boldsymbol{\nabla}\boldsymbol{U}_1 \right) 
		\,\mathrm{d}V \,\mathrm{d}t
		- \int\limits_{T} \int\limits_{S} 
		\tilde{p}_1 \,\boldsymbol{\tilde{V}}_1 \cdot \boldsymbol{n} 
		\,\mathrm{d}S \,\mathrm{d}t \\
		+ \int\limits_{T} \int\limits_{V} 
		\boldsymbol{\tilde{V}}_1 \cdot 
		\left(\boldsymbol{\nabla} \cdot \boldsymbol{\tilde{T}}_{\!S} \right) 
		\,\mathrm{d}V \,\mathrm{d}t
		+ \int\limits_{T} \int\limits_{V} 
		\boldsymbol{\tilde{V}}_1 \cdot 
		\left(\boldsymbol{\nabla} \cdot \boldsymbol{\tilde{T}}_{\!P} \right) 
		\,\mathrm{d}V \,\mathrm{d}t,
	\end{aligned}
\end{equation}
where $e = \boldsymbol{\tilde{V}}_1 \cdot \boldsymbol{\tilde{V}}_1 / 2$, 
$S$ denotes the surface of the control volume, and $\boldsymbol{n}$ is the outward unit normal vector. 
By applying Gauss’s theorem together with the incompressibility condition, 
and incorporating the boundary conditions at $r=1$, 
the equation reduces to the compact budget form
\begin{equation}\label{eq:2.23}
	\begin{aligned}
		\mathrm{KE} 
		= \mathrm{REY} + \mathrm{PRL} + \mathrm{SHL} + \mathrm{NVL} 
		+ \mathrm{PRG} + \mathrm{NVG} + \mathrm{SHG} + \mathrm{ATG} \\
		+ \mathrm{TEL} + \mathrm{NEL} + \mathrm{SUT} + \mathrm{SHB} 
		+ \mathrm{DIS} + \mathrm{NE} + \mathrm{DIP},
	\end{aligned}
\end{equation}

Explicit expressions of each term are given in Appendix~\ref{app:energy_terms}. 
Within the spatial instability analysis, the disturbance kinetic energy KE 
represents the streamwise rate of change of perturbation energy. 
REY quantifies the energy exchange between the base flow and the disturbance 
via Reynolds stresses, while DIS accounts for viscous dissipation in the solvent. 
Elastic effects are characterized by NE, the work done by the unrelaxed axial elastic stress on the perturbed interface, and DIP, which represents the additional energy transfer mediated by polymeric stresses.  
Unlike temporal instability analysis, where disturbances are assumed periodic in the 
streamwise direction and boundary fluxes vanish, 
spatial instability analysis requires explicit treatment of fluxes at the upstream and downstream boundaries. 
Accordingly, PRL, SHL, and NVL denote, respectively, the work of pressure, tangential viscous stress, and normal viscous stress
across the upstream and downstream control surfaces of the liquid domain.
Similarly, TEL and NEL correspond to the work associated with tangential and normal elastic stresses across these control surfaces.
Additional contributions arise from the gas phase at the liquid–gas interface: 
PRG represents the work of gas pressure fluctuations, NVG represents the effect of 
normal viscous stress, and SHG and ATG correspond to the axial and azimuthal shear stresses, respectively. 
Surface-tension effects are represented by SUT, 
while SHB accounts for axial shear stress induced by base-flow distortion 
due to interfacial deformation. The physical interpretations of all energy-budget terms are summarized in Table~\ref{tab:1}.

\begin{table}
	\centering
	\footnotesize
	\begin{tabular}{ll @{\hspace{0.6cm}} ll}
		KE  & \makecell[l]{Rate of change of disturbance kinetic \\ \quad energy} 
		& PRG & \makecell[l]{Rate of work done by the gas pressure \\ \quad fluctuation on the liquid jet}
		\\
		REY & \makecell[l]{Energy exchange between the \\ \quad disturbance and basic flow} 
		& SHG & \makecell[l]{Rate of work done by axial gas shear \\ \quad stress}
		\\
		DIS & \makecell[l]{Rate of work done by the solvent \\ \quad viscous dissipation} 
		& ATG & \makecell[l]{Rate of work done by azimuthal gas shear \\ \quad stress}
		\\
		NE  & \makecell[l]{Rate of work done by the unrelaxed \\ \quad axial elastic stress} 
		& DIP & \makecell[l]{Rate of work done by the elastic stress}
		\\
		SUT & \makecell[l]{Rate of work done by the surface tension} & SHB & \makecell[l]{Rate of work done by the shear stress \\
			\quad associated with the basic flow distortion}
		\\
		NVL & \makecell[l]{Rate of work done by the normal viscous \\ \quad stresses at the top and the bottom of the \\ \quad control volume} & PRL & \makecell[l]{Rate of work done by the tangential \\ \quad viscous stresses at the top and the bottom \\ \quad of the control volume}
		\\
		NEL & \makecell[l]{Rate of work done by the normal elastic \\ \quad stresses at the top and the bottom of the \\ \quad control volume} & TEL & \makecell[l]{Rate of work done by the tangential \\ \quad elastic stresses at the top and the bottom \\ \quad of the control volume}
	\end{tabular}
	\caption{Physical interpretation of terms from the energy analysis.}
	\label{tab:1}
\end{table}

To remove the arbitrariness of individual terms, 
it is convenient to normalize by the relative rate of energy change 
\citep{Ye2016,Qiao2020}. 
For spatial instability analysis, the normalization scale is defined as 
\begin{equation}\label{eq:2.24}
	\mathrm{EK} = \int\limits_{T} \int\limits_{V} U_1 \,
	\boldsymbol{\tilde{V}}_1 \cdot \boldsymbol{\tilde{V}}_1 \,\mathrm{d}V \,\mathrm{d}t,
\end{equation}
so that all contributions can be expressed in normalized form,
\begin{equation}\label{eq:2.25}
	\overline{\mathrm{KE}} = \mathrm{KE}/\mathrm{EK}, \quad
	\overline{\mathrm{REY}} = \mathrm{REY}/\mathrm{EK}, \quad
	\overline{\mathrm{NE}} = \mathrm{NE}/\mathrm{EK}, \;\text{etc.}
\end{equation}
Since $\omega$ is real and $k$ is complex, 
the real part of $\overline{\mathrm{KE}}$ is equal to the growth rate $-k_i$. 
For brevity, overbars on normalized terms are omitted hereafter. 
A positive contribution from a given term indicates that the corresponding mechanism 
promotes instability, whereas a negative contribution implies a stabilizing effect. 
Finally, the $\mathrm{SHB}$ term vanishes because
\begin{equation}\label{eq:2.26}
	\left.\frac{\mathrm{d}^2 U_1}{\mathrm{d}r^2}\right|_{r=1} 
	= \left.\frac{\mathrm{d}^2 U_2}{\mathrm{d}r^2}\right|_{r=1} = 0 .
\end{equation}

\section{Results and discussion}\label{Sec:results and discussion}

\subsection{Reference state}\label{sec:reference_state}

To investigate the influence of elasticity on viscoelastic flow focusing, it is essential to establish a well-defined reference state derived from experimental observations. The liquid phase is a dilute polymer solution, consisting of polyethylene oxide (PEO) dissolved in a glycerol--water mixture. This solution is characterized by a density of $\rho_1 = 1.0186 \times 10^3\ \mathrm{kg/m^3}$, a dynamic viscosity of $\mu_1 = 1.16 \times 10^{-3}\ \mathrm{Pa{\cdot}s}$, a relaxation time of $\lambda = 4.72 \times 10^{-4}\ \mathrm{s}$, and a centerline velocity of $U_0 = 2.4\ \mathrm{m/s}$. These fluid properties can be experimentally tuned by adjusting the molecular weight and concentration of PEO in the solvent. The surrounding gas is nitrogen, with a density of $\rho_2 = 1.29\ \mathrm{kg/m^3}$ and a dynamic viscosity of $\mu_2 = 2 \times 10^{-5}\ \mathrm{Pa{\cdot}s}$. The surface tension at the liquid--gas interface is $\sigma = 61.16\ \mathrm{mN/m}$, and the radius of the liquid jet is $R_1 = 0.07\ \mathrm{mm}$. Based on these parameters, the characteristic dimensionless groups are calculated as $\Rey = 150$, $\We = 7$, $\El = 0.1$, $Q = 0.0013$, $N = 0.0172$, and $X = 0.9$.

As for the basic velocity profile, $U_{\!s}$ is larger than unity and represents the interfacial velocity normalized by the jet centerline velocity, while $K$ characterizes the slope of the velocity profile at the liquid–gas interface. 
In the present reference state, the values $U_{\!s} = 1.33$ and $K = 1.2$ are determined by fitting theoretical predictions to experimental phase boundaries (see \S\ref{sec:phase}).
These parameters define a hyperbolic-tangent base profile that represents the experimentally realized shear-layer structure.

\begin{figure}
	\centering
	\subfigure
	{
		\label{fig:3a}		
		\includegraphics[width=0.48\textwidth]{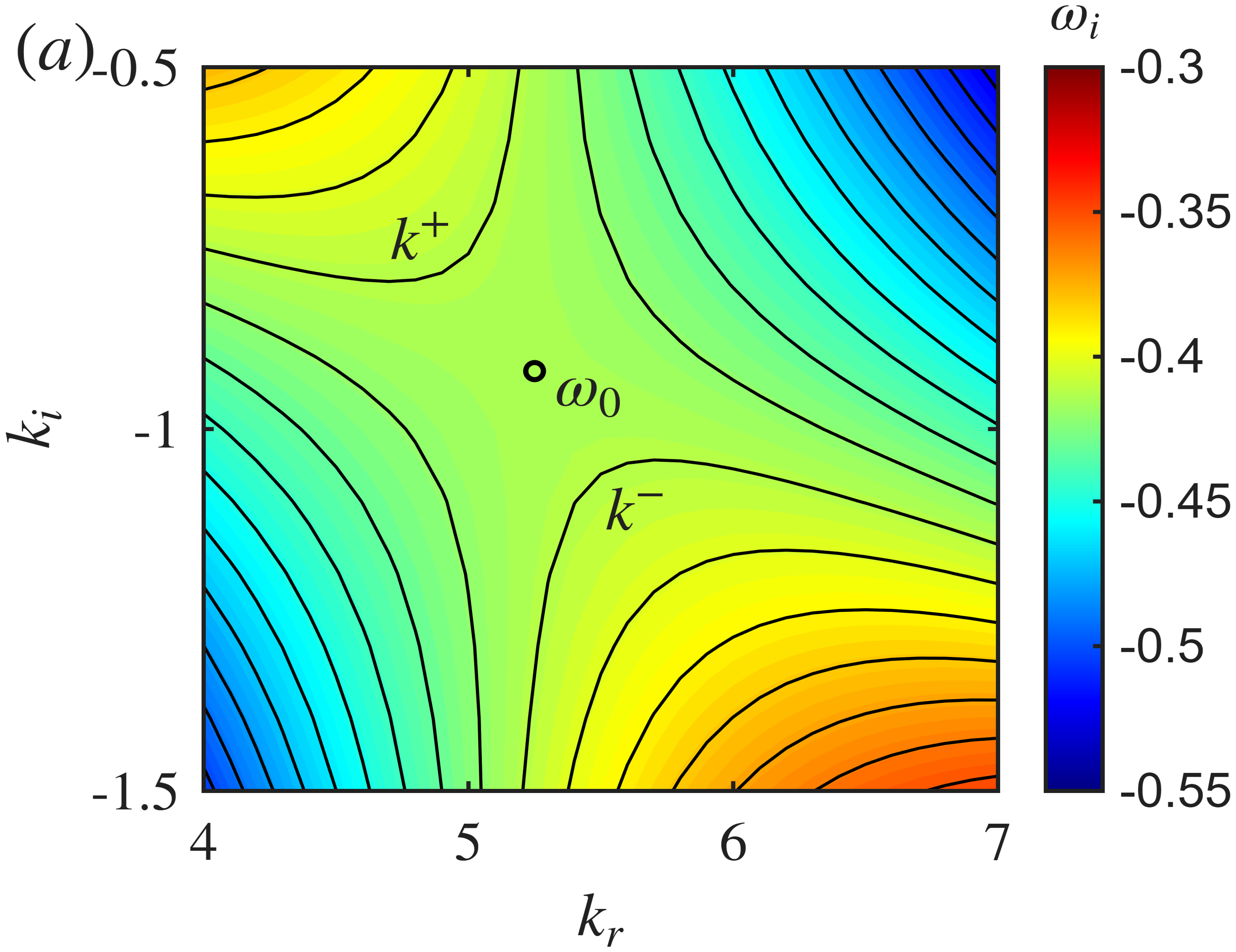}
	}
	{
		\label{fig:3b}		
		\includegraphics[width=0.48\textwidth]{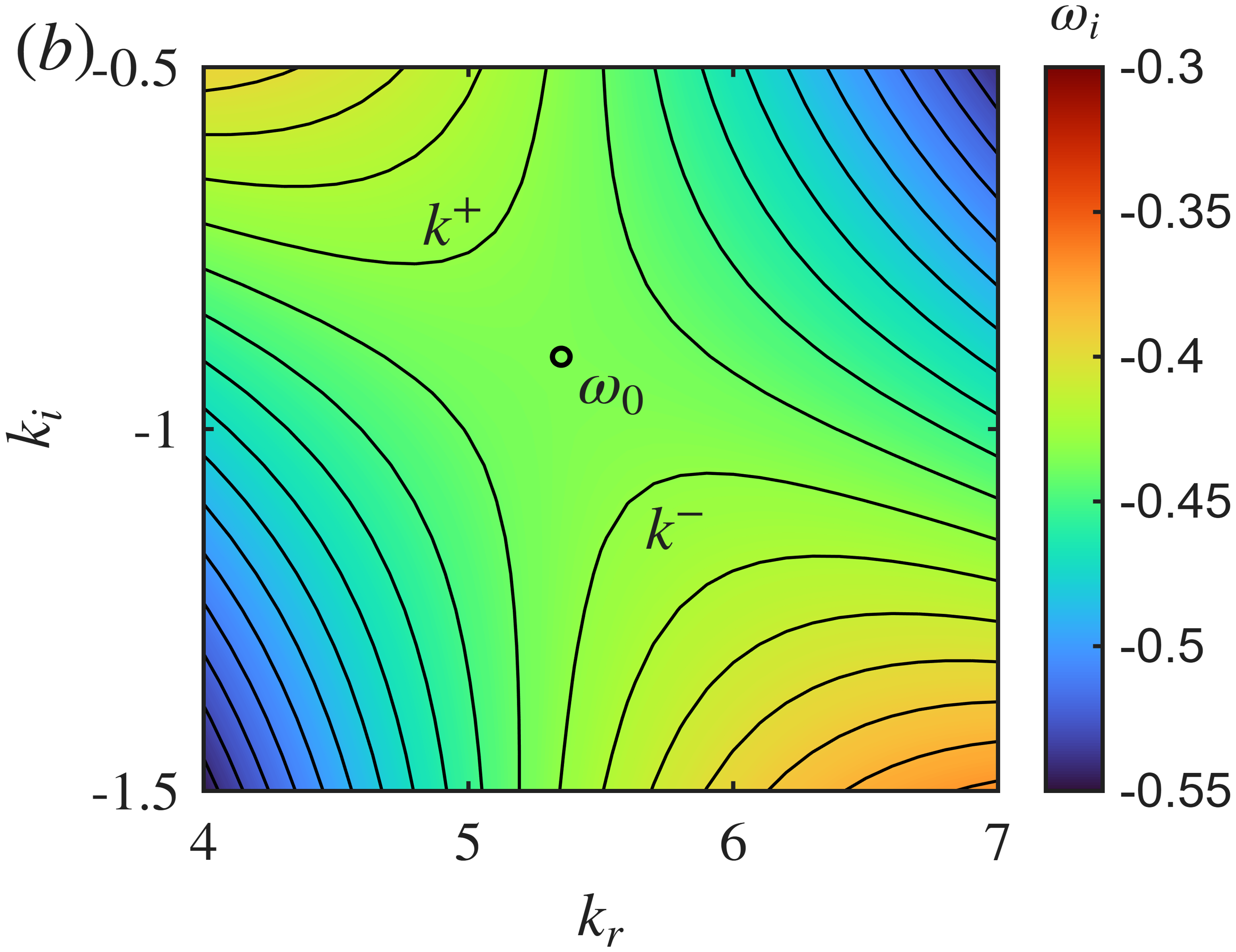}
	}
	\caption{\label{fig:3} 
		Saddle points in the complex $k$--plane for the axisymmetric and helical modes under the reference state. 
		The intersecting spatial branches arise from the downstream-propagating branch $k^{+}$ and the upstream-propagating branch $k^{-}$. 
		($a$) Axisymmetric mode: the negative imaginary part of the saddle-point frequency, $\omega_{0i} = -0.409134 < 0$, indicates that the instability is convective. 
		($b$) Helical mode: the negative imaginary part of the saddle-point frequency, $\omega_{0i} = -0.430328 < 0$, likewise indicates a convective instability.
	}
\end{figure}

To validate the suitability of this reference state for spatial linear instability analysis, it is necessary to confirm that the base flow is convectively unstable, since spatial instability analysis is only applicable to convectively unstable configurations. We therefore perform a spatio–temporal instability analysis for the present reference state.

The analysis is based on the dispersion relation
\begin{equation}\label{eq:3.1}
	D(n,\omega,k;\Rey,\We,\El,U_{\!s},K,Q,N,X)=0,
\end{equation}
where both the axial wavenumber $k$ and the perturbation frequency $\omega$ are complex quantities,
$k = k_r + \mathrm{i}k_i$ and $\omega = \omega_r + \mathrm{i}\omega_i$.
The instability type is determined by locating a saddle point $(k_0,\omega_0)$ satisfying the zero–group–velocity condition
\begin{equation}\label{eq:3.2}
	\frac{\partial\omega}{\partial k}=0.
\end{equation}

In the complex $k$–plane, the solutions of $D=0$ form two spatial branches which, as $\omega_i$ decreases, approach each other and eventually coalesce at a saddle point. According to the Briggs–Bers criterion \citep{Briggs1964,Bers1973,HuerreRossi1998}, only collisions between a downstream–propagating branch $k^{+}(\omega)$ and an upstream–propagating branch $k^{-}(\omega)$ are physically relevant. The sign of the imaginary part of the saddle–point frequency then distinguishes the instability type: $\omega_{0i}>0$ corresponds to absolute instability, whereas $\omega_{0i}<0$ indicates convective instability.

For the present reference state, the saddle point for the axisymmetric mode ($n=0$) is located at
$k_0 = 5.229015 - 0.921874\,\mathrm{i}$ with the corresponding frequency
$\omega_0 = 2.379980 - 0.409134\,\mathrm{i}$, as shown in figure~\ref{fig:3}($a$).
The negative imaginary part of the saddle-point frequency, $\omega_{0i}<0$, indicates that the instability is convective: perturbations are amplified while being advected downstream, without giving rise to self-sustained upstream-propagating disturbances. For the helical mode ($n=1$), the saddle point converges to
$k_0 = 5.355895 - 0.918524\,\mathrm{i}$ with the corresponding frequency
$\omega_0 = 2.255092 - 0.430328\,\mathrm{i}$, as shown in figure~\ref{fig:3}($b$).
The negative value of $\omega_{0i}$ again confirms the convective instability. These results demonstrate that, under the present reference state, both the axisymmetric and helical modes are convectively unstable. The absence of absolute instability justifies the use of a spatial instability framework in the following analysis.

\subsection{Effects of liquid elasticity}\label{effects_elasticity}

\begin{figure}
	\centerline{\includegraphics[width=\linewidth]{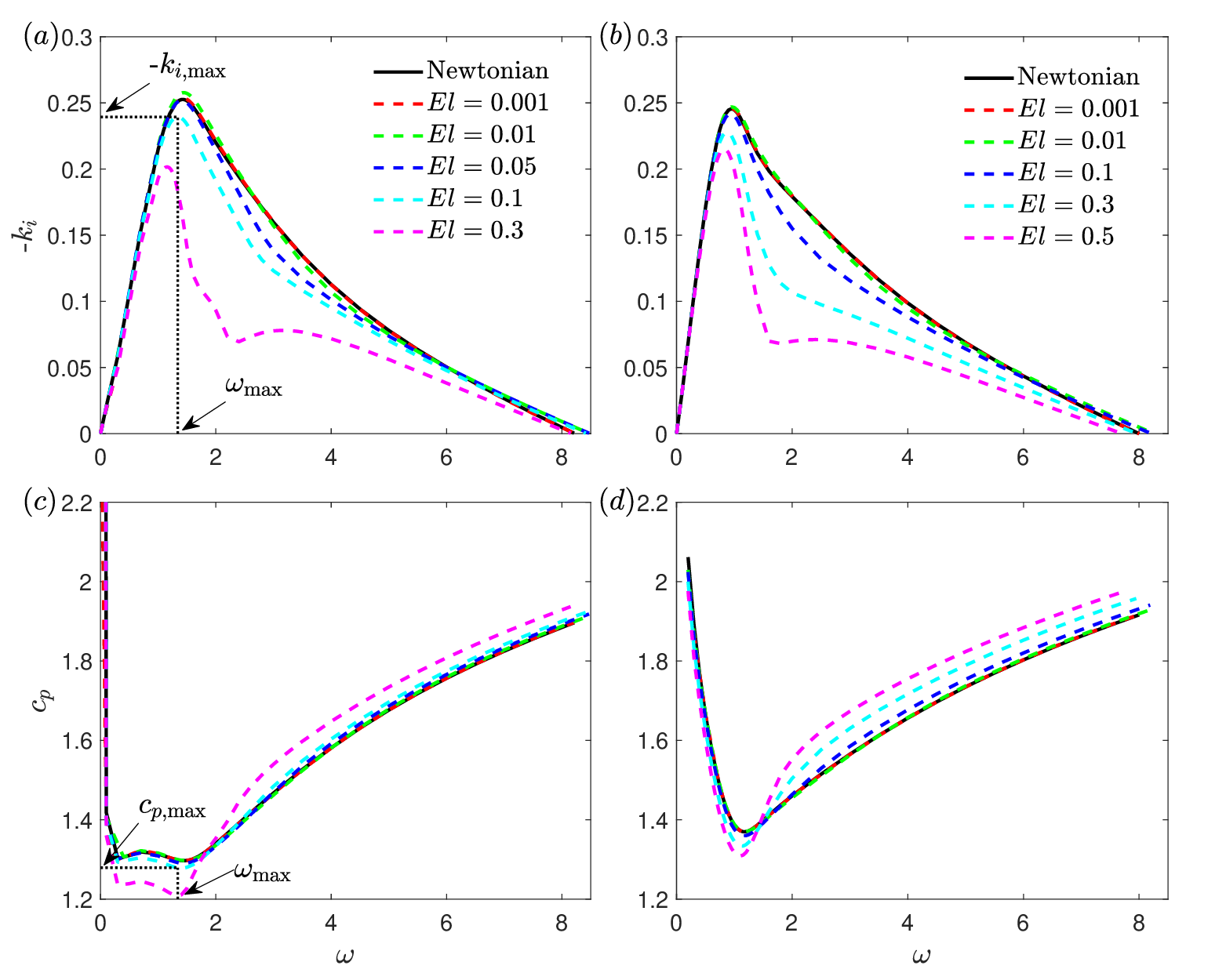}}
	\caption{
		Variations of the growth rate $-k_i$ and the phase speed $c_{\!p}$ with frequency $\omega$ 
		for different elasticity numbers $\El$, at fixed parameters $\Rey = 150$, $\We = 7$, $U_{\!s} = 1.33$, 
		$K = 1.2$, $Q = 0.0013$, $N = 0.0172$, and $X = 0.9$: 
		($a,c$) axisymmetric mode; ($b,d$) helical mode. 
		For $\El = 0.1$, the maximum growth rate $(-k_i)_{\max}$ and the most unstable phase speed $(c_{\!p})_{\max}$ are marked in ($a$) and ($c$), respectively.
		For convenience, the maximum growth rate $(-k_i)_{\max}$ and the most unstable phase speed $(c_{\!p})_{\max}$ 
		are denoted as $-k_{i,\max}$ and $c_{\!p,\max}$, respectively.
	}
	\label{fig:4}
\end{figure}

Elasticity is the most distinctive feature of viscoelastic fluids, and its influence on the flow-focused jet is first examined. In viscoelastic coflow systems, the degree of elasticity is governed by both the intrinsic relaxation properties of the polymer chains and their concentration relative to the Newtonian solvent. The relaxation process, quantified through the elasticity number $\El$, characterizes the competition between elastic and viscous time scales. A larger $\El$ indicates that elastic stresses relax more slowly, allowing elastic effects to dominate the flow dynamics. The second controlling factor is the solvent-polymer composition ratio, represented by $X$. A decrease in $X$ corresponds to an increase in the polymer fraction, thereby enhancing elastic contributions while reducing viscous damping from the solvent. Together, these two parameters determine how strongly elasticity influences the jet instability. In the present work, we focus on dilute polymer solutions and isolate the role of elasticity by varying $\El$ while holding $X = 0.9$ fixed.

Figure~\ref{fig:4} shows the spatial growth rate $-k_i$ and the phase speed $c_{\!p} = \omega/k_r$ as functions of $\omega$ for the axisymmetric and helical modes, highlighting the dependence of the dispersion relation on $El$. 
	In spatial instability theory, the frequency $\omega$ is prescribed to be real; therefore, $c_{\!p}=\omega/k_r$ represents the propagation speed of a wave of single real frequency $\omega$.
	The spatial growth rates $-k_i$ are obtained by solving the quadratic eigenvalue problem~(\ref{eq:2.21}) formulated in \S\ref{sec:linear_instability}. Consequently, each point on the dispersion curves corresponds to an eigenvalue of this problem. For a given real frequency $\omega$, the criterion for identifying the spatial growth rate $-k_i$ is explained in Appendix~\ref{app:method_validation}. By repeating this identification over the entire range of $\omega$, the dispersion curves in figures~\ref{fig:4}($a$,$c$) are constructed. Meanwhile, the eigenspectrum presented in Appendix~\ref{app:eigenvalue spectrum} display the full distribution of eigenvalues in the complex $k$-plane and their variation with elasticity. Further discussion of the rationale for the selection criterion, as well as of the structure of the eigenspectrum, can be found in Appendices~\ref{app:method_validation} and~\ref{app:eigenvalue spectrum}.
Instability arises only within a finite frequency band, outside which disturbances decay exponentially. For each mode, the spatial growth rate reaches a maximum $-k_{i,\max}$ at the most unstable frequency $\omega_{\max}$, corresponding to a resonance between the disturbance time scale and the dominant instability mechanism. This most amplified disturbance is therefore expected to dictate the jet dynamics in the unstable regime. The frequency $\omega_{\max}$ is associated with the axial wavenumber $k_{r,\max}$, which sets the disturbance structure, while the wavelength $\lambda_{\max} = 2\pi/k_{r,\max}$ characterizes the droplet size. The corresponding phase speed $c_{\!p,\max} = \omega_{\max}/k_{r,\max}$ defines the convective speed of the most unstable disturbance, thereby setting the characteristic propagation rate of perturbations along the jet. For both the axisymmetric mode (figure~\ref{fig:4}$a$) and the helical mode (figure~\ref{fig:4}$b$), $-k_{i,\max}$ initially increases slightly with $\El$ and remains higher than that of the Newtonian case. At larger $\El$ (e.g., $\El=0.3$), however, the growth rate decreases sharply, ultimately rendering the viscoelastic jet more stable than its Newtonian counterpart. This trend is consistent with the temporal instability analysis of \citet{Ding2022}, who showed that jet instability reflects a competition between unrelaxed elastic stresses, which destabilize the flow, and nonlinear viscoelastic effects, which stabilize it. In addition, the right cut-off frequency shows a weak dependence on $\El$.

As shown in figure~\ref{fig:4}($c,d$), the phase speed $c_{\!p}$ exhibits a non-monotonic dependence on frequency $\omega$.
At low frequencies, $c_{\!p}$ decreases with increasing $\omega$ for the axisymmetric mode.
With further increase in frequency, the system enters a transitional regime in which $c_{\!p}$ varies only weakly with $\omega$.
At higher frequencies, $c_{\!p}$ increases monotonically with $\omega$.
In contrast, although the phase speed $c_{\!p}$ of the helical mode also decreases at low frequencies and increases again at high frequencies, it does not display a pronounced transitional regime as observed in the axisymmetric mode.
At low frequencies, increasing the elasticity number $\El$ reduces the phase speed, bringing it closer to the jet centerline velocity.
At higher frequencies, however, increasing elasticity leads to a higher phase speed, indicating that elasticity enhances the propagation of high-frequency disturbances.

\begin{figure}
	\centerline{\includegraphics[width=\linewidth]{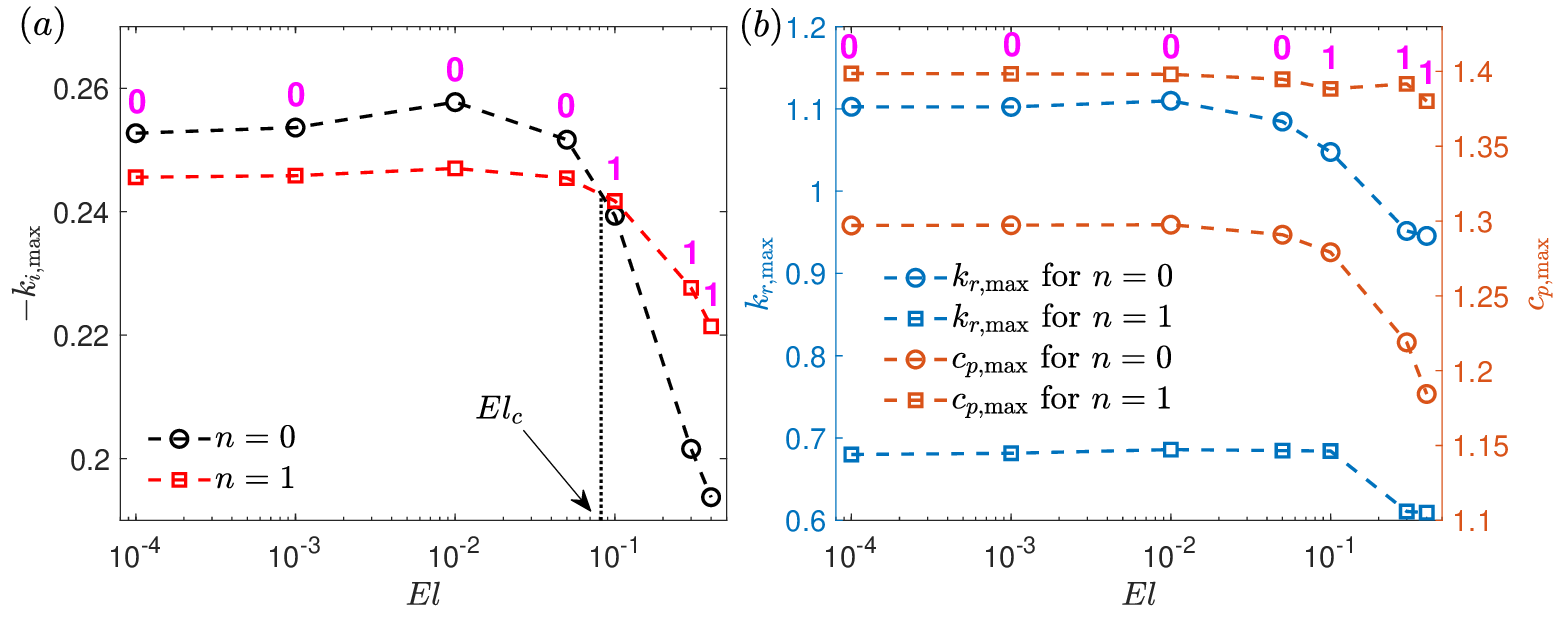}}
	\caption{
		Variations of the maximum growth rate $-k_{i,\max}$, 
		the most unstable axial wavenumber $k_{r,\max}$, and the most unstable phase speed $c_{\!p,\max}$ 
		with elasticity number $\El$, at fixed parameters $\Rey = 150$, $\We = 7$, $U_{\!s} = 1.33$, 
		$K = 1.2$, $Q = 0.0013$, $N = 0.0172$, and $X = 0.9$: 
		($a$) the maximum growth rate $-k_{i,\max}$; 
		($b$) the most unstable axial wavenumber $k_{r,\max}$ (left axis) and the corresponding phase speed $c_{\!p,\max}$ (right axis).
		The numbers 0 and 1 indicate that the predominant mode under the current parameter set is the axisymmetric mode and the helical mode, respectively.
	}
	\label{fig:5}
\end{figure}

Figure~\ref{fig:5}($a$) presents the maximum growth rate $-k_{i,\max}$ for the axisymmetric and helical modes as functions of $\El$. Consistent with figure~\ref{fig:4}, elasticity exerts a dual influence: it first enhances instability at small $\El$ (e.g., $\El \leq 0.01$) but suppresses it at larger $\El$. The mode with the larger $-k_{i,\max}$ is regarded as predominant, and the transition is clearly revealed by comparing the two curves, with the predominant mode at each $\El$ marked by purple numbers. When $\El$ is below approximately $0.08$, the axisymmetric mode dominates, whereas increasing elasticity gradually shifts the predominance toward the helical mode. The critical elasticity number $El_{c} (\approx 0.08)$, at which the two modes achieve equal spatial growth rates, signifies the onset of this mode transition. This highlights the role of elasticity in sustaining the axisymmetric mode at small $\El$ and in promoting helical disturbances as $\El$ increases.

Figure~\ref{fig:5}($b$) shows the corresponding axial wavenumber $k_{r,\max}$ and phase speed $c_{\!p,\max}$. The variation of $k_{r,\max}$ mirrors that of figure~\ref{fig:5}($a$), increasing initially and then decreasing as $\El$ grows. In contrast, the phase speed exhibits a mode-dependent trend: for the axisymmetric mode, $c_{\!p,\max}$ decreases with increasing $\El$, while for the helical mode it shows only a slight decline. In other words, increasing elasticity drives the phase speed closer to the jet centerline velocity (i.e. $c_{\!p,\max}\!\to\!1$), indicating a tendency toward a ‘center mode’. This feature is analogous to the centre-mode instability reported in viscoelastic Poiseuille flow \citep{Garg2018}. Notably, this inward shift of the phase speed is consistent with the evolution of the eigenfunctions shown in figure~\ref{fig:9}, which also exhibit a progressive concentration of disturbance energy toward the jet core. The above trends of the growth rate and phase speed with respect to $\El$ for the axisymmetric and helical modes are summarized in Table~\ref{tab:2}.

\subsection{Effects of Weber number}\label{sec:effects of Weber number}

\begin{figure}
	\centerline{\includegraphics[width=\linewidth]{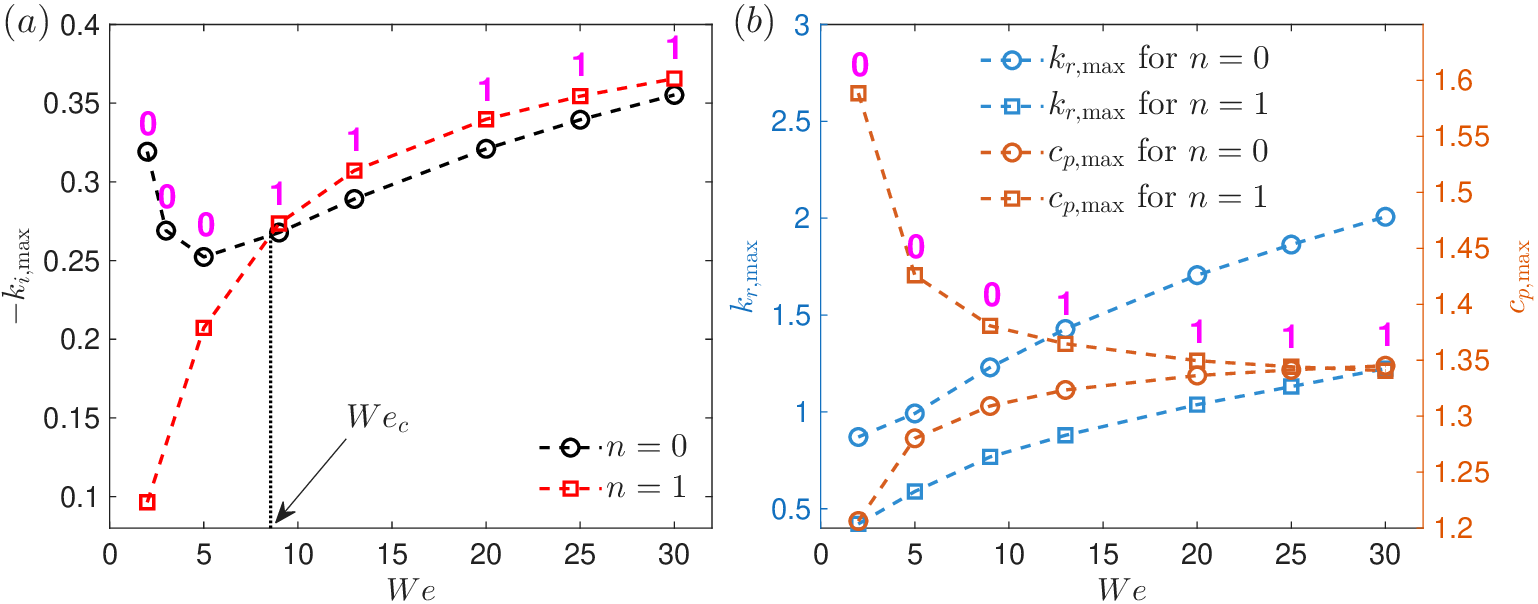}}
	\caption{Variations of the maximum growth rate $-k_{i,\max}$, 
		the maximum axial wavenumber $k_{r,\max}$, and the most unstable phase speed $c_{\!p,\max}$ 
		with Weber number $\We$, where $\El = 0.01$, $\Rey = 150$, $U_{\!s} = 1.33$, 
		$K = 1.2$, $Q = 0.0013$, $N = 0.0172$, and $X = 0.9$: 
		($a$) the maximum growth rate $-k_{i,\max}$; 
		($b$) the most unstable axial wavenumber $k_{r,\max}$ (left axis) and the corresponding phase speed $c_{\!p,\max}$ (right axis).
		The numbers 0 and 1 indicate that the predominant mode under the current parameter set is the axisymmetric mode and the helical mode, respectively.
		}
	\label{fig:6}
\end{figure}

To identify the parameter regimes in which the axisymmetric and helical modes dominate, we examine the influence of the Weber number ($\We$) on the key instability characteristics: the maximum spatial growth rate $-k_{i,\max}$, the most unstable axial wavenumber $k_{r,\max}$, and the corresponding phase speed $c_{\!p,\max}$. Figure~\ref{fig:6}($a$) shows the variation of $-k_{i,\max}$ for the axisymmetric and helical modes as $\We$ increases. 
As in figure~\ref{fig:5}, the predominant mode at each Weber number is indicated by the purple numbers, with $0$ and $1$ denoting the axisymmetric and helical modes, respectively.
For the axisymmetric mode, $-k_{i,\max}$ decreases at low to moderate $\We$ (i.e. $\We < 6$), where the most unstable wavenumber satisfies $k_{r,\max}\!<\!1$, indicating a long-wave regime (figure~\ref{fig:6}$b$) in which surface tension promotes instability. As $\We$ increases further, the most unstable wavenumber shifts beyond unity ($k_{r,\max} \! > \!1$), and $-k_{i,\max}$ rises again, marking a transition to the short-wave regime (figure~\ref{fig:6}$b$), where surface tension instead exerts a stabilizing influence. By contrast, the helical mode displays a monotonic increase of $-k_{i,\max}$ with $\We$. This behavior defines a critical Weber number $\We_c$: for $\We\!<\!\We_c$, the axisymmetric mode dominates, whereas for $\We\!>\!\We_c$, the helical mode becomes predominant. The underlying predominant mode transition arises from the interplay among surface tension, interfacial pressure fluctuations, interfacial shear, and energy transfer between disturbances and the base flow, and will be further clarified in \S\ref{sec:mechanisms of modal transition}. Notably, this trend is qualitatively consistent with theoretical predictions and experimental observations of Newtonian flow focusing \citep{Ganan-Calvo1999,Rosell-Llompart2008,Si2009,Si2010}.

\begin{table}
	\centering
	\small
	\setlength{\tabcolsep}{5pt}
	\renewcommand{\arraystretch}{1.15}

	\begin{tabular}{p{2.2cm} p{5.2cm} p{5.2cm}}
		
		\textbf{Aspect} & \textbf{Axisymmetric mode ($n=0$)} & \textbf{Helical mode ($n=1$)} \\[4pt]
		
		\textbf{Phase speed} \newline
		$c_{p,\max}$ &
		$\El$ increased: suppressed \newline
		$\We$ increased: promoted &
		$\El$ increased: suppressed \newline
		$\We$ increased: suppressed \\[14pt]
		
		\textbf{Growth rate} \newline
		$-k_{i,\max}$ &
		Low $\El$ (high $\We$): promoted \newline
		High $\El$ (low $\We$): suppressed &
		Low $\El$ (increasing $\We$): promoted \newline
		High $\El$: suppressed \\[14pt]
		
		\textbf{Instability mechanism} &
		Low $\We$: capillary instability \newline
		High $\We$: Kelvin--Helmholtz instability &
		Low $\El$: Kelvin--Helmholtz instability \newline
		High $\El$: elasticity-enhanced shear-driven instability \\[4pt]
	\end{tabular}
    
	\caption{Summary of the effects of elasticity number ($\El$) and Weber number ($\We$) on the maximum growth rate $-k_{i,\max}$, phase speed $c_{\!p,\max}$, and predominant instability mechanism for the axisymmetric and helical modes. As discussed in \S\ref{sec:phase}, the capillary instability corresponds to surface-tension-dominated behavior, the Kelvin--Helmholtz instability is driven primarily by gas-pressure fluctuations or interfacial shear, and the elasticity-enhanced shear-driven instability arises from energy transfer between the base-flow shear and the perturbation velocities.}
	\label{tab:2}
\end{table}

Figure~\ref{fig:6}($b$) presents the most unstable axial wavenumber $k_{r,\max}$ and the corresponding phase speed $c_{\!p,\max}$ for the axisymmetric and helical modes. The axial wavenumber $k_{r,\max}$ increases monotonically with $\We$, indicating that higher $\We$ shifts the instability toward shorter wavelengths. This trend reflects the competition between inertia and capillarity: at large $\We$, inertia dominates over surface tension, making the system more susceptible to short-wave disturbances. The phase speed, however, shows mode-dependent behavior. For the axisymmetric mode, $c_{\!p,\max}$ increases with $\We$, whereas for the helical mode it decreases initially and then levels off. At sufficiently high $\We$, the phase speeds of both modes converge toward the same limiting value, close to the interfacial velocity $U_{\!s}$. This convergence highlights a key feature: under strong inertial conditions, the propagation of disturbances—irrespective of symmetry—is primarily governed by interfacial dynamics rather than by the bulk liquid or gas alone, consistent with the interpretation from the energy budget analysis (see figure~\ref{fig:12}), and can therefore be identified as an ‘interface mode’. The above features of the growth rate and phase speed with respect to $\We$ for the axisymmetric and helical modes are summarized in Table~\ref{tab:2}.

\subsection{Mechanisms of predominant mode transition}\label{sec:mechanisms of modal transition}

\begin{figure}
	\centering
	\includegraphics[width=\textwidth]{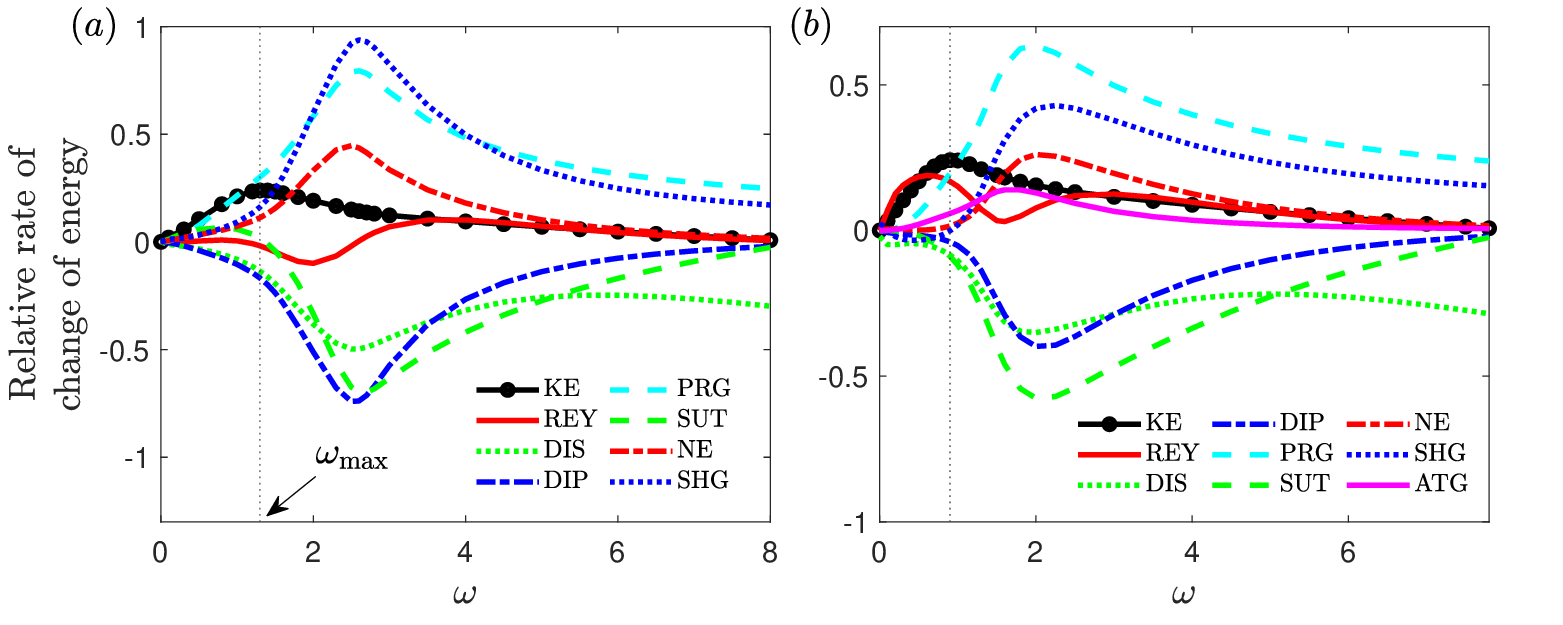}
	\caption{
		Energy budget of different azimuthal modes under the reference state 
		at $\El = 0.1$, $\Rey = 150$, $\We = 7$, $U_{\!s} = 1.33$, $K = 1.2$, $Q = 0.0013$, $N = 0.0172$, and $X = 0.9$:
		($a$) axisymmetric mode;
		($b$) helical mode.
		The dashed lines denote the relative rate of change of energy at the most unstable frequency $\omega_{\max}$.
		The definitions of the individual energy terms are given in equation~(\ref{eq:2.23}).
		The physical interpretation of energy terms are given in Table~\ref{tab:1}.
	}
	\label{fig:7}
\end{figure}

To elucidate the physical mechanisms responsible for the transition of the predominant modes, energy budget analysis was conducted. Figure~\ref{fig:7} displays the energy budgets of the axisymmetric mode and the helical mode obtained from the spatial linear instability analysis. The contributions from PRL, NVL, SHL, TEL, NEL, and NVG are found to be small in magnitude, each accounting for less than 5\% of the kinetic energy (KE) over the entire parameter range considered. Therefore, their influence on the jet instability is limited, and they are omitted from the figure for clarity.

\begin{figure}
	\centering
	\subfigure
	{
		\label{fig:10a}		
		\includegraphics[width=0.48\textwidth]{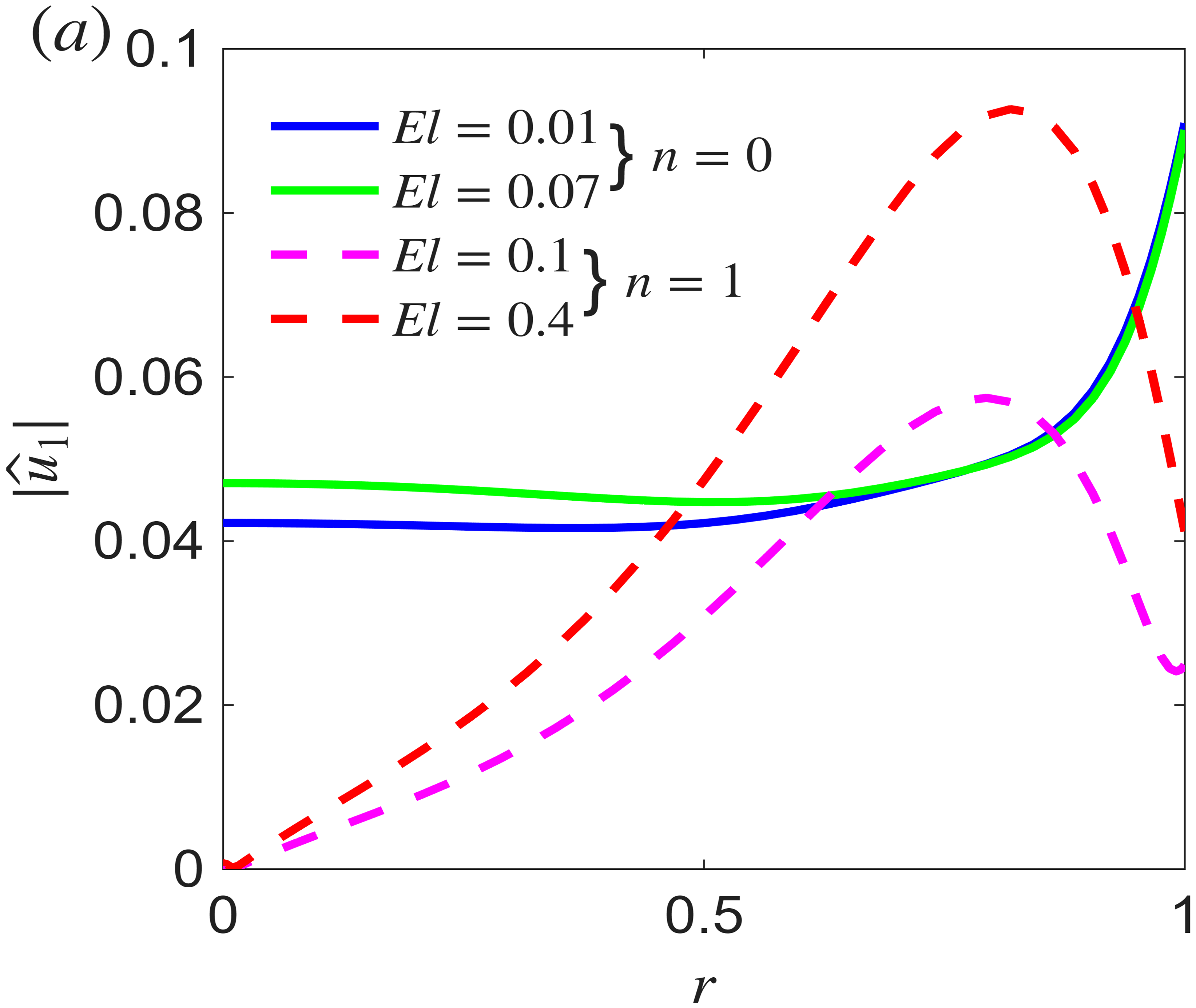}
	}
	{
		\label{fig:10b}		
		\includegraphics[width=0.48\textwidth]{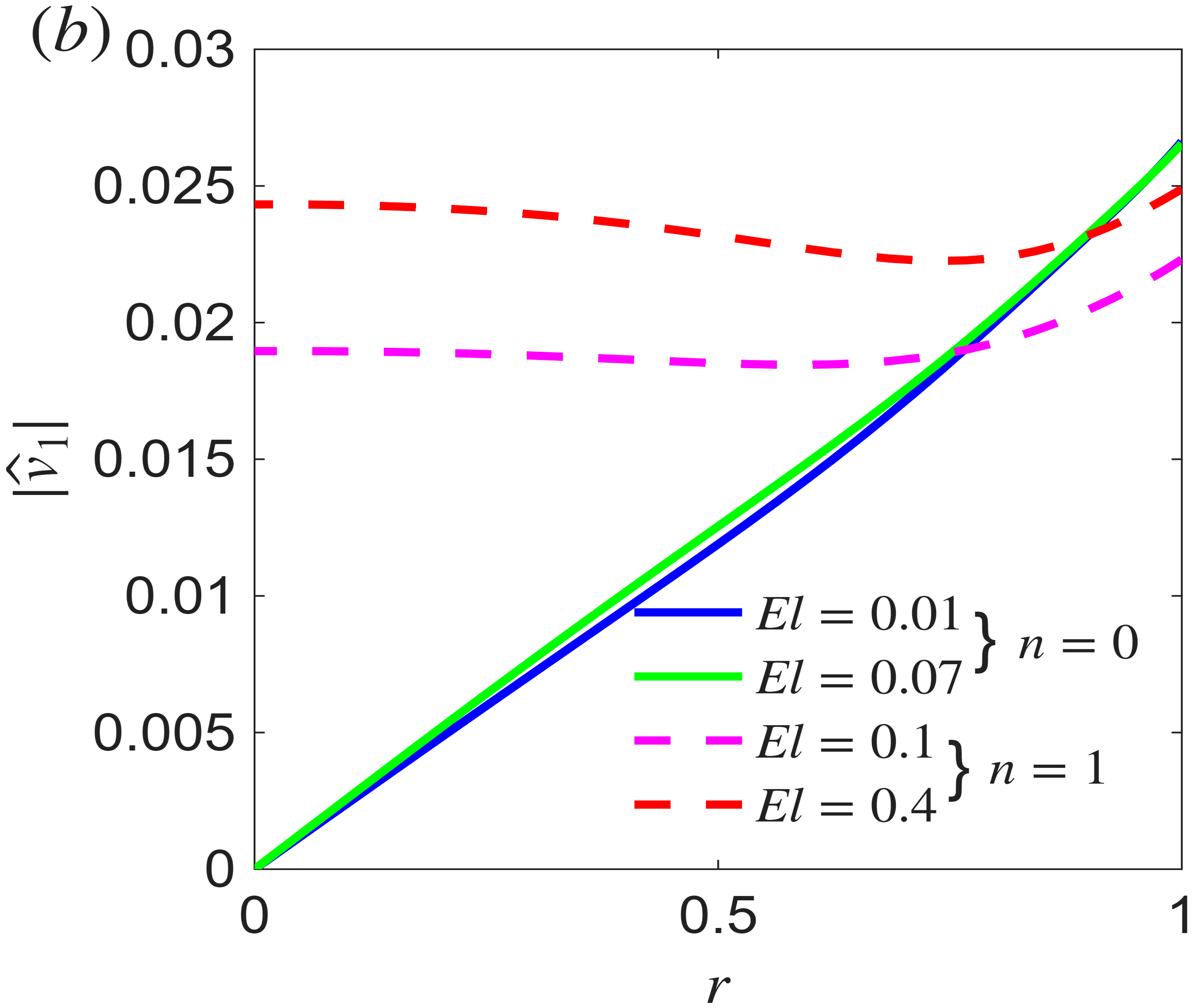}
	}
	{
		\label{fig:10c}		
		\includegraphics[width=0.48\textwidth]{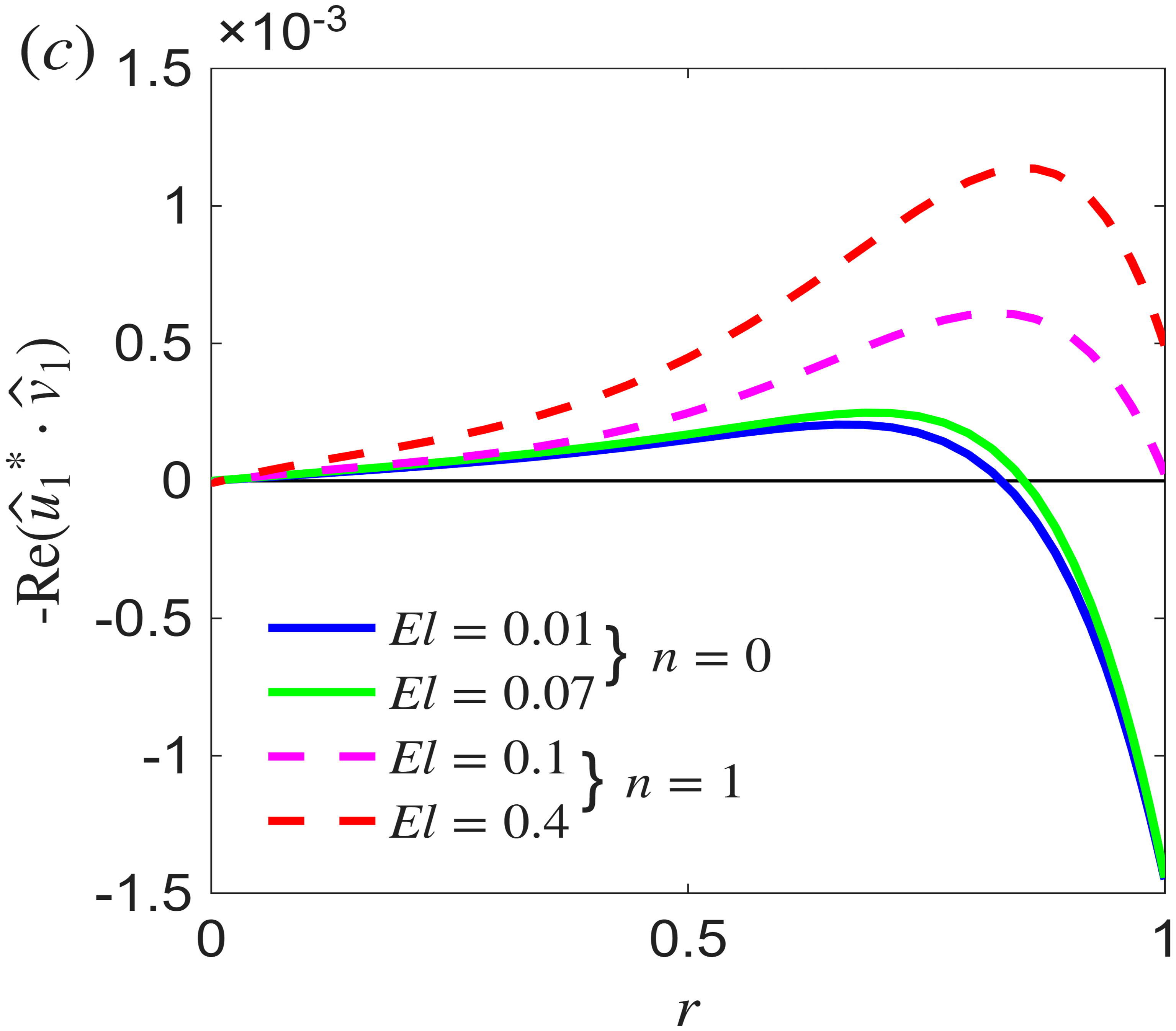}
	}
	\caption{
		Disturbance eigenfunctions of predominant mode at the most unstable frequency $\omega_{\max}$ for different elasticity numbers $\El$. 
		All other parameters are identical to those of the reference state.
		The predominant mode is axisymmetric at small elasticity 
		($\El = 0.01$ and $0.07$), whereas it becomes helical at larger elasticity 
		($\El = 0.1$ and $0.4$).
		($a$) Axial velocity $\hat{u}_1$ for the predominant mode;
		($b$) Radial velocity $\hat{v}_1$ for the predominant mode;
		($c$) Real part of the cross-term $-\hat{u}_1^{*} \cdot \hat{v}_1$ for the predominant mode at the most unstable frequency $\omega_{\max}$ for different elasticity numbers $\El$.
		Here, $(\cdot)^{*}$ denotes complex conjugation, and $\Real(\cdot)$ denotes the real part.
	}
	\label{fig:10}
\end{figure}

For the axisymmetric mode (figure~\ref{fig:7}$a$), the surface-tension term (SUT) is positive in the low-frequency (long-wave) regime, showing that surface tension promotes instability, whereas at higher frequencies (short waves) SUT turns negative, indicating stabilization. In contrast, for the helical mode (figure~\ref{fig:7}$b$), SUT remains negative, i.e. consistently stabilizing. The pressure-related term (PRG) is positive, indicating that interfacial pressure fluctuations enhance instability, while the dissipation term (DIS) is negative, reflecting the stabilizing effect of viscous damping. The shear-related terms (SHG and ATG) are positive, confirming that shear drives destabilization. 
The Reynolds-stress term (REY) is weak and slightly stabilizing for the axisymmetric mode, but destabilizing for the helical mode, at the most unstable frequency $\omega_{\max}$. 
This contrast originates from different momentum-exchange regions: the axisymmetric mode dissipates energy near the interface, whereas the helical mode extracts energy from the liquid shear layer (see figure~\ref{fig:10}($c$)), as further discussed below. Regarding elasticity, the unrelaxed axial tension (NE) is positive and destabilizing, whereas the polymeric elastic stress (DIP) is negative and stabilizing, indicating that elasticity exerts competing effects on instability.

\begin{figure}
	\centerline{\includegraphics[width=\textwidth]{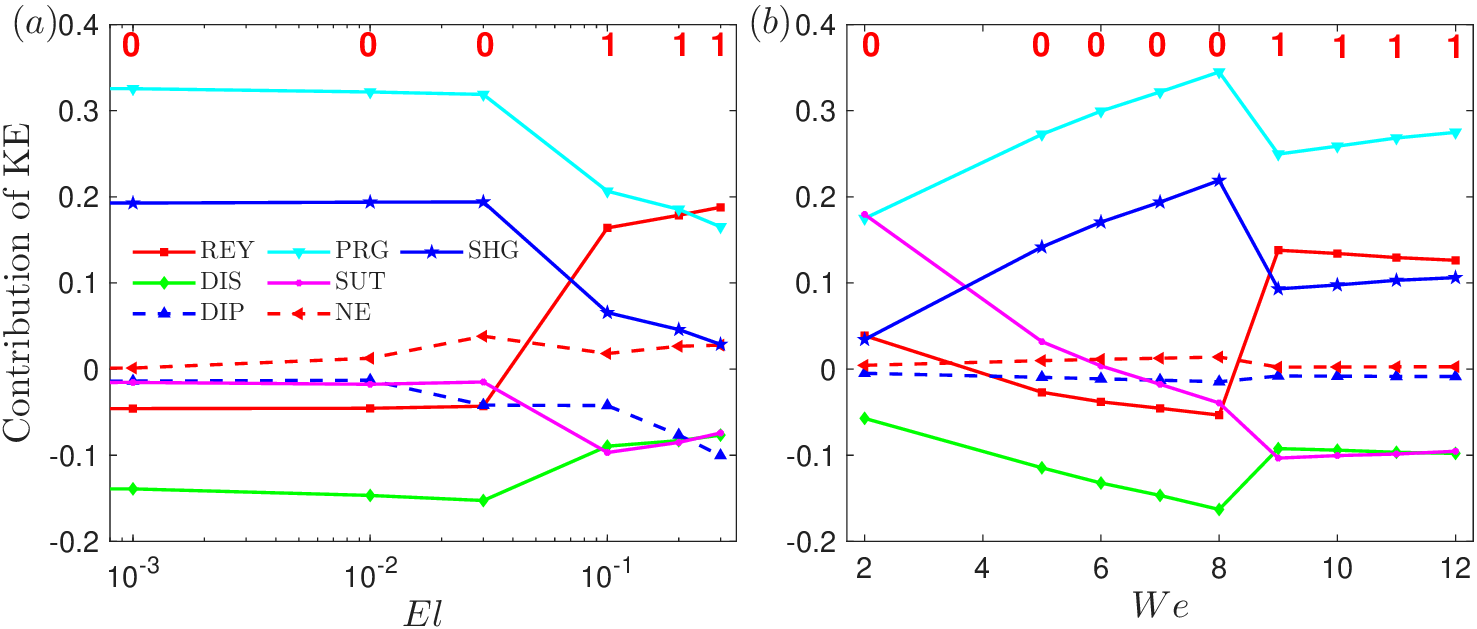}}
	\caption{Energy budget of the predominant mode at the most unstable frequency: ($a$) varying elasticity number $\El$, where $\We = 7$; ($b$) varying Weber number $\We$, where $\El = 0.01$. All other parameters are the same as in the reference case. Red numbers indicate the predominant mode.}
	\label{fig:8}
\end{figure}

Figure~\ref{fig:8} present the energy budget at the most unstable frequency $\omega_{\max}$ for varying $\El$ and $\We$. Since the predominant mode dictates the instability dynamics of the liquid jet, the energy budget is evaluated only for this mode, highlighted by the red numbers in the figure. Note that ATG and SHG both represent interfacial shear contributions; therefore, they are combined and shown as SHG. In figure~\ref{fig:8}($a$), when elasticity is weak ($\El\! <\! 0.03$), the pressure-related term (PRG) provides the largest positive input, indicating that gas-pressure fluctuations are the predominant driver of the axisymmetric instability. Interfacial shear (SHG) also contributes positively, reinforcing the growth. The instability in this regime is therefore governed jointly by gas-pressure fluctuations and interfacial shear, consistent with a Kelvin–Helmholtz (KHI) mechanism. As $\El$ increases to $0.1$, however, the predominant mode shifts from the axisymmetric to the helical mode. During this transition, the elastic contributions (NE and DIP) both increase in absolute value but remain much smaller than the other energy terms, suggesting that elasticity does not directly dominate the energy exchange. Instead, elasticity modifies other energy pathways—most notably by enhancing Reynolds-stress transfer (REY). At sufficiently large $\El$ ($=\! 0.3$), REY may even exceed PRG, indicating that instability is then primarily driven by energy exchange between the liquid shear layer and perturbation velocities, representing a new predominant instability mechanism---elasticity-enhanced shear-driven instability (ESI) distinct from that in Newtonian jets. A further discussion of how elasticity governs this mode transition will be presented in a subsequent section.

In figure~\ref{fig:8}($b$), the elasticity number is fixed at $\El = 0.01$, corresponding to weak elastic effects and behavior close to that of a Newtonian fluid, thereby allowing the influence of $\We$ to be examined independently. For a very small Weber number, $\We = 2$, SUT attains the largest positive value among all terms, indicating that surface-tension work dominates the instability and the mechanism is capillary instability, corresponding to a varicose (axisymmetric) mode. As $\We$ increases, both PRG and SHG gradually surpass SUT and become the leading positive contributions, signifying a transition to a Kelvin–Helmholtz instability driven by gas-pressure fluctuations (PRG) and interfacial shear (SHG). As $\We$ increases toward a critical value around $6$, SUT weakens and decreases, explaining the initial drop in the maximum spatial growth rate reported in \S\ref{sec:effects of Weber number}. With further increase in $\We$, SUT changes sign and becomes negative, i.e. stabilizing, while SHG and PRG intensify due to stronger interfacial shear and gas–inertia coupling, leading to renewed growth of the axisymmetric mode. At higher $\We$ (e.g., $\We = 9$), REY also rises as the predominant mode shifts to the helical mode, but this increase is mainly a by-product of the predominant mode transition rather than the emergence of a new predominant mechanism. In fact, the dominant source of instability remains aerodynamic, with PRG consistently maintaining the largest magnitude. Moreover, as $\We$ increases from $9$ to $12$, the relative contribution of REY diminishes, while PRG and SHG continue to grow, underscoring that the instability is still governed primarily by aerodynamic effects. This redistribution of energetic pathways coincides with the transition of the predominant mode to the helical mode.
A concise summary of the predominant instability mechanisms for the axisymmetric and helical modes identified across the parameter space is provided in Table~\ref{tab:2}.

To clarify how elasticity influences the instability mechanisms, we complement the energy budget analysis with a structural examination of the disturbance eigenfunctions. While the budget terms indicate where growth is generated or dissipated, they do not fully account for the predominant mode transition on their own. We therefore inspect the eigenfunctions at the most unstable frequency $\omega_{\max}$ across varying $\El$, in order to expose the redistribution of amplitudes and the associated structural changes.

\begin{figure}
	\centering
	\subfigure
	{
		\label{fig:9a}		
		\includegraphics[width=0.48\textwidth]{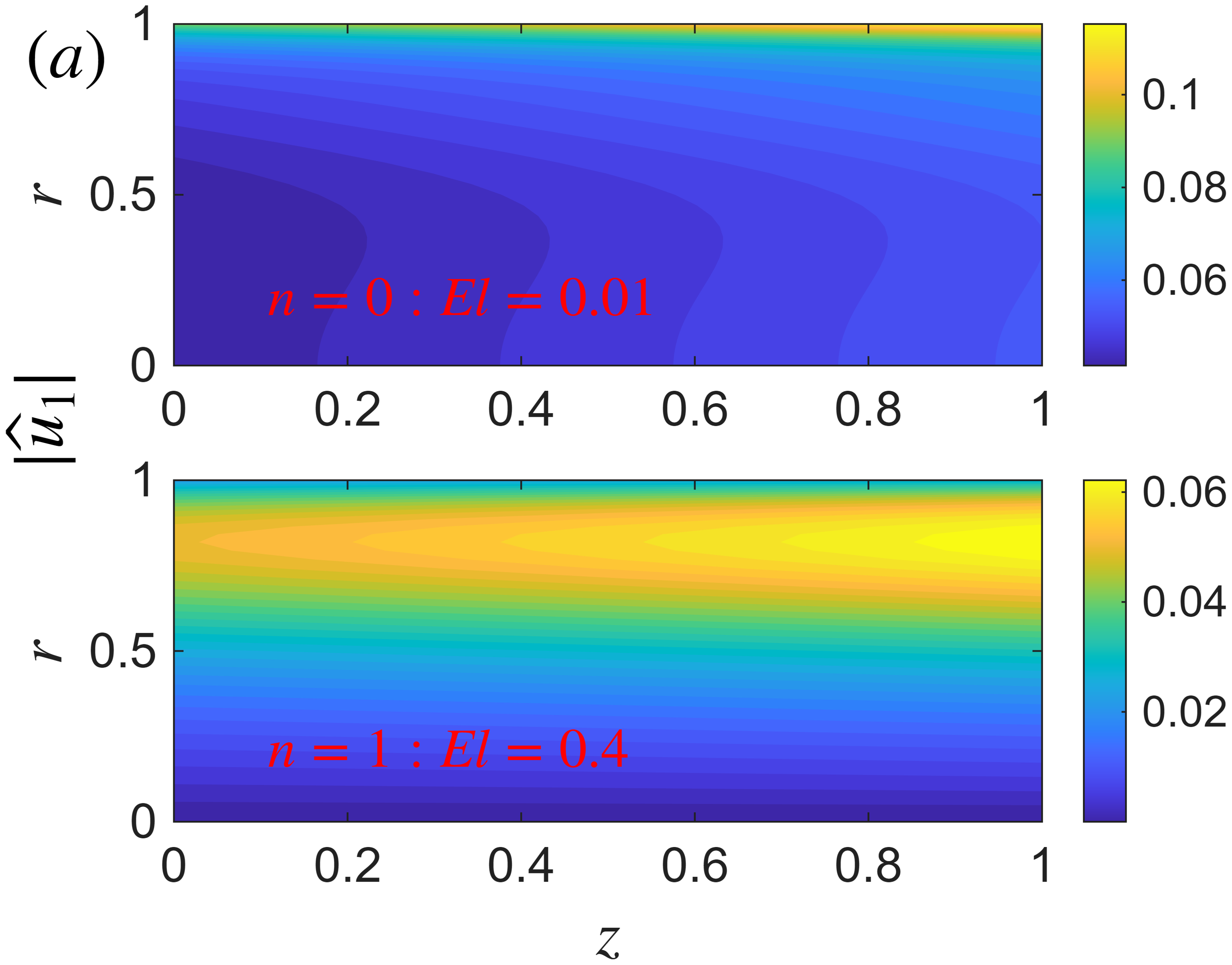}
	}
	{
		\label{fig:9b}		
		\includegraphics[width=0.48\textwidth]{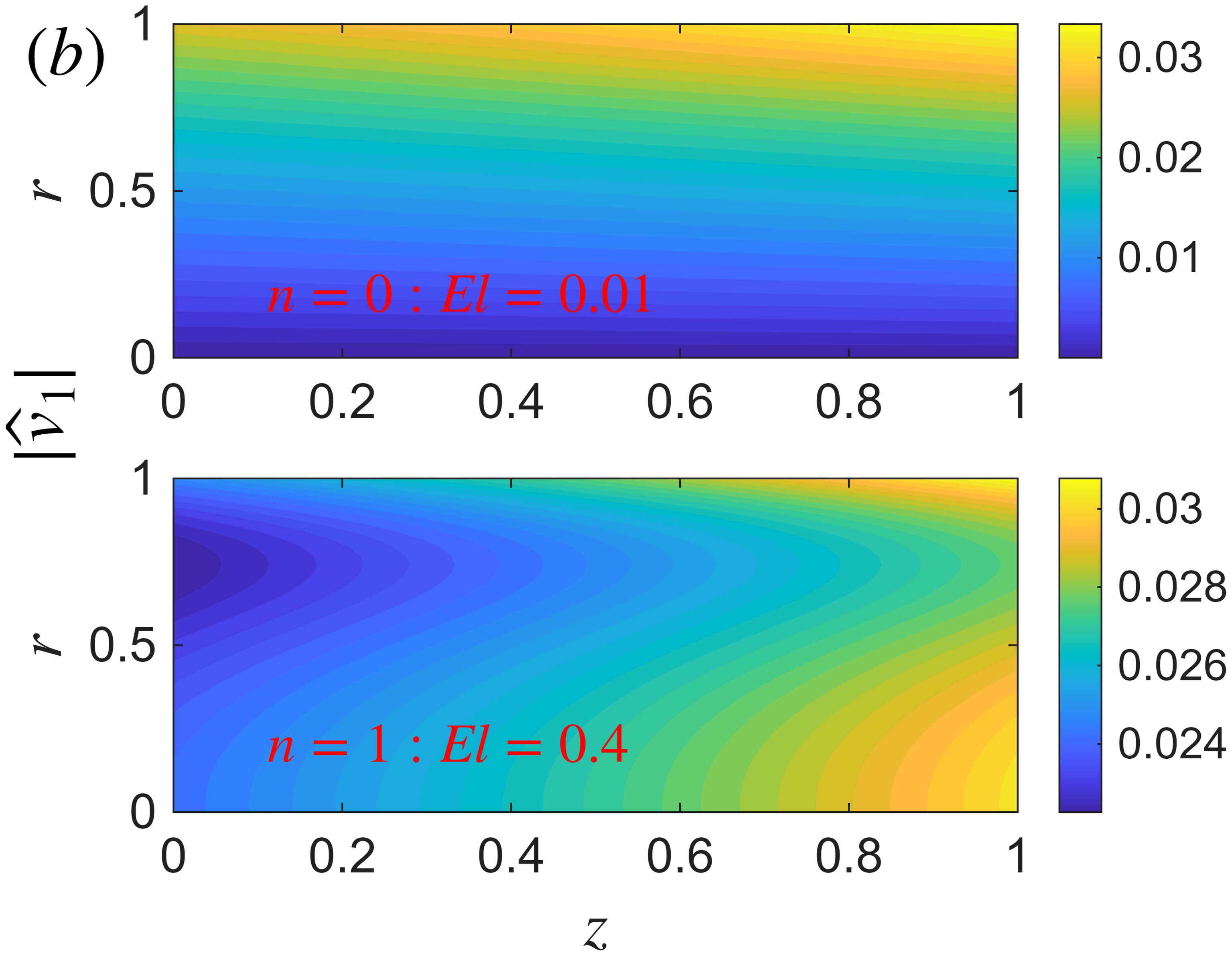}
	}
	\caption{
		Spatial structure of the disturbance eigenfunctions of the predominant mode at the most unstable frequency $\omega_{\max}$ for different elasticity numbers $\El$. 
		($a$) shows the axial disturbance velocity $|\hat{u}_1|$ exhibiting its radial distribution and streamwise evolution in the $(r,z)$–plane, 
		while ($b$) presents the corresponding radial disturbance velocity $|\hat{v}_1|$ with its radial and streamwise variations. 
		All other parameters are identical to those of the reference state.
		Here, $|\cdot|$ denotes the modulus.
		}
		\label{fig:9}
\end{figure}

Figure~\ref{fig:9} illustrates the spatial structure of the disturbance eigenfunctions of the predominant mode at the most unstable frequency $\omega_{\max}$, highlighting their radial distribution and streamwise evolution. 
As the disturbances evolve downstream along the $z$-direction, their amplitudes increase progressively, indicating spatial growth of the instability. 
When the elasticity number is small ($\El=0.01$), the amplitude of the predominant mode at $\omega_{\max}$ attains its maximum at the interface $r=1$ and is mainly concentrated in the vicinity of the interface, while its influence is relatively weak away from the interface. 
However, as the elasticity number increases to a larger value (e.g., $\El=0.4$), the structure of the predominant mode at $\omega_{\max}$ gradually shifts from being interface-localized to extending further into the jet core, reflecting a redistribution of disturbance energy from the interfacial region towards the interior of the jet. 
This trend parallels the observation in \S\ref{effects_elasticity} that the most unstable phase speed $c_{\!p,\max}$ approaches the centerline velocity as $\El$ increases, indicating a gradual transition from an interface-dominated mode to a center-mode structure.

The disturbance eigenfunctions of the predominant mode display qualitatively similar radial structures at different axial positions $z$. Therefore, in order to concentrate on the radial characteristics of the mode, we fix the axial coordinate at $z=0$ without loss of generality and present the corresponding radial profiles of the eigenfunctions at this location. Figure~\ref{fig:10}($a$,$b$) show the axial and radial disturbance velocities of the predominant mode at the most unstable frequency $\omega_{\max}$ for different elasticity numbers $\El$. The predominant mode is axisymmetric for small elasticity numbers ($\El = 0.01$ and $0.07$), whereas it becomes helical at larger elasticity ($\El = 0.1$ and $0.4$).
For the predominant axisymmetric mode at small elasticity ($\El = 0.01$ and $0.07$), the axial disturbance velocity $\hat{u}_1$ (see figure~\ref{fig:10}$(a)$) is nearly uniform in the core and attains its maximum at the interface, characteristic of an interfacial mode. As elasticity increases and the predominant mode becomes helical ($\El = 0.1$ and $0.4$), the axial disturbance velocity develops a distinct peak in the jet interior, whose amplitude increases as $\El$ is further raised. A similar restructuring occurs in the radial disturbance velocity $\hat{v}_1$ (see figure~\ref{fig:10}$(b)$). For the predominant axisymmetric mode at small $\El$, $\hat{v}_1$ resembles the Newtonian limit and increases monotonically with radius. As elasticity increases and the predominant mode switches to the helical one, $\hat{v}_1$ develops a central maximum whose amplitude can even exceed the interfacial value, highlighting the strong core localization of the disturbance.

Among the energy-budget terms, the Reynolds-stress contribution (REY) exhibits the most pronounced variation across the predominant mode transition. Figure~\ref{fig:10}$(c)$ shows the cross-correlation $-\Real\!\left(\hat{u}_1^{*} \cdot \hat{v}_1\right)$ of the predominant mode as a function of $\El$, which directly links to the Reynolds-stress exchange identified in the energy budget. For $\El=0.01$ and $\El=0.07$, the predominant mode is axisymmetric, whereas at larger $\El$ it shifts to the helical mode. Here, $(\cdot)^{*}$ denotes complex conjugation, and $\Real(\cdot)$ denotes the real part. In flow-focusing configurations, the liquid-phase velocity gradient $\mathrm{d}U_1/\mathrm{d}r$ is positive, and thus the sign of REY follows directly from $-\Real\!\left(\hat{u}_1^{*} \cdot \hat{v}_1\right)$. 
At low elasticity (e.g., $\El \le 0.07$), the axisymmetric mode exhibits negative values near the interface, with only a limited region of positive values inside the liquid, indicating that REY acts as an energy sink; its growth therefore relies primarily on interfacial mechanisms such as interfacial shear (SHG) and gas-pressure fluctuations (PRG). As $\El$ increases to $0.1$, however, $-\Real\!\left(\hat{u}_1^{*} \cdot \hat{v}_1\right)$ becomes positive within the liquid shear layer, enabling the helical mode to efficiently extract energy from the velocity gradient. Quantitatively, the distribution shifts markedly toward the jet interior as $\El$ increases from $0.07$ to $0.1$, confirming that the predominant region of momentum exchange migrates from the interface to the liquid core. This indicates that the liquid shear layer (REY) becomes the primary source of instability.

\subsection{Phase diagram of predominant mode transition}\label{sec:phase}

\begin{figure}
	\centerline{\includegraphics[width=0.55\textwidth]{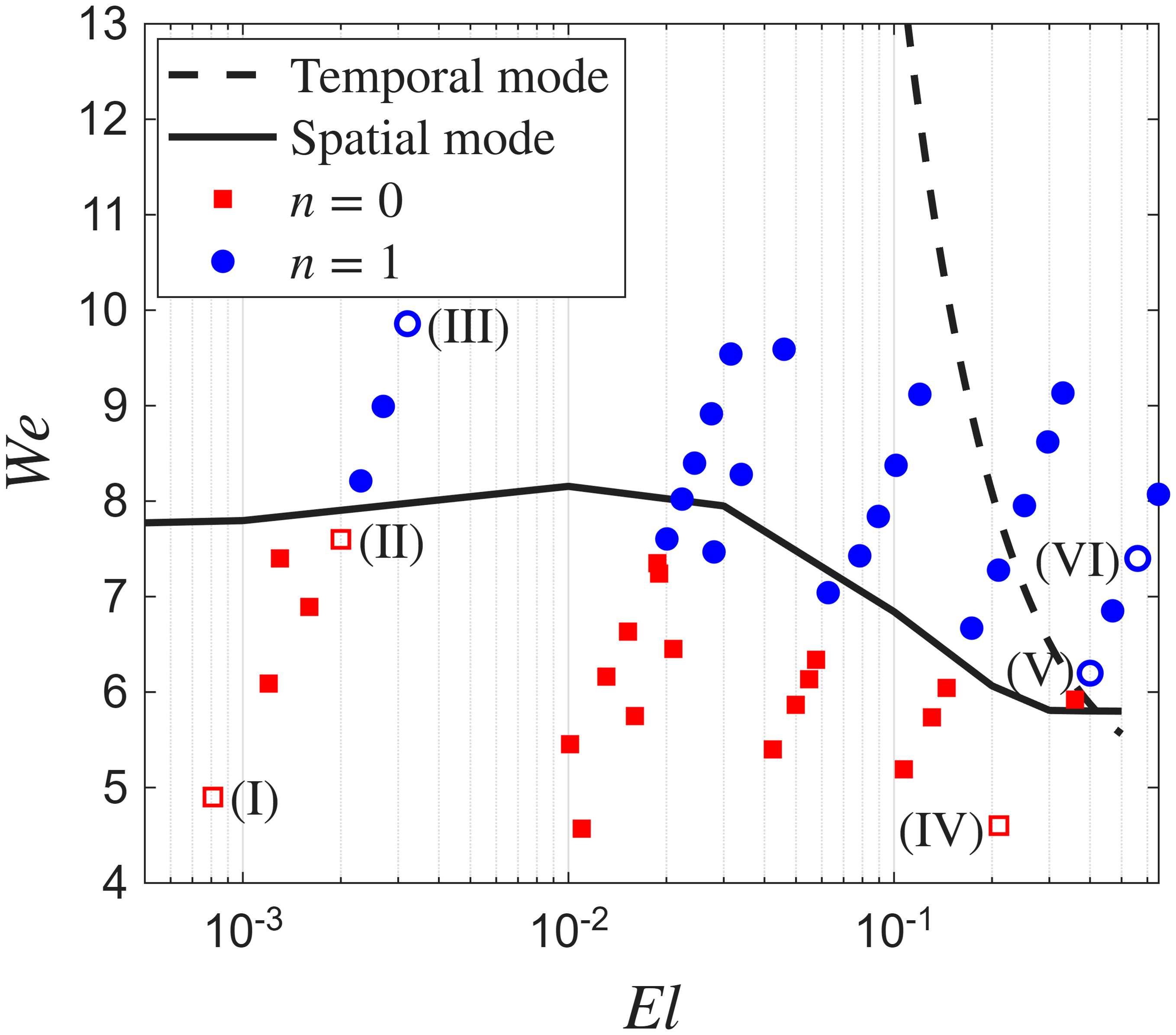}}
	\caption{
		Phase diagram of the predominant modes in the $\We$--$\El$ plane at the reference parameters 
		($\Rey = 150$, $U_{\!s}=1.33$, $K=1.2$, $Q=0.0013$, $N=0.0172$, and $X=0.9$).
		The diagram compares the predicted boundaries of predominant-mode transition obtained from the spatial 
		(solid line) and temporal (dashed line) instability analyses with the experimentally observed mode distribution (symbols).
		The experimental cases are labeled in the order of increasing elasticity number as (I)--(VI), and these labels are indicated next to the corresponding data points in the present figure. 
		The same labels (I)--(VI) are also used in figure~\ref{fig:14} to denote the corresponding experimental images, so that each experimental condition in figure~\ref{fig:14} can be directly located in the $\We$--$\El$ plane here. 
		Red open squares denote axisymmetric modes, corresponding to cases (I), (II), and (IV), whereas blue open circles denote helical modes, corresponding to cases (III), (V), and (VI).
	}
	\label{fig:11}
\end{figure}

To elucidate how the predominant instability mode varies with $\El$ and $\We$, we construct a transition map in the $\We$--$\El$ plane at the reference parameters and compare theoretical predictions with experimental results. Figure~\ref{fig:11} compares the predicted boundaries of predominant-mode transition obtained from the spatial and temporal instability analyses with the experimentally observed mode distribution. The detailed description of the experimental setup, measurement procedures, and parameter calibration is presented separately in \S\ref{sec:experiment}, while the focus here is on the quantitative comparison between theory and experiment. The spatial instability analysis reveals a clear partitioning of modes in the $\We$--$\El$ space: at relatively large $\We$, the helical mode dominates, consistent with figure~\ref{fig:6}, whereas at smaller $\We$ the predominant mode depends on $\El$---the axisymmetric mode dominates at low $\El$, while increasing $\El$ gradually shifts the dominance to the helical mode, in line with figure~\ref{fig:5}.

In contrast, the temporal instability analysis exhibits notable deficiencies: at small $\El$, it fails to capture the correct transition boundary, instead producing an approximately inverse relation in the $\We$--$\El$ plane, manifested as a sustained axisymmetric dominance over a broad $\We$ range. This behavior stands in clear contradiction to spatial instability analysis and experimental observations, which reveal the emergence of the helical mode once $\We$ exceeds a finite threshold $\We_c$. Consequently, the boundary obtained from spatial growth-rate analysis agrees much more closely with the experiments. 
More quantitatively, when $\El = 0.2$ the temporal analysis predicts a critical Weber number that is approximately $33\%$ larger than that obtained from the spatial analysis, and at $\We = 7$ the critical elasticity number from the temporal analysis exceeds the spatial prediction by more than $200\%$.
For completeness, a detailed examination of the temporal framework is presented in Appendix~\ref{Comparison with temporal instability analysis}. These discrepancies are consistent with the known limitations of the temporal (real-$k$, complex-$\omega$) formulation in strongly convective, non-parallel jet configurations such as flow focusing: temporal instability analysis emphasizes local amplification in time at a fixed station, while the instability observed in flow focusing is predominantly convective and manifests as downstream spatial growth.
Consequently, the spatial (real-$\omega$, complex-$k$) analysis provides a more faithful prediction of predominant mode boundaries and agrees more closely with experiments. Physically, the transition map highlights the interplay of inertia, capillarity, and elasticity, while clarifying the limitations of temporal instability theory in jet instabilities.

\begin{figure}
	\centerline{\includegraphics[width=0.6\textwidth]{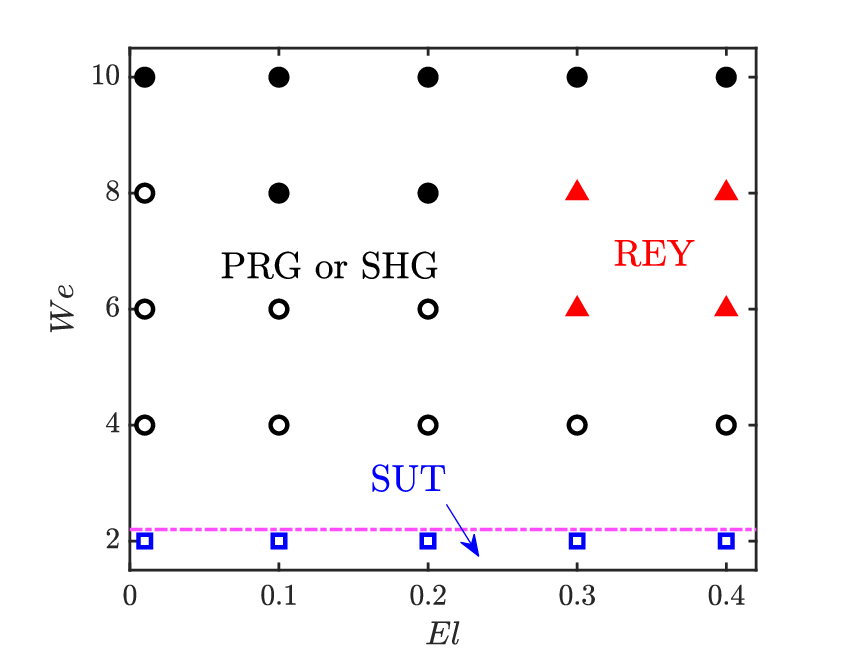}}
	\caption{Phase diagram of primary instability mechanisms in the $\We$--$\El$ plane, with $\Rey = 150$, $U_{\!s} = 1.33$, 
		$K = 1.2$, $Q = 0.0013$, $N = 0.0172$, and $X = 0.9$. 
		Open symbols denote the axisymmetric mode ($n=0$), and solid symbols denote the helical mode ($n=1$).}
	\label{fig:12}
\end{figure}

Since the transition of the predominant mode reflects a competition among distinct instability mechanisms, we further perform an energy budget analysis of the predominant mode. Figure~\ref{fig:12} presents the phase diagram of the primary instability mechanisms in the $\We$--$\El$ plane, corresponding to the operating conditions shown in figure~\ref{fig:11}. Open symbols denote the axisymmetric mode, while solid symbols denote the helical mode.
These mechanisms are identified according to the dominant energy transfer term in the budget analysis. Specifically, SUT characterizes the instability driven by surface tension, corresponding to the classical capillary instability (CPI).  
PRG and SHG are associated with the Kelvin--Helmholtz instability (KHI), reflecting the destabilizing effects of gas pressure fluctuations and velocity shear at the liquid--gas interface, respectively.  
In contrast, REY denotes an elasticity-enhanced shear-driven instability (ESI), arising from the energy exchange between the disturbance and the base flow through Reynolds stress, typically linked to shear-driven amplification within the liquid core. From a physical perspective, at low $\We$ ($\leq 2$), surface tension governs the jet dynamics, and disturbances are amplified through the interplay between interface curvature and capillary forces, such that SUT dominates. As $\We$ increases, the influence of surface tension weakens while gas pressure fluctuations and interfacial shear become more pronounced, leading to a transition toward PRG and SHG, characteristic of KHI. With a further increase in $\We$, viscoelastic jets with small elasticity behave similarly to Newtonian jets, where KHI remains the dominant mechanism. At higher elasticity (e.g., $\El = 0.3$), however, REY may emerge as the prevailing mechanism, indicating that instability is mainly driven by shear amplification inside the liquid, thus revealing a new predominant instability mechanism that differs fundamentally from that in Newtonian jets. Eventually, at sufficiently large $\We$ (e.g., $\We=10$), interfacial dynamics regain dominance, and the instability mechanism shifts back to KHI. 

\subsection{Experimental setup}\label{sec:experiment}

\begin{figure}
	\centerline{\includegraphics[width=\textwidth]{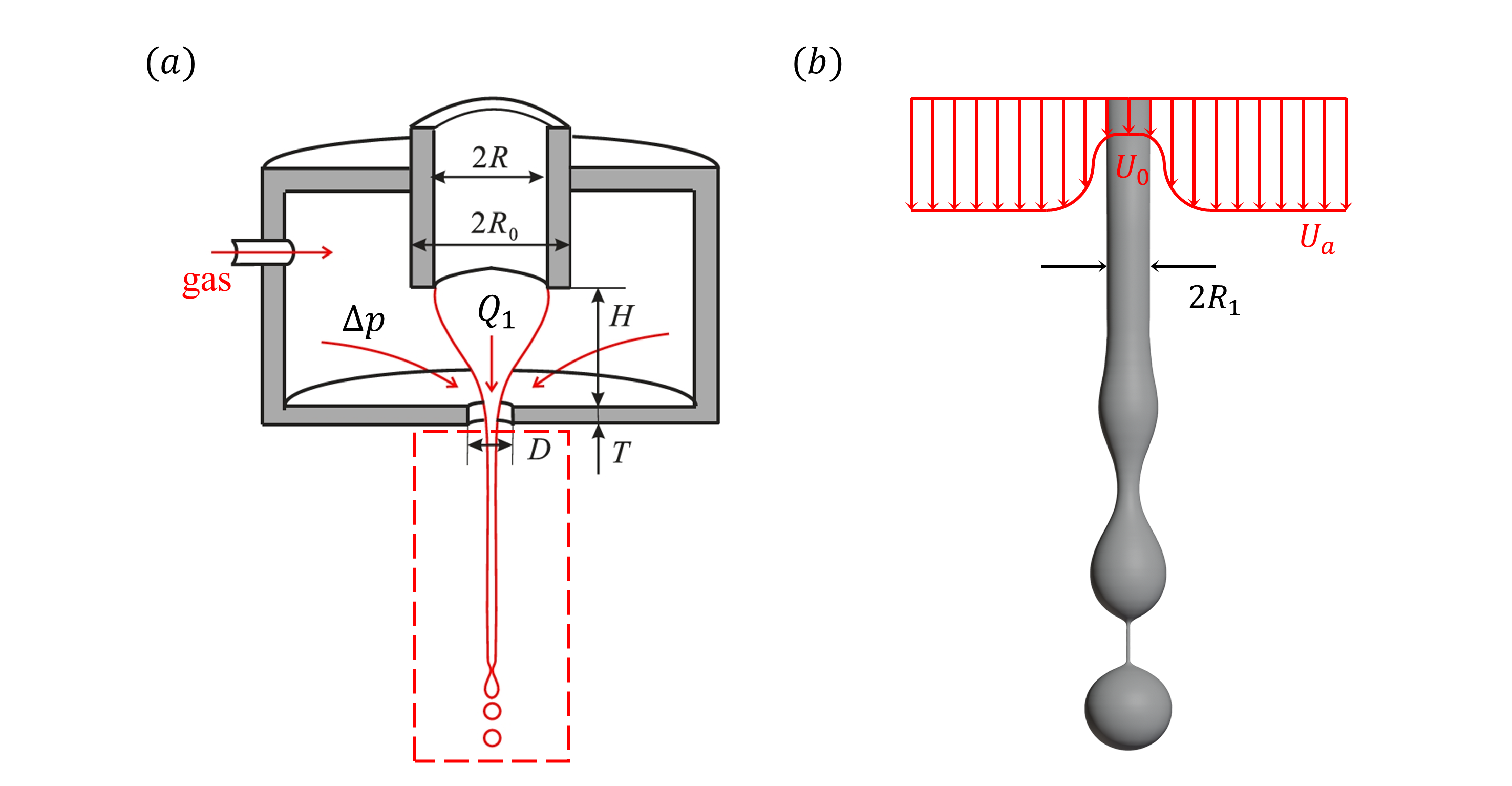}}
	\caption{($a$) Sketch of the flow focusing system considering the viscoelastic focused phase and the focusing gas phase, where $Q_1$ represent the flow rates of the focused liquid and the gas flow rate is determined by the pressure difference $\Delta p$, $R$ and $R_0$ denote the inner and outer radii of the capillary tube, $D$ and $T$ represent the diameter and thickness of the orifice, and $H$ denotes the distance between the capillary tube and the orifice, respectively.
	($b$) Jet of the $n = 0$ mode downstream of the orifice exit is shown in an enlarged view. $U_1$ and $U_{\!a}$ denote the jet and the focusing gas velocity, respectively, and $R_1$ represents the jet radius.
	$U_0$ and $U_{\!a}$ denote the base flow velocity at the jet centerline and gas flow velocity at $r=D/2$, respectively, and $R_1$ represents the jet radius.}
	\label{fig:13}
\end{figure}

This section supplements the experimental results presented in \S\ref{sec:phase} by providing a detailed description of the measurement procedures, flow-focusing setup, and data interpretation.
Rather than repeating the results, it aims to specify the experimental conditions under which the data in figure~\ref{fig:11} were obtained. The following describes the flow-focusing configuration and measurement procedures used to obtain these data.
It should be noted that experimental observations show that, in the region where the instability emerges and develops, the jet diameter varies only weakly along the axial direction and no noticeable tilt of the jet core is observed. Therefore, the assumption of a constant-radius, axisymmetric base flow adopted in the theoretical analysis is justified in this region.

To present quantitative experimental validations with the theoretical results, a flow-focusing device is used to generate a coaxially gas-driven viscoelastic jet. In flow-focusing process, the liquid jet is driven by co-flowing high-speed gas stream, presenting either the axisymmetric or the helical mode of instability as it evolves downstream.
The experimental platform is constituted by a capillary tube (with inner radius $R$ and outer radius $R_0$) and a sealed chamber with an orifice (with diameter $D$ and thickness $T$) at the bottom, coaxial with the capillary tube, as shown by the sketch map of figure~\ref{fig:13}($a$) \citep{Wang2024}.
The distance between the exit of the tube and the focusing orifice is denoted by $H$.
In our experiments, the geometric parameters are maintained constant with $R=525\, \upmu$m, $R_0=630\, \upmu$m, $H=700\, \upmu$m, $D=800\, \upmu$m and $T=700\, \upmu$m, respectively.
The sealed chamber is assembled by transparent glass windows on all directions for the convenience of observation.
A syringe pump is utilized to supply the viscoelastic focused liquid (with density $\rho_1$, viscosity $\mu_1$, flow rate $Q_1$, surface tension coefficient $\sigma$, and relaxation time $\lambda$) into the capillary tube and compressed nitrogen gas (with density $\rho_2 = 1.29\ \mathrm{kg/m^3}$, viscosity $\mu_2 = 1.81 \times 10^{-5}\ \mathrm{Pa{\cdot}s}$) is supplied by a gas cylinder connected to a flow meter, serving as the focusing gas.	One probe of the differential pressure transducer is inserted into the chamber and used to measure the gas pressure difference $\Delta p$.
The viscoelastic focused liquids employed in the experiments are polyethylene oxide (PEO) solutions dissolved in glycerol-water mixtures (with solvent viscosity $\mu_s$ and molecular weight of polymer $M_w$).
The mass fraction of PEO is fixed at $c = 0.05 \% $ for all solutions, while the mass fraction of glycerol, $c_G$, varies among different samples.
Their physical properties are listed in detail in Table~\ref{tab:3}.

\begin{table}
	\begin{center}
		\def~{\hphantom{0}}
		\begin{tabular}{lccccccc}
			& $M_w$ (g/mol) & $c_G$ (\% wt.) & $\rho_1$ (kg/m$^3$) & $\mu_s$ (mPa$\cdot$s) & $\mu_1$ (mPa$\cdot$s) & $\lambda$ ($\upmu$s) & $\sigma$ (mN/m) \\[3pt]
			& $10^5$ (100k) & 10.0 & 1024 & 1.168 & 1.238 & 8.87 & 59.70 \\
			& 200k & 9.0 & 1021 & 1.121 & 1.236 & 89.1 & 59.83 \\
			PEO solution & 300k & 8.0 & 1019 & 1.092 & 1.216 & 130 & 60.43 \\
			& 400k & 7.0 & 1019 & 1.084 & 1.138 & 294 & 61.16 \\
			& 500k & 7.0 & 1019 & 1.084 & 1.160 & 805 & 60.72 \\
			& 600k & 6.0 & 1014 & 1.037 & 1.268 & 2381 & 61.90 \\
		\end{tabular}
		\caption{Physical properties of the focused liquids.}
		\label{tab:3}
	\end{center}
\end{table}

The dimensionless parameters and characteristic scales employed in the experiment are consistent with those presented in \S\ref{sec:theoretical model}, and are given by:
\begin{equation}\label{eq:3.3}
	\Rey = \frac{\rho_1 U_0 R_1}{\mu_1}, \We = \frac{\rho_1 U_0^2 R_1}{\sigma}, \El = \frac{\lambda \mu_1}{\rho_1 R_1^{2}}. 
\end{equation}

It is well known that changing the mass fraction $c$ or molecular weight of polymer $M_w$ in PEO solutions leads to changes in both the solution viscosity $\mu_1$ and the relaxation time $\lambda$ together.
Although different solution formulations are attempted to achieve approximately identical $\mu_1$, it remains difficult to obtain a sufficient number of data points while keeping $\Rey$ constant.
However, the following characteristics are present in flow focusing and have been experimentally verified\citep{GananCalvo1998,Herrada2008,Si2009}:

(i) The gas flow velocities upstream and downstream of the focused orifice are governed by the Bernoulli equation $\Delta p = \rho_2 U_{\!a}^2 /2$, so the external focusing gas velocity $U_{\!a} = \sqrt{2 \Delta p /\rho_2} $ is obtained.

(ii) The gas-to-liquid momentum ratio is very close to unity in flow focusing, i.e. $\frac{1}{2} \rho_{1} U_{0}^{2}=\frac{1}{2} \rho_{2} U_{a}^{2}$, which is determined by the simplified averaged momentum equation in the axial direction in which the viscous terms are neglected. Therefore, the jet velocity can be given by $U_0 = \sqrt{\rho_2 /\rho_1} \ U_{\!a}$.

(iii) The axial velocity of the jet is approximately uniform, so that $Q_1 = \pi R_1^2 U_0$.

From the above relationships, the jet radius is obtained as $R_{1}=\left(\rho_{1}/\left({2 \pi^{2} \Delta p}\right)\right)^{1 / 4} Q_{1}^{1 / 2}$, which is in good quantitative agreement with experimental measurements \citep{GananCalvo1998,Si2009}. Accordingly, the dimensionless parameters $\Rey$, $\We$, and $\El$ can be rewritten as follows:
    \begin{equation}\label{eq:3.4}
		\Rey = \left( \frac{2 \rho_{1}^{3}}{\pi^{2} \mu_{1}^{4}} \right)^{1/4}
		\left(Q_{1}^{2} \Delta p \right)^{1/4},
	\end{equation}
	\begin{equation}\label{eq:3.5}
		\We = \left( \frac{8 \rho_{1}}{\pi^{2} \sigma^{4}} \right)^{1/4}
		\left(Q_{1}^{2} \Delta p \right)^{1/4}
		(\Delta p)^{1/2},
	\end{equation}
	\begin{equation}\label{eq:3.6}
		\El = \left( \frac{2 \pi^{2} \mu_{1}^{2} \lambda^{2}}{\rho_{1}^{3}} \right)^{1/2}
		\left(Q_{1}^{2} \Delta p \right)^{-1/2}
		\Delta p .
\end{equation}
Therefore, by controlling $\left(Q_{1}^{2} \Delta p\right)$ constant and varying only $\Delta p$ in the experiments, $Re$ $( \approx 150)$ can be conveniently maintained unchanged, while both $\We$ $(4\sim10)$ and $\El$ $(10^{-3}\sim1)$ can be simultaneously adjusted.
Using this method, a set of data points in the $\We$--$\El$ plane can be obtained with a single viscoelastic solution, and these points will form a sloped trajectory in figure~\ref{fig:11}.

\begin{figure}
	\centering
	\includegraphics[width=0.8\textwidth]{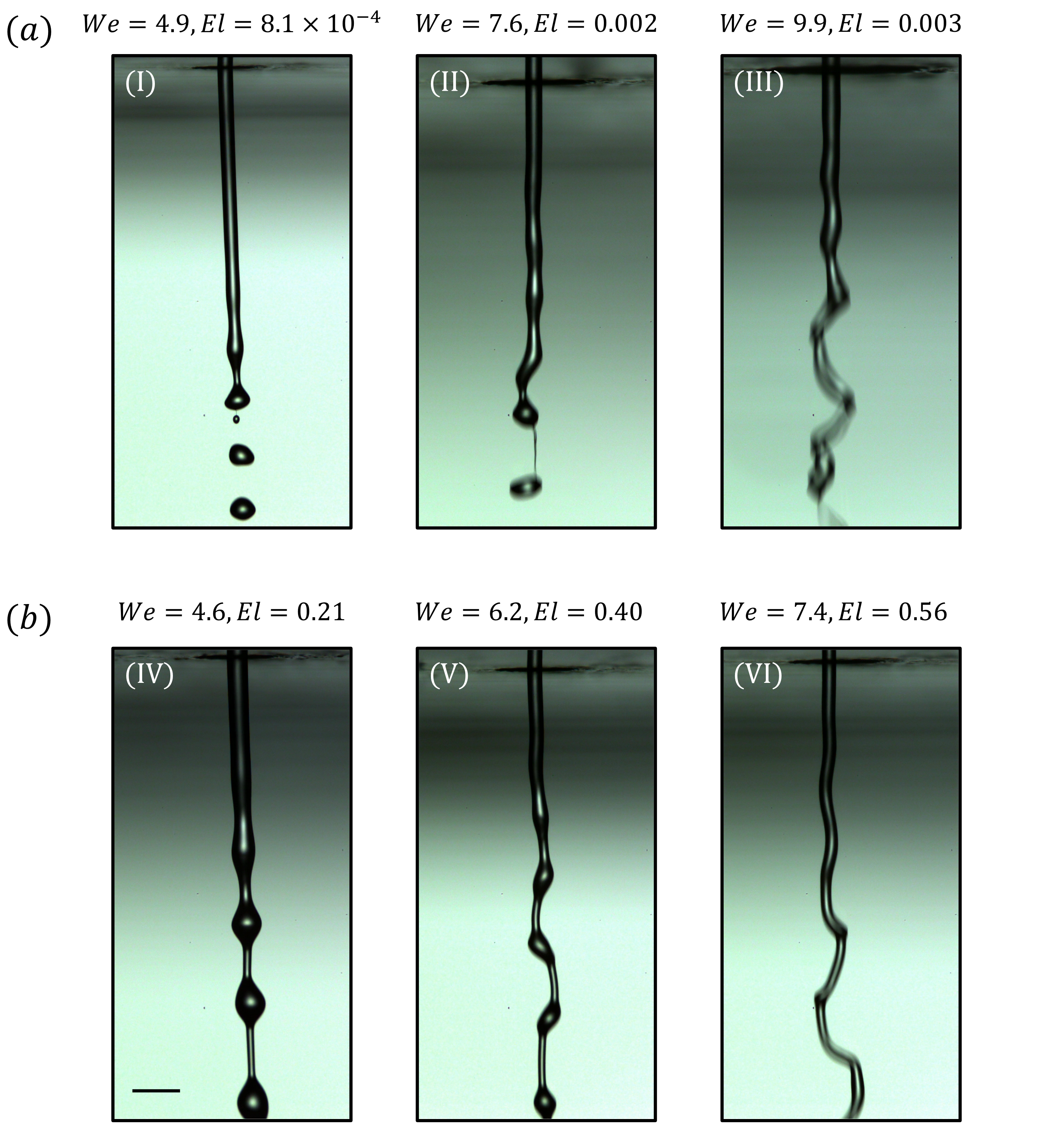}
	\caption{The transition of the jets from $n = 0$ mode to the $n = 1$ mode is observed in the experiments with $\Rey \approx 150$. Scale bars of 400 $\mu$m are provided in the figure.
		($a$) For the polymer with a molecular weight of $M_w=100$k g/mol, where elastic effects are relatively weak, the non-axisymmetric instability of the jet is primarily induced by an increase in the $\We$ number.
		($b$) For the polymer with a molecular weight of $M_w=600$k g/mol, where elastic effects are stronger, the helical instability is mainly caused by an increase in the $\El$ number. The corresponding experimental results are displayed in the $\We$--$\El$ plane using open symbols (see figure~\ref{fig:11}).}
	\label{fig:14}
\end{figure}

As predicted by theory, with increasing $\We$ and $\El$, the dynamic evolution of the jet transitioning from the axisymmetric mode to the helical mode is clearly demonstrated in figure~\ref{fig:14}. The corresponding experimental results are displayed in the $\We$--$\El$ plane using hollow symbols (see figure~\ref{fig:11}).
When the solution exhibits weak elasticity, as shown in figure~\ref{fig:14}($a$), the variation in $\El$ achieved in the experiments is relatively small. With the increase in $\We$, the dominant mode gradually transforms from the axisymmetric mode to the helical mode, which validates the theoretical analysis presented in figure~\ref{fig:6}.
In contrast, when the elasticity of the solution becomes strong, as depicted in figure~\ref{fig:14}($b$), the increase in $\El$ has a more pronounced influence on the mode transition: even at comparable $\We$ values, a higher $\El$ promotes the earlier emergence of the helical mode, which can be seen by comparing figure~\ref{fig:14}($a$) at $\We=7.6$ with figure~\ref{fig:14}($b$) at $\We=7.4$.
This observation is consistent with the trends shown in figure~\ref{fig:5}.

\section{Conclusions and outlook}\label{Sec:conclusion}

This study investigates the instability and predominant mode transition of viscoelastic jets in a flow-focusing configuration by combining spatial linear instability analysis with controlled experiments. In the theoretical study, a non-uniform hyperbolic-tangent base profile is employed to represent the realistic core–sheath shear. The Oldroyd-B constitutive model, coupled with Chebyshev spectral discretization, enables resolution of the growth characteristics for both the axisymmetric and helical modes. Energy budget analyses are utilized to study the underlying physical mechanisms. The analysis shows that at low Weber numbers, the jet evolves in an axisymmetric manner, where interfacial pressure fluctuations dominate the instability. As the Weber number increases, the growing inertia drives the predominant instability to shift from the axisymmetric to the helical mode, during which the jet core exhibits a helical oscillation. By varying the elasticity number $\El$, the influence of elasticity is further clarified. When elasticity is weak, the instability is primarily governed by gas-pressure fluctuations. As elasticity increases, the disturbance structures migrate from the interface toward the jet interior, where elastic effects enhance the coupling between velocity perturbations and the base-flow shear. This coupling gives rise to a new predominant instability mechanism---elasticity-enhanced shear-driven instability mechanism (ESI), distinct from the capillary instability (CPI) or Kelvin--Helmholtz instability (KHI) in Newtonian jets. In this regime, elasticity does not directly provide energy but alters how kinetic energy is extracted from the mean shear, thereby promoting the predominance of the helical mode. A phase diagram of predominant modes in the $\We$--$\El$ space is further plotted, providing a comprehensive framework that integrates the effects of capillarity, inertia, and elasticity, and quantitatively delineates the transition boundary between the axisymmetric and helical modes. We also investigate the transitions of the predominant instability mechanisms across the $\We$--$\El$ space.

Complementary experiments performed in a flow-focusing device validated these theoretical predictions. By systematically varying the solution elasticity through PEO molecular weight and concentration, parameter variations across the $\We$--$\El$ space were realized and compared with theoretical phase boundaries. The observed jet morphologies confirmed that increasing $\We$ induces mode transition in weakly elastic solutions, while strongly elastic solutions exhibit earlier emergence of the helical mode due to enhanced elastic effects. The agreement between theory and experiment underscores the robustness of the spatial instability analysis and highlights the limitations of temporal approaches, which consistently overpredict axisymmetric instability. Taken together, the constructed $\We$--$\El$ phase diagram and the unified theoretical-experimental framework provide mechanistic guidance for controlling breakup and fiber formation in viscoelastic jets, with implications for microfluidics, inkjet printing, and spray technologies. 

Finally, several possible avenues for future research are outlined and briefly discussed in the following subsection.

\subsection{Extension to highly elastic regimes}\label{sec:outlook}
The present analysis has focused on dilute polymer solutions and on a range of moderate elasticity numbers, within which the Oldroyd--B model is widely employed and is generally regarded as providing a reasonable description of the flow. In this parameter regime, the elasticity-enhanced shear-driven instability identified in this study is primarily governed by the interaction between elastic stresses and the shear layer, rather than by extreme extensional stresses. At higher levels of elasticity, in particular within the strongly extensional region near the focusing point, constitutive equations incorporating finite polymer extensibility (e.g. FENE-P) are expected to be more appropriate. A natural continuation of the present work is therefore to extend the spatial-stability and energy-budget analyses to such models, so as to examine how bounded extensional viscosity alters the quantitative structure of the $\We$–$\El$ phase diagram and to evaluate the robustness of the ESI mechanism and the associated mode-transition scenario over a broader elastic regime.

\subsection{From local to global instability analysis}
The present study has been conducted within a local linear stability framework, in which the base flow is assumed to be parallel in the streamwise direction and the viscoelastic stresses are taken from the corresponding steady solution. This approach has proved effective in identifying the predominant instability mechanisms of the flow-focused viscoelastic jet and in quantifying the competition among elasticity, inertia and surface tension in the selection of the most amplified modes. Nevertheless, in real flow-focusing configurations the jet experiences downstream acceleration and thinning after passing through the contraction, so that its radius, velocity and viscoelastic stresses vary along the axial direction. In particular, because of stress memory and the persistence of unrelaxed elastic tension in viscoelastic liquids, the deformation history may play an important role in the overall instability dynamics \citep{PonceTorres2016}. These observations suggest that streamwise non-parallelism and stress-history effects may influence the global mode selection mechanism in a fundamental way.

In Newtonian fluids, the theories of local and global instability have been systematically developed and applied to spatially developing flows such as jets, wakes and boundary layers \citep{Huerre1990,Chomaz2005,Theofilis2011}. Global stability analyses have demonstrated that, when a region of local absolute instability exists, the system may sustain a self-excited global mode, whose frequency and spatial structure are selected in a compact ``wavemaker'' region. Jet and thread flows provide representative examples. For instance, global analyses of gravitationally stretched jets \citep{Sauter2005,RubioRubio2013}, low-density jets \citep{Coenen2017} and surfactant-laden falling jets \citep{Sun2024} have quantitatively predicted self-excited oscillation frequencies in good agreement with experiments. Closely related to the present configuration, a global stability analysis of axisymmetric liquid--liquid flow focusing has recently been carried out \citep{Cabezas2021}, showing that both the contraction region and the downstream slender jet can act as sources of global instability, again in agreement with experimental observations.

On the other hand, studies of viscoelastic jets and threads have revealed that axially varying unrelaxed elastic tension and stress evolution may affect instability and breakup dynamics \citep{PonceTorres2016,Xie2019}. These works highlight the importance of stress history and streamwise evolution, and indicate that local analyses based on the parallel-flow assumption may not be sufficient to fully capture the global dynamics of viscoelastic jets.

A natural continuation of the present work is therefore to construct a non-parallel viscoelastic base flow for the flow-focusing configuration, in which the jet radius, axial velocity and polymeric stresses evolve jointly along the streamwise coordinate, for example within a slender-jet approximation consistent with the present two-phase formulation. On top of such a base state, a global eigenvalue analysis could be carried out to determine the globally selected modes and to reassess the onset of absolute versus convective instability in the presence of elasticity. Given the success of global-mode analyses in Newtonian jets and in flow-focusing systems, their extension to viscoelastic flow-focused jets appears both feasible and promising. One may expect that stress memory and axially evolving elastic tension will enrich the global-instability scenarios far beyond those encountered in Newtonian flows, and a comprehensive investigation of these effects is left for future work.

\backsection[Funding]{
	This work was supported by the Nation Natural Science Foundation of China (Grants No. 12272188, No. 12272372, No. 12388101), the Youth Innovation Promotion Association CAS (Grants No. 2018491, No. 2023477), the Natural Science Foundation of Inner Mongolia Autonomous Region (Grant No. 2025ZDLH007), the Nation Science Foundation for Distinguished Young Scholars of Inner Mongolia Autonomous Region of China (Grant No. 2023JQ16), USTC Tang Scholar and the New Cornerstone Science Foundation through the XPLORER PRIZE.}

\backsection[Declaration of interests]{
	The authors declare no conflict of interest.}

\backsection[Author ORCIDs]{\\
	Jiawei Li https://orcid.org/0009-0003-1022-565X;\\
	Ming Wang https://orcid.org/0009-0003-3123-8476;\\
	Kai Mu https://orcid.org/0000-0002-4743-2332;\\
	Zhaodong Ding https://orcid.org/0000-0003-3560-9003;\\
	Ting Si https://orcid.org/0000-0001-9071-8646.}

%
%
%
%
%

\appendix

\section{Linearized equations for disturbances}\label{app:linearized_equations}

\begin{equation}\label{eq:A1}
	\boldsymbol{\nabla} \cdot \tilde{\boldsymbol{V}}_1 = 0,
\end{equation}
\begin{equation}\label{eq:A2}
	\frac{\partial \tilde{\boldsymbol{V}}_1}{\partial t} 
	+ \tilde{\boldsymbol{V}}_1 \cdot \boldsymbol{\nabla} \boldsymbol{U}_1 
	+ \boldsymbol{U}_1 \cdot \boldsymbol{\nabla} \tilde{\boldsymbol{V}}_1
	= -\boldsymbol{\nabla} \tilde{p}_1 
	+ \boldsymbol{\nabla} \cdot \tilde{\boldsymbol{T}}_{\!S} 
	+ \boldsymbol{\nabla} \cdot \tilde{\boldsymbol{T}}_{\!P},
\end{equation}

where
\begin{equation}\label{eq:A3}
	\tilde{\boldsymbol{T}}_{\!S} = \frac{X}{\Rey} 
	\left[
	\boldsymbol{\nabla} \tilde{\boldsymbol{V}}_1 
	+ \left(\boldsymbol{\nabla} \tilde{\boldsymbol{V}}_1\right)^\mathrm{T}
	\right].
\end{equation}
\begin{equation}\label{eq:A4}
	\begin{aligned}
		\tilde{\boldsymbol{T}}_{\!P} 
		&+ \Wi \Big[
		\frac{\partial \tilde{\boldsymbol{T}}_{\!P}}{\partial t}
		+ \boldsymbol{U}_1 \cdot \boldsymbol{\nabla} \tilde{\boldsymbol{T}}_{\!P}
		- \left(\boldsymbol{\nabla}\boldsymbol{U}_1\right)^{\mathrm{T}} \cdot \tilde{\boldsymbol{T}}_{\!P}
		+ \tilde{\boldsymbol{V}}_1 \cdot \boldsymbol{\nabla} \boldsymbol{\Gamma}
		- \left(\boldsymbol{\nabla} \tilde{\boldsymbol{V}}_1\right)^{\mathrm{T}} \cdot \boldsymbol{\Gamma}
		\Big] \\
		&= \frac{1 - X}{\Rey} \Big[
		\boldsymbol{\nabla} \tilde{\boldsymbol{V}}_1 
		+ \left(\boldsymbol{\nabla} \tilde{\boldsymbol{V}}_1\right)^{\mathrm{T}}
		\Big],
	\end{aligned}
\end{equation}
\begin{equation}\label{eq:A5}
	\boldsymbol{\nabla} \cdot \tilde{\boldsymbol{V}}_2 = 0,
\end{equation}
\begin{equation}\label{eq:A6}
	\frac{\partial \tilde{\boldsymbol{V}}_2}{\partial t}
	+ \left(\tilde{\boldsymbol{V}}_2 \cdot \boldsymbol{\nabla}\right) \tilde{\boldsymbol{U}}_2
	+ \left(\tilde{\boldsymbol{U}}_2 \cdot \boldsymbol{\nabla}\right) \tilde{\boldsymbol{V}}_2
	= -\frac{1}{Q} \boldsymbol{\nabla} \tilde{p}_2
	+ \frac{N}{Q \Rey} \boldsymbol{\nabla}^2 \tilde{\boldsymbol{V}}_2,
\end{equation}

For the boundary conditions, at the centerline \(r=0\), we have:
\begin{equation}\label{eq:A7}
	\frac{\partial \tilde{u}_1}{\partial \theta} 
	= \frac{\partial \tilde{v}_1}{\partial \theta} - \tilde{w}_1 
	= \tilde{v}_1 + \frac{\partial \tilde{w}_1}{\partial \theta} 
	= \frac{\partial \tilde{p}_1}{\partial \theta} 
	= 0.
\end{equation}

On the perturbed interface \(r=1+\eta\), we impose:
\begin{equation}\label{eq:A8}
	\tilde{u}_1 - \tilde{u}_2 
	+ \left( \frac{\mathrm{d}U_1}{\mathrm{d}r} - \frac{\mathrm{d}U_2}{\mathrm{d}r} \right) \eta = 0,
\end{equation}

\begin{equation}\label{eq:A9}
	\tilde{v}_1 = \tilde{v}_2,
\end{equation}

\begin{equation}\label{eq:A10}
	\tilde{w}_1 = \tilde{w}_2,
\end{equation}

\begin{equation}\label{eq:A11}
	\tilde{v}_1 = \left( \frac{\partial}{\partial t} + U_1 \frac{\partial}{\partial z} \right) \eta,
\end{equation}
\begin{equation}\label{eq:A12}
	\tilde{p}_1 - \tilde{p}_2 
	+ \frac{2N}{\Rey} \frac{\partial \tilde{v}_2}{\partial r} 
	- \frac{2X}{\Rey} \frac{\partial \tilde{v}_1}{\partial r} 
	- \tilde{\tau}_{rr}
	+ \frac{1}{\We} \left( 1 + \frac{\partial^2}{\partial \theta^2} + \frac{\partial^2}{\partial z^2} \right) \eta = 0,
\end{equation}
\begin{equation}\label{eq:A13}
	\frac{N}{\Rey} 
	\left(
	\frac{\partial \tilde{w}_2}{\partial r} 
	+ \frac{\partial \tilde{v}_2}{\partial \theta} 
	- \tilde{w}_2 
	\right)
	- \frac{X}{\Rey} 
	\left(
	\frac{\partial \tilde{w}_1}{\partial r} 
	+ \frac{\partial \tilde{v}_1}{\partial \theta} 
	- \tilde{w}_1 
	\right)
	- \tilde{\tau}_{r \theta} = 0,
\end{equation}
\begin{equation}\label{eq:A14}
	\frac{N}{\Rey} \left(
	\frac{\partial \tilde{u}_2}{\partial r} 
	+ \frac{\partial \tilde{v}_2}{\partial z}
	+ \frac{\mathrm{d}^2 U_2}{\mathrm{d} r^2}\,\eta 
	\right)
	+ \frac{X}{\Rey} \left(
	\frac{\partial \tilde{u}_1}{\partial r} 
	+ \frac{\partial \tilde{v}_1}{\partial z}
	+ \frac{\mathrm{d}^2 U_1}{\mathrm{d} r^2}\,\eta 
	\right) 
	- \tilde{\tau}_{rz}
	- \frac{\mathrm{d} \Gamma_{rz}}{\mathrm{d} r}\,\eta
	+ \Gamma_{zz} \frac{\partial \eta}{\partial z} = 0.
\end{equation}

At the outer boundary \(r=a\), we apply:
\begin{equation}\label{eq:A15}
	\tilde{u}_2 = \tilde{v}_2 = \tilde{w}_2 = \tilde{p}_2 = 0.
\end{equation}

We formulate the instability problem by substituting Eq.~(\ref{eq:2.18}) into Eq.~(\ref{eq:A1})--Eq.~(\ref{eq:A15}), which leads to the following equations:
\begin{equation}\label{eq:A16}
	\frac{1}{r}\hat{v}_1 + \frac{\partial \hat{v}_1}{\partial r} + \frac{\mathrm{i}n}{r}\hat{w}_1
	= -\mathrm{i}k \hat{u}_1,
\end{equation}
\begin{equation}\label{eq:A17}
	\begin{aligned}
		\frac{X}{\Rey}\frac{\partial^2 \hat{u}_1}{\partial r^2} 
		+ \frac{X}{\Rey}\frac{1}{r}\frac{\partial \hat{u}_1}{\partial r}
		- \left(\frac{X}{\Rey}\frac{n^2}{r^2} - \mathrm{i}\omega \right)\hat{u}_1
		- \frac{\mathrm{d}U_1}{\mathrm{d}r}\hat{v}_1
		+ \frac{\partial \hat{\tau}_{rz}}{\partial r}
		+ \frac{1}{r}\hat{\tau}_{rz} + \frac{\mathrm{i}n}{r}\hat{\tau}_{\theta z} \\
		= k\left(\mathrm{i}U_1\hat{u}_1 + \mathrm{i}\hat{p}_1 - \mathrm{i}\hat{\tau}_{zz} + \frac{X}{\Rey}k\hat{u}_1\right),
	\end{aligned}
\end{equation}
\begin{equation}\label{eq:A18}
	\begin{aligned}
		\frac{X}{\Rey}\frac{\partial^2 \hat{v}_1}{\partial r^2}
		+ \frac{X}{\Rey}\frac{1}{r}\frac{\partial \hat{v}_1}{\partial r}
		- \left(\frac{X}{\Rey}\frac{1+n^2}{r^2} - \mathrm{i}\omega \right)\hat{v}_1
		- \frac{X}{\Rey}\frac{2\mathrm{i}n}{r^2}\hat{w}_1
		- \frac{\partial \hat{p}_1}{\partial r}
		+ \frac{\partial \hat{\tau}_{rr}}{\partial r}
		+ \frac{1}{r}\hat{\tau}_{rr} \\
		- \frac{\mathrm{i}n}{r}\hat{\tau}_{\theta z}
		- \frac{1}{r}\hat{\tau}_{\theta\theta}
		= k\left(\mathrm{i}U_1\hat{v}_1 - \mathrm{i}\hat{\tau}_{rz} + \frac{X}{\Rey}k\hat{v}_1\right),
	\end{aligned}
\end{equation}
\begin{equation}\label{eq:A19}
	\begin{aligned}
		\frac{X}{\Rey}\frac{2\mathrm{i}n}{r^2}\hat{v}_1
		+ \frac{X}{\Rey}\frac{\partial^2 \hat{w}_1}{\partial r^2}
		+ \frac{X}{\Rey}\frac{1}{r}\frac{\partial \hat{w}_1}{\partial r}
		- \left(\frac{X}{\Rey}\frac{1+n^2}{r^2} - \mathrm{i}\omega \right)\hat{w}_1
		- \frac{\mathrm{i}n}{r}\hat{p}_1
		+ \frac{\partial \hat{\tau}_{r\theta}}{\partial r}
		+ \frac{2}{r}\hat{\tau}_{r\theta} \\
		+ \frac{\mathrm{i}n}{r}\hat{\tau}_{\theta\theta}
		= k\left(\mathrm{i}U_1 \hat{v}_1 - \mathrm{i}\hat{\tau}_{\theta z} + \frac{X}{\Rey}k\hat{w}_1\right),
	\end{aligned}
\end{equation}
\begin{equation}\label{eq:A20}
	\begin{aligned}
		\frac{2(1-X)}{\Rey^2}\frac{\partial \hat{v}_1}{\partial r}
		- \frac{1}{\Rey}\left(1 - \mathrm{i}\omega \El\Rey\right)\hat{\tau}_{rr}
		= k\left(-\mathrm{i}\frac{2(1-X)}{\Rey}\El \frac{\mathrm{d}U_1}{\mathrm{d}r}\hat{v}_1
		+ \mathrm{i}\El U_1\hat{\tau}_{rr}\right),
	\end{aligned}
\end{equation}
\begin{equation}\label{eq:A21}
	\begin{aligned}
		-\mathrm{i}\frac{1-X}{\Rey^2}\frac{\partial \hat{v}_1}{\partial r}
		+ \frac{1-X}{\Rey^2}\frac{\partial \hat{w}_1}{\partial r}
		- \frac{1}{\Rey}\left(1 - \mathrm{i}\omega \El\Rey\right)\hat{\tau}_{r\theta}
		= k\Bigg(
		-\frac{1-X}{\Rey^2}\hat{u}_1 \\
		- \mathrm{i}\frac{1-X}{\Rey}\El\frac{\mathrm{d}U_1}{\mathrm{d}r}\hat{w}_1
		+ \mathrm{i}\El U_1\hat{\tau}_{r\theta}
		\Bigg),
	\end{aligned}
\end{equation}
\begin{equation}\label{eq:A22}
	\begin{aligned}
		\frac{1-X}{\Rey^2}\frac{\partial \hat{u}_1}{\partial r}
		+ \frac{1-X}{\Rey}\El\frac{\mathrm{d}U_1}{\mathrm{d}r}\frac{\partial \hat{v}_1}{\partial r}
		- \frac{1-X}{\Rey}\El\frac{\mathrm{d}^2 U_1}{\mathrm{d}r^2}\hat{v}_1
		+ \El\frac{\mathrm{d}U_1}{\mathrm{d}r}\hat{\tau}_{rr} \\
		- \frac{1}{\Rey}\left(1 - \mathrm{i}\omega \El\Rey\right)\hat{\tau}_{rz}
		= k\Bigg[
		-\mathrm{i}\frac{1-X}{\Rey}\El\frac{\mathrm{d}U_1}{\mathrm{d}r}\hat{u}_1
		- \mathrm{i}\frac{1-X}{\Rey^2}\hat{v}_1 \\
		- 2\mathrm{i}(1-X)\El^2\left(\frac{\mathrm{d}U_1}{\mathrm{d}r}\right)^2\hat{v}_1
		+ \mathrm{i}\El U_1\hat{\tau}_{rz}
		\Bigg],
	\end{aligned}
\end{equation}
\begin{equation}\label{eq:A23}
	\begin{aligned}
		-\frac{2(1-X)}{\Rey^2}\frac{\partial \hat{v}_1}{\partial r}
		- \frac{1}{\Rey}\left(1 - \mathrm{i}\omega \El\Rey\right)\hat{\tau}_{\theta\theta}
		= k\left(\mathrm{i}\frac{2(1-X)}{\Rey^2}\hat{u}_1
		+ \mathrm{i}\El U_1 \hat{\tau}_{\theta\theta}\right),
	\end{aligned}
\end{equation}
\begin{equation}\label{eq:A24}
	\begin{aligned}
		\frac{1-X}{\Rey^2}\frac{\mathrm{i}n}{r}\hat{u}_1
		+ \frac{1-X}{\Rey}\El\frac{\mathrm{d}U_1}{\mathrm{d}r}\frac{\partial \hat{w}_1}{\partial r}
		- \frac{1-X}{\Rey}\frac{1}{r}\El\frac{\mathrm{d}U_1}{\mathrm{d}r}\hat{w}_1 
		+ \El\frac{\mathrm{d}U_1}{\mathrm{d}r}\hat{\tau}_{r\theta} 
		- \frac{1}{\Rey}\left(1 - \mathrm{i}\omega \El\Rey\right)\hat{\tau}_{\theta z} \\
		= k\Bigg[
		-\mathrm{i}\frac{1-X}{\Rey^2}\hat{w}_1
		- 2\mathrm{i}(1-X)\El^2\left(\frac{\mathrm{d}U_1}{\mathrm{d}r}\right)^2\hat{w}_1 
		+ \mathrm{i}\El U_1 \hat{\tau}_{\theta z}
		\Bigg],
	\end{aligned}
\end{equation}
\begin{equation}\label{eq:A25}
	\begin{aligned}
		\frac{2(1-X)}{\Rey}\El \frac{\mathrm{d}U_1}{\mathrm{d}r}\hat{u}_1
		- 4(1-X)\El^2\frac{\mathrm{d}U_1}{\mathrm{d}r}\frac{\mathrm{d}^2 U_1}{\mathrm{d}r^2}\hat{v}_1
		+ 2\El\frac{\mathrm{d}U_1}{\mathrm{d}r}\hat{\tau}_{rz}
		- \frac{1}{\Rey}\left(1 - \mathrm{i}\omega \El\Rey\right)\hat{\tau}_{zz} \\
		= k\Bigg[
		-\mathrm{i}\frac{2(1-X)}{\Rey^2}\hat{u}_1
		- 4\mathrm{i}(1-X)\El^2\left(\frac{\mathrm{d}U_1}{\mathrm{d}r}\right)^2\hat{u}_1
		+ \mathrm{i}\El U_1 \hat{\tau}_{zz}
		\Bigg],
	\end{aligned}
\end{equation}
\begin{equation}\label{eq:A26}
	\frac{1}{r}\hat{v}_2 + \frac{\partial \hat{v}_2}{\partial r} + \frac{\mathrm{i}n}{r}\hat{w}_2
	= -\mathrm{i}k \hat{u}_2,
\end{equation}
\begin{equation}\label{eq:A27}
	\begin{aligned}
		\frac{N}{Q\Rey}\frac{\partial^2 \hat{u}_2}{\partial r^2}
		+ \frac{N}{Q\Rey}\frac{1}{r}\frac{\partial \hat{u}_2}{\partial r}
		- \left(\frac{N}{Q\Rey}\frac{n^2}{r^2} - \mathrm{i}\omega \right)\hat{u}_2
		- \frac{\mathrm{d}U_2}{\mathrm{d}r}\hat{v}_2 
		= k\Bigg(
		\mathrm{i}U_2\hat{u}_2 \\
		+ \frac{\mathrm{i}}{Q}\hat{p}_2
		+ \frac{N}{Q\Rey}k\hat{u}_2
		\Bigg),
	\end{aligned}
\end{equation}
\begin{equation}\label{eq:A28}
	\begin{aligned}
		\frac{N}{Q\Rey}\frac{\partial^2 \hat{v}_2}{\partial r^2}
		+ \frac{N}{Q\Rey}\frac{1}{r}\frac{\partial \hat{v}_2}{\partial r}
		- \left(\frac{N}{Q\Rey}\frac{1+n^2}{r^2} - \mathrm{i}\omega \right)\hat{v}_2
		- \frac{N}{Q\Rey}\frac{2\mathrm{i}n}{r^2}\hat{w}_2
		- \frac{1}{Q}\frac{\partial \hat{p}_2}{\partial r} \\
		= k\left(
		\mathrm{i}U_2 \hat{v}_2
		+ \frac{N}{Q\Rey}k\hat{v}_2
		\right),
	\end{aligned}
\end{equation}
\begin{equation}\label{eq:A29}
	\begin{aligned}
		\frac{N}{Q\Rey}\frac{2\mathrm{i}n}{r^2}\hat{v}_2
		+ \frac{N}{Q\Rey}\frac{\partial^2 \hat{w}_2}{\partial r^2}
		+ \frac{N}{Q\Rey}\frac{1}{r}\frac{\partial \hat{w}_2}{\partial r}
		- \left(\frac{N}{Q\Rey}\frac{1+n^2}{r^2} - \mathrm{i}\omega \right)\hat{w}_2
		- \frac{1}{Q}\frac{\mathrm{i}n}{r}\hat{p}_2 \\
		= k\left(
		\mathrm{i}U_2\hat{w}_2
		+ \frac{N}{Q\Rey}k\hat{w}_2
		\right).
	\end{aligned}
\end{equation}

At the axis of symmetry ($r=0$), the boundary conditions are
\begin{subequations}\label{eq:A30}
	\begin{align}
		\frac{\partial \hat{u}_1}{\partial r} = \hat{v}_1 = \hat{w}_1 
		= \frac{\partial \hat{p}_1}{\partial r} = 0,
		\qquad &\text{for } n=0, \label{eq:2.33a}\\
		\hat{u}_1 = \hat{p}_1 = 0, \quad 
		\hat{v}_1 + \mathrm{i}\hat{w}_1 = 0, \quad 
		2\frac{\mathrm{d}\hat{v}_1}{\mathrm{d}r} + \mathrm{i}\frac{\mathrm{d}\hat{w}_1}{\mathrm{d}r} = 0,
		\qquad &\text{for } n=1, \label{eq:2.33b}\\
		\hat{u}_1 = \hat{v}_1 = \hat{w}_1 = \hat{p}_1 = 0,
		\qquad &\text{for } n>1. \label{eq:2.33c}
	\end{align}
\end{subequations}

At the liquid--gas interface ($r=1$):

\begin{equation}\label{eq:A31}
	\hat{u}_1 - \hat{u}_2 
	+ \left(\frac{\mathrm{d}U_1}{\mathrm{d}r} - \frac{\mathrm{d}U_2}{\mathrm{d}r}\right)\eta = 0,
\end{equation}

\begin{equation}\label{eq:A32}
	\hat{v}_1 - \hat{v}_2 = 0,
\end{equation}

\begin{equation}\label{eq:A33}
	\hat{w}_1 - \hat{w}_2 = 0,
\end{equation}

\begin{equation}\label{eq:A34}
	\begin{aligned}
		\frac{X}{\Rey}\frac{\partial \hat{u}_1}{\partial r}
		+ \hat{\tau}_{rz} 
		- \frac{N}{\Rey}\frac{\partial \hat{u}_2}{\partial r} 
		+ \frac{1}{\Rey}\!\left(
		\frac{\mathrm{d}^2 U_1}{\mathrm{d}r^2} 
		- N\frac{\mathrm{d}^2 U_2}{\mathrm{d}r^2}
		\right)\eta
		= k\Bigg[
		-\mathrm{i}\frac{X}{\Rey}\hat{v}_1 
		+ \mathrm{i}\frac{N}{\Rey}\hat{v}_2 \\
		+ 2\mathrm{i}(1-X)\El\!\left(
		\frac{\mathrm{d}U_1}{\mathrm{d}r}
		\right)^2 \eta
		\Bigg],
	\end{aligned}
\end{equation}
\begin{equation}\label{eq:A35}
	-\frac{2X}{\Rey}\frac{\partial \hat{v}_1}{\partial r} + \hat{p}_1 - \hat{\tau}_{rr}
	+ \frac{2N}{\Rey}\frac{\partial \hat{v}_2}{\partial r} - \hat{p}_2
	+ \frac{1}{\We}(1-n^2)\eta = \frac{k^2}{\We}\,\eta,
\end{equation}
\begin{equation}\label{eq:A36}
	\frac{X}{\Rey}\,\mathrm{i}n\hat{v}_1 + \frac{X}{\Rey}\frac{\partial \hat{w}_1}{\partial r}
	- \frac{X}{\Rey}\hat{w}_1 + \hat{\tau}_{r\theta}
	- \frac{N}{\Rey}\,\mathrm{i}n\hat{v}_2 - \frac{N}{\Rey}\frac{\partial \hat{w}_2}{\partial r}
	+ \frac{N}{\Rey}\hat{w}_2 = 0.
\end{equation}

At the outer interface ($r=a$), the boundary condition is
\begin{equation}\label{eq:A37}
	\frac{\partial \hat{u}_2}{\partial r} = \hat{v}_2 = \hat{w}_2 = \frac{\partial \hat{p}_2}{\partial r} = 0.
\end{equation}

\section{Numerical method and validation}\label{app:method_validation}

\begin{figure}
	\centerline{\includegraphics[width=\textwidth]{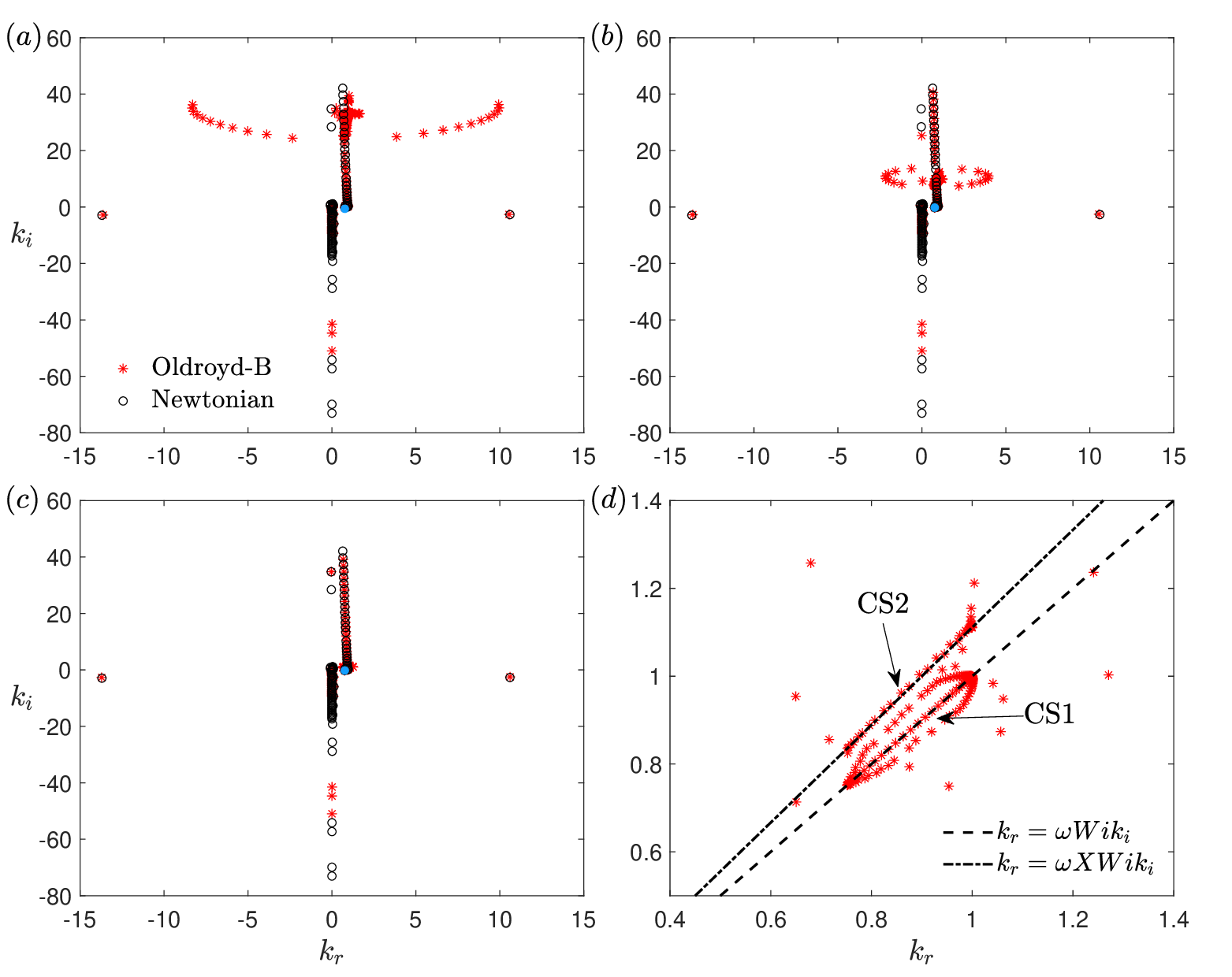}}
	\caption{Eigenvalue spectra of Oldroyd-B and Newtonian jets from spatial stability analysis at $\omega=1$, $n=0$, $\Rey=150$, $\We=7$, $U_{\!s}=1.33$, $K=1.2$, $Q=0.0013$, $N=0.0172$, $X=0.9$, and $a=11$, with $\Wi$ ranging from $10^{-2}$ to $10^{0}$. 
		($a$) $\Wi=0.03$, ($b$) $\Wi=0.1$, ($c$) $\Wi=1$, and ($d$) a magnified view of the spectrum at $\Wi=1$. 
		Blue markers indicate discrete eigenvalues obtained from the numerical spatial stability analysis. 
		The solid lines correspond to the analytically predicted continuous spectra (CS1 and CS2), as discussed in Appendix~\ref{app:eigenvalue spectrum}, given by $k_r = \omega \Wi\, k_i$ and $k_r = \omega X \Wi\, k_i$, with $U_1(r)\in[1,1.33]$ (note that $U_{\!s}=1.33$).}
	\label{fig:15}
\end{figure}

Equation~(\ref{eq:2.21}) involves 30 dependent variables due to the companion
linearization. In practice, however, the quadratic dependence on the wavenumber
$k$ arises only from a subset of variables. Therefore, the system can be
efficiently reduced by introducing only the auxiliary variables
\begin{equation}\label{eq:B1}
	\hat{F}_i = k\hat{u}_i,\quad
	\hat{G}_i = k\hat{v}_i,\quad
	\hat{H}_i = k\hat{w}_i,\quad
	\xi = k\eta,
\end{equation}
which reduces the number of dependent variables to 22.

The liquid region $r \in [0,1]$ is mapped onto the computational domain $y \in [-1,1]$ using a linear transformation
\begin{equation}\label{eq:B2}
	r = \frac{1+y}{2},
\end{equation}
whereas the gas region $r \in [1,a]$ is mapped via
\begin{equation}\label{eq:B3}
	r = \frac{y(1-a) + (1+a)}{2}.
\end{equation}

For discretization, the Chebyshev--Gauss--Lobatto collocation method is adopted, with collocation points defined as
\[
y_j = \cos\left(\frac{j\pi}{N}\right), \quad j=0,1,\ldots,N.
\]
Through this discretization, the continuous eigenvalue problem is reduced to a finite-dimensional generalized eigenvalue problem of the form
\begin{equation}\label{eq:B4}
	[\boldsymbol{M}]\,\boldsymbol{s} = k [\boldsymbol{N}]\,\boldsymbol{s},
\end{equation}
where $[\boldsymbol{M}]$ and $[\boldsymbol{N}]$ are coefficient matrices assembled from the discretized governing equations and boundary conditions. 
The unknown vector $\boldsymbol{s}$ is expressed as
\[
\boldsymbol{s} = (\hat{u}_1,\hat{v}_1,\hat{w}_1,\hat{p}_1,\hat{\tau}_{rr},\hat{\tau}_{r\theta},\hat{\tau}_{rz},\hat{\tau}_{\theta \theta},\hat{\tau}_{\theta z},\hat{\tau}_{zz},\hat{F}_1,\hat{G}_1,\hat{H}_1,\hat{u}_2,\hat{v}_2,\hat{w}_2,\hat{p}_2,\hat{F}_2,\hat{G}_2,\hat{H}_2,\eta,\xi)^{\mathrm{T}}.
\]
The resulting coefficient matrices have dimensions of $(13N_1 + 7N_2 + 22) \times (13N_1 + 7N_2 + 22)$, 
where $N_1+1$ and $N_2+1$ denote the numbers of Gauss--Lobatto collocation points in the liquid and gas regions, respectively.

A MATLAB code has been developed to solve the generalized eigenvalue problem in equation~(\ref{eq:B4}). 
The use of augmented eigenvectors, combined with spectral discretization, inevitably generates spurious eigenvalues. 
These spurious modes are sensitive to the collocation point distribution and can be eliminated by systematically varying $N_1$ and $N_2$. 
Figure~\ref{fig:15} presents the eigenspectrum computed for $\Wi$ between $10^{-2}$ and $10^{-1}$, with the Newtonian counterpart under the same conditions included for reference. Although many modes with $k_i\!<\!0$ appear in figure~\ref{fig:15}($a$), they do not represent physical instabilities but rather the discretized representation of the continuous spectrum. 
Only the mode with $k=0.761086 - 214963\,\mathrm{i}$ (highlighted by the blue marker) exhibits instability, while the other modes correspond to upstream-propagating disturbances \citep{Ashpis1990,Tumin2003}. 
A more detailed discussion of eigenspectrum characteristics is provided in Appendix~\ref{app:eigenvalue spectrum}.

\begin{table}
	\begin{center}
		\begin{tabular}{cccc}
			$N_1$ & $N_2$ & $k_r$ & $-k_i$ \\[3pt]
			40 & 50 & 0.771812 & 0.211541 \\
			40 & 60 & 0.771812 & 0.211542 \\
			50 & 60 & 0.771812 & 0.211542 \\
			60 & 70 & 0.771812 & 0.211542 \\
		\end{tabular}
		\caption{Eigenvalues corresponding to different collocation points $N_1$ and $N_2$, where $\omega =1$, $n=0$, $\Rey=150$, $\We=7$, $\El=0.1$, $U_{\!s}=1.33$, $K=1.2$, $Q=0.0013$, $N=0.0172$, $X=0.9$, and $a=11$. 
			The results demonstrate convergence with respect to discretization.}
		\label{tab:4}
	\end{center}
\end{table}

\begin{table}
	\begin{center}
		\begin{tabular}{ccc}
			$a$ & $k_r$ & $-k_i$ \\[3pt]
			10 & 0.771812 & 0.211542 \\
			11 & 0.771812 & 0.211542 \\
			12 & 0.771812 & 0.211542 \\
			13 & 0.771812 & 0.211542 \\
		\end{tabular}
		\caption{Eigenvalues corresponding to different outer radius $a$, where $\omega =1$, $n=0$, $\Rey=150$, $\We=7$, $\El=0.1$, $U_{\!s}=1.33$, $K=1.2$, $Q=0.0013$, $N=0.0172$, $X=0.9$, $N_1=50$, and $N_2=60$. 
			The results indicate negligible sensitivity to $a$.}
		\label{tab:5}
	\end{center}
\end{table}

A convergence study has been performed by systematically varying the outer boundary radius~$a$ and the number of collocation points~$N_1$ and~$N_2$. The computed eigenvalues, including the real part of the wavenumber~$k_r$ and the spatial growth rate~$-k_i$, are presented in Tables~\ref{tab:4} and~\ref{tab:5}. It is found that the parameter set $(a, N_1, N_2) = (11, 50, 60)$ yields consistent values of both~$k_r$ and~$-k_i$ with accuracy exceeding three significant digits. This configuration is therefore adopted in all subsequent computations. 

\begin{figure}
	\centerline{\includegraphics[width=0.6\textwidth]{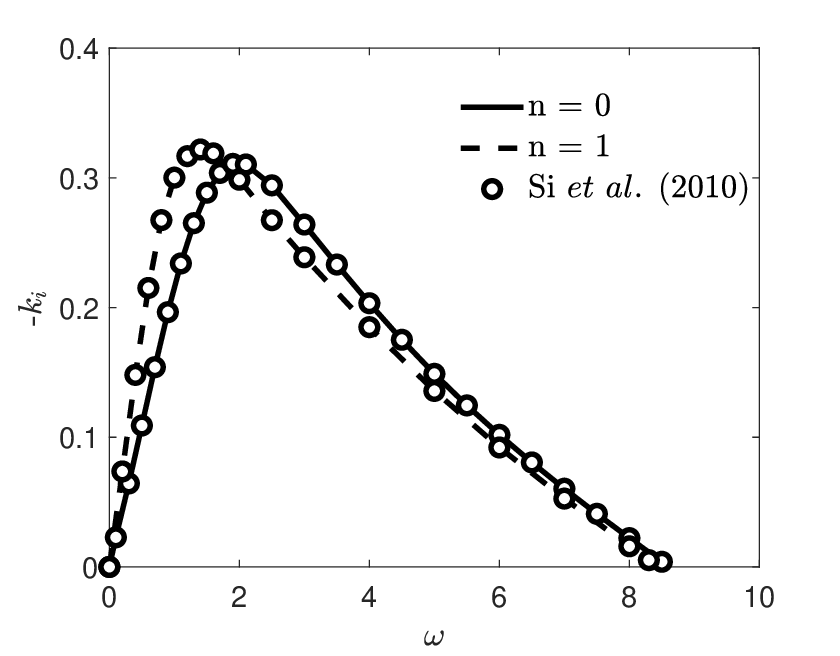}}
	\caption{Validation of the numerical code by comparison with \citet{Si2010} in the Newtonian limit ($\El \to 0$), where $\El=0.001$, $\Rey=200$, $\We=30$, $U_{\!s}=1.2$, $K=1$, $Q=0.0013$, $N=0.018$, $X=0.9$, and $a=11$.}
	\label{fig:16}
\end{figure}

To further validate the accuracy and reliability of the numerical implementation, comparisons are made with the results reported by \citet{Si2010}. As shown in figure~\ref{fig:16}, excellent agreement is observed in the Newtonian limit, demonstrating the correctness and robustness of the present numerical scheme.

\section{Specific forms of each term in the energy budget analysis}\label{app:energy_terms}

\begin{equation}
	\mathrm{KE} = \frac{1}{2} \int_{0}^{T} \int_{0}^{\lambda} \int_{0}^{2\pi} \int_{0}^{1} 
	\left( \partial_t + U_1 \partial_z \right) 
	\left(\tilde{u}_1^2 + \tilde{v}_1^2 + \tilde{w}_1^2 \right) 
	r \,\mathrm{d}r \,\mathrm{d}\theta \,\mathrm{d}z \,\mathrm{d}t ,
	\label{eq:KE}
\end{equation}
\begin{equation}
	\mathrm{REY} = -\int_{0}^{T} \int_{0}^{\lambda} \int_{0}^{2\pi} \int_{0}^{1} 
	\tilde{u}_1 \tilde{v}_1 \frac{\mathrm{d}U_1}{\mathrm{d}r} 
	\, r \,\mathrm{d}r \,\mathrm{d}\theta \,\mathrm{d}z \,\mathrm{d}t ,
	\label{eq:REY}
\end{equation}
\begin{equation}
	\mathrm{PRL} = \int_{0}^{T} \int_{0}^{2\pi} \int_{0}^{1} 
	\bigl[\tilde{p}_1 \tilde{v}_1 \bigr]_{z=0}^{z=\lambda} 
	\, r \,\mathrm{d}r \,\mathrm{d}\theta \,\mathrm{d}t ,
	\label{eq:PRL}
\end{equation}
\begin{equation}
	\mathrm{SHL} = \frac{X}{\Rey} \int_{0}^{T} \int_{0}^{2\pi} \int_{0}^{1} 
	\Biggl[ 
	\tilde{v}_1 \left(\frac{\partial \tilde{v}_1}{\partial z} + \frac{\partial \tilde{u}_1}{\partial r} \right) 
	+ \tilde{w}_1 \left(\frac{\partial \tilde{w}_1}{\partial z} + \frac{1}{r}\frac{\partial \tilde{u}_1}{\partial \theta} \right) 
	\Biggr]_{z=0}^{z=\lambda} 
	r \,\mathrm{d}r \,\mathrm{d}\theta \,\mathrm{d}t .
	\label{eq:SHL}
\end{equation}
\begin{equation}
	\mathrm{NVL} = \frac{X}{\Rey} \int_{0}^{T} \int_{0}^{2\pi} \int_{0}^{1} 
	\left[2\tilde{u}_1 \frac{\partial \tilde{u}_1}{\partial z}\right]_{z=0}^{z=\lambda} 
	r \,\mathrm{d}r \,\mathrm{d}\theta \,\mathrm{d}t ,
	\label{eq:NVL}
\end{equation}
\begin{equation}
	\mathrm{NEL} = \int_{0}^{T} \int_{0}^{2\pi} \int_{0}^{1} 
	\left[\tilde{u}_1 \tilde{\tau}_{zz}\right]_{z=0}^{z=\lambda} 
	r \,\mathrm{d}r \,\mathrm{d}\theta \,\mathrm{d}t ,
	\label{eq:NEL}
\end{equation}
\begin{equation}
	\mathrm{TEL} = \int_{0}^{T} \int_{0}^{2\pi} \int_{0}^{1} 
	\left[\tilde{v}_1 \tilde{\tau}_{rz} + \tilde{w}_1 \tilde{\tau}_{\theta z}\right]_{z=0}^{z=\lambda} 
	r \,\mathrm{d}r \,\mathrm{d}\theta \,\mathrm{d}t ,
	\label{eq:TEL}
\end{equation}
\begin{equation}
	\mathrm{SUT} = \frac{1}{\We} \int_{0}^{T} \int_{0}^{\lambda} \int_{0}^{2\pi} 
	\left[\tilde{v}_1 \left(1+\frac{\partial^2}{\partial \theta^2} + \frac{\partial^2}{\partial z^2} \right)\eta \right]_{r=1} 
	\mathrm{d}\theta \,\mathrm{d}z \,\mathrm{d}t ,
	\label{eq:SUT}
\end{equation}
\begin{equation}
	\mathrm{PRG} = - \int_{0}^{T} \int_{0}^{\lambda} \int_{0}^{2\pi} 
	\left[\tilde{p}_2 \tilde{v}_1\right]_{r=1} 
	\mathrm{d}\theta \,\mathrm{d}z \,\mathrm{d}t .
	\label{eq:PRG}
\end{equation}
\begin{equation}
	\mathrm{SHG} = \frac{N}{\Rey} \int_{0}^{T} \int_{0}^{\lambda} \int_{0}^{2\pi} 
	\left[\tilde{u}_1 \left(\frac{\partial \tilde{u}_2}{\partial r} + \frac{\partial \tilde{v}_2}{\partial z} \right)\right]_{r=1} 
	\,\mathrm{d}\theta \,\mathrm{d}z \,\mathrm{d}t ,
	\label{eq:SHG}
\end{equation}
\begin{equation}
	\mathrm{ATG} = \frac{N}{\Rey} \int_{0}^{T} \int_{0}^{\lambda} \int_{0}^{2\pi} 
	\left[\tilde{w}_1 \left(\frac{\partial \tilde{w}_2}{\partial r} + \frac{\partial \tilde{v}_2}{\partial \theta} - \tilde{w}_2 \right)\right]_{r=1} 
	\,\mathrm{d}\theta \,\mathrm{d}z \,\mathrm{d}t ,
	\label{eq:ATG}
\end{equation}
\begin{equation}
	\mathrm{NVG} = \frac{2N}{\Rey} \int_{0}^{T} \int_{0}^{\lambda} \int_{0}^{2\pi} 
	\left(\tilde{v}_1 \frac{\partial \tilde{v}_2}{\partial r}\right)_{r=1} 
	\,\mathrm{d}\theta \,\mathrm{d}z \,\mathrm{d}t ,
	\label{eq:NVG}
\end{equation}
\begin{equation}
	\mathrm{SHB} = \frac{1}{\Rey} \int_{0}^{T} \int_{0}^{\lambda} \int_{0}^{2\pi} 
	\left[\tilde{u}_1 \left(N \frac{\mathrm{d}^2 U_2}{\mathrm{d}r^2} - \frac{\mathrm{d}^2 U_1}{\mathrm{d}r^2}\right)\right]_{r=1} 
	\,\mathrm{d}\theta \,\mathrm{d}z \,\mathrm{d}t ,
	\label{eq:SHB}
\end{equation}
\begin{equation}
	\mathrm{NE} = \int_{0}^{T} \int_{0}^{\lambda} \int_{0}^{2\pi} 
	\frac{\partial \eta}{\partial z} \left(\tilde{u}_1 \Gamma_{zz}\right)_{r=1} 
	\,\mathrm{d}\theta \,\mathrm{d}z \,\mathrm{d}t ,
	\label{eq:NE}
\end{equation}
\begin{equation}
	\mathrm{DIS} = - \int_{0}^{T} \int_{V} 
	\boldsymbol{\tilde{T}}_{\!S} : \boldsymbol{\nabla} \boldsymbol{\tilde{V}}_1 
	\,\mathrm{d}V \,\mathrm{d}t ,
	\label{eq:DIS}
\end{equation}
\begin{equation}
	\mathrm{DIP} = - \int_{0}^{T} \int_{V} 
	\boldsymbol{\tilde{T}}_{\!P} : \boldsymbol{\nabla} \boldsymbol{\tilde{V}}_1 
	\,\mathrm{d}V \,\mathrm{d}t .
	\label{eq:DIP}
\end{equation}

\section{Eigenspectrum in spatial stability analysis}\label{app:eigenvalue spectrum}

We analyze the eigenspectra of Oldroyd-B jets obtained from spatial stability analysis, highlighting how viscoelasticity fundamentally alters the spectral structure relative to the Newtonian case. Figure~\ref{fig:15} shows the eigenspectra for $\Wi \in [10^{-2},10^{-1}]$, computed using the numerical scheme described in Appendix~\ref{app:method_validation}, together with the Newtonian spectrum under identical conditions for reference. At small elasticity, parts of the viscoelastic spectrum overlap with the Newtonian branches, while additional features emerge that are absent in the Newtonian case. As $\Wi$ (or equivalently $\El$) increases, the spectrum evolves into a semi-circular arc that gradually closes into a ring-like structure. With further increase in $\Wi$, the ring contracts and eventually collapses, disappearing once $\Wi$ reaches $O(1)$ or higher, as illustrated in figure~\ref{fig:15}($c$). A magnified view at $\Wi=1$ in figure~\ref{fig:15}($d$) reveals fine spectral details that are not visible in the full spectrum.  

A distinctive feature of the viscoelastic eigenspectrum is the presence of two continuous spectra (CS) near the center of the ring. These originate from the local formulation of polymeric stress in the Oldroyd-B constitutive model \citep{Chaudhary2021,Roy2022} and are associated with a singularity in the governing stability equation when the coefficient of the highest-order derivative vanishes, namely \([1 + \text{i} \Wi (k U_1 - \omega)] [1 + \text{i} X \Wi (k U_1 - \omega)]\) \citep{Chaudhary2021,Sterza2025}, with $U_1(r) \in [1,1.33]$ (note that $U_{\!s}=1.33$). Within the spatial framework, this singularity produces two lines in the $k_r$--$k_i$ plane: $k_r = \omega \Wi k_i$ and $k_r = \omega X \Wi k_i$, denoted as CS1 and CS2, respectively. CS1 persists even in the absence of solvent ($X=0$), whereas CS2 appears only when $X \neq 0$.  

Since $\Wi$ characterizes the elastic response of viscoelastic fluids \citep{Ding2021}, the ring structure observed for $\Wi \!<\! O(1)$ can be attributed primarily to the fluid’s linear elastic properties \citep{Ding2022}. In contrast, when $\Wi \!=\! O(1)$ or higher, nonlinear elastic effects dominate, leading to the collapse of the ring and fundamentally reshaping the eigenspectrum relative to the Newtonian case.

\section{Comparison with temporal instability analysis}\label{Comparison with temporal instability analysis}

\begin{figure}
	\centerline{\includegraphics[width=1\linewidth]{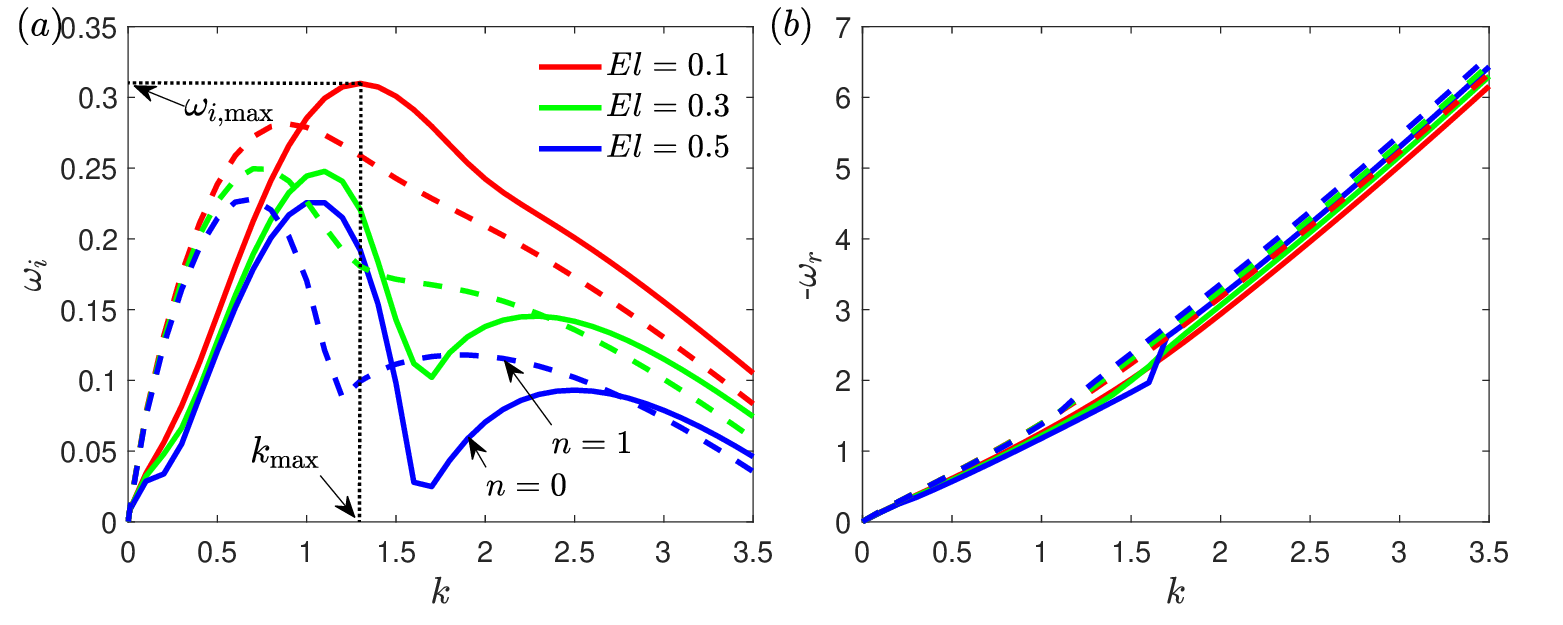}}
	\caption{\label{fig:17} 
		Dispersion relations from temporal instability analysis for different elasticity numbers $\El$, 
		at fixed parameters $\Rey = 150$, $\We = 8$, $U_{\!s} = 1.33$, $K = 1.2$, $Q = 0.0013$, $N = 0.0172$, and $X = 0.9$. 
		($a$) Temporal growth rate $\omega_i$ as a function of the axial wavenumber $k$; 
		($b$) frequency $-\omega_r$ as a function of $k$. 
		Solid lines denote the axisymmetric mode ($n=0$), while dashed lines denote the helical mode ($n=1$).}
\end{figure}

\begin{figure}
	\centerline{\includegraphics[width=1\linewidth]{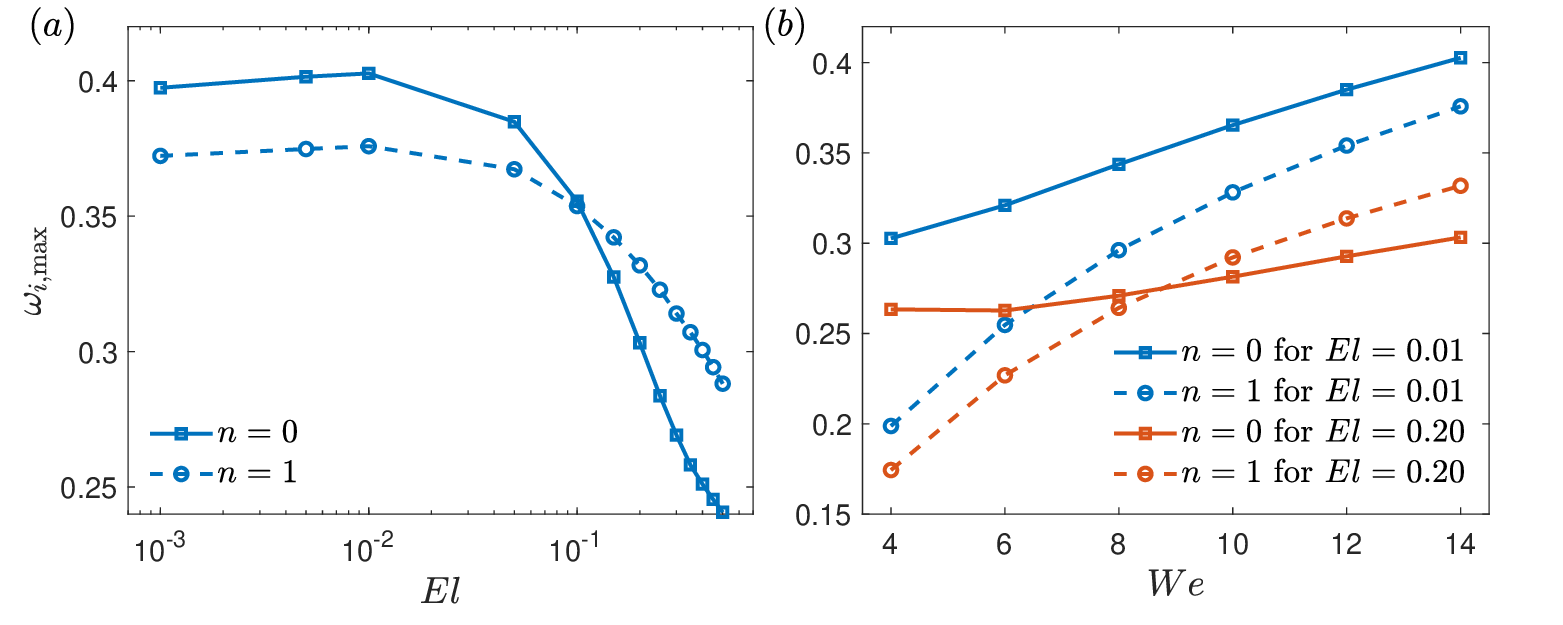}}
	\caption{\label{fig:18} 
		($a$) Variations of the maximum temporal growth rate $\omega_{i,\max}$ with elasticity number $\El$ at $\We = 10$; 
		($b$) Variations of $\omega_{i,\max}$ with Weber number $\We$ at $\El = 0.01$ and $\El = 0.2$. 
		All other parameters are the same as in the reference case.}
\end{figure}

We next compare our results with the temporal framework.
Figure~\ref{fig:17} shows the dispersion relations from temporal instability analysis: figure~\ref{fig:17}($a$) gives the temporal growth rate $\omega_i$ versus axial wavenumber $k$, and figure~\ref{fig:17}($b$) the corresponding real frequency $\omega_r$. Solid and dashed lines denote the axisymmetric ($n=0$) and the helical ($n=1$) modes, respectively. In the temporal stability analysis, the maximum growth rate $\omega_{i,\max}$ together with the corresponding most unstable wavenumber $k_{\max}$ characterize the dominant instability, where $\omega_{i,\max}$ reflects the amplification intensity and $k_{\max}$ determines the disturbance structure and associated droplet size. Figure~\ref{fig:17}($a$) further shows that as the elasticity number $\El$ increases, the temporal growth rates $\omega_i$ of both the axisymmetric and helical modes decrease noticeably, indicating that under stronger elasticity, the instability is suppressed. Meanwhile, the most unstable wavenumber $k_{\max}$ shifts toward the long-wave region, suggesting that stronger elasticity tends to favor longer-wavelength disturbances.
Figure~\ref{fig:17}($b$) shows that, similar to the spatial stability results, elasticity exerts a wavenumber-selective modulation on the phase speed $-\omega_r/k$: increasing elasticity reduces the phase speed in the low-wavenumber regime while enhancing it in the high-wavenumber regime, corresponding to the decrease at low frequencies and increase at high frequencies observed in the spatial analysis.

Figure~\ref{fig:18} further summarizes the variations of the maximum temporal growth rate $\omega_{i,\max}$ with elasticity number $\El$ and Weber number $\We$ under the reference conditions. As shown in figure~\ref{fig:18}$(a)$, with increasing $\El$, the temporal analysis also predicts a transition of the predominant mode from $n=0$ to $n=1$, qualitatively similar to the spatial results in figure~\ref{fig:5}($a$). Quantitatively, however, the crossover in the temporal framework occurs at significantly larger $\El$; correspondingly, the temporal transition curve in the $\We$--$\El$ plane is shifted to the right relative to the spatial boundary and the experiments (see figure~\ref{fig:11}). Regarding $\We$ (see figure~\ref{fig:18}$b$), the temporal instability analysis captures a $\We$-induced transition at $\El=0.2$, but at small $\El$ ($= 0.01$) it retains axisymmetric dominance across the explored $\We$ range, in contrast to the spatial instability analysis and experiments which show emergence of the $n=1$ mode beyond a finite $\We_c$.

\bibliographystyle{jfm}
\bibliography{jfm}



\end{document}